\documentclass[superscriptaddress,nofootinbib,11pt]{revtex4-1}

\usepackage{graphicx,amsmath,array,verbatim,setspace}
\usepackage{hyperref}       % hyperlinks
\usepackage{url}            % simple URL typesetting
\usepackage{amsfonts}       % blackboard math symbols
\usepackage[italicdiff]{physics}
\usepackage{color}
\usepackage{amssymb,amsthm}
\usepackage{tikz}
\usepackage[subrefformat=parens]{subcaption}
\usetikzlibrary{positioning}
\usetikzlibrary{shapes.geometric}
\usetikzlibrary{shapes.misc}
\usepackage{tikz-cd}

\allowdisplaybreaks

\newcommand{\be}{\begin{eqnarray}}
\newcommand{\ee}{\end{eqnarray}}
\newcommand{\nn}{\nonumber}
\newcommand{\nl}{\nonumber \\}
\newcommand{\pd}{\partial}

\newcommand{\bul}{\overset{\underset{\bullet}{}}}

\def\Z{{\mathbb Z}}

\def\coeff#1#2{{\textstyle {\frac {#1}{#2}}}}

\def\half{\coeff 12}

\definecolor{dgreen}{rgb}{0.0, 0.5, 0.0}
\definecolor{dred}{rgb}{0.82, 0.1, 0.26}
\definecolor{dblue}{rgb}{0.0, 0.0, 1.0}

\graphicspath{{./figures/}}

\begin{document}
\title{Exact-WKB analysis for SUSY and quantum deformed potentials:\\
Quantum mechanics with Grassmann fields and Wess-Zumino terms
}

\author{Syo Kamata}
\email{skamata11phys@gmail.com}
\affiliation{National Centre for Nuclear Research, 02-093 Warsaw, Poland}

\author{Tatsuhiro Misumi}
\email{misumi@phys.kindai.ac.jp}
\affiliation{Department of Physics, Kindai University,  Osaka 577-8502, Japan}
\affiliation{Hiyoshi Department of Physics, Keio University, Kanagawa 223-8521, Japan}

\author{Naohisa Sueishi}
\email{sueishi@eken.phys.nagoya-u.ac.jp}
\affiliation{Hiyoshi Department of Physics, Keio University, Kanagawa 223-8521, Japan}

\author{Mithat \"{U}nsal}
\email{unsal.mithat@gmail.com}
\affiliation{Department of Physics, North Carolina State University, Raleigh, NC 27607, USA}

\begin{abstract}
Quantum  deformed potentials arise naturally in quantum mechanical systems of 
one bosonic coordinate coupled to  $N_f$ Grassmann valued fermionic coordinates, or  to a    topological Wess-Zumino term.
These systems  decompose into sectors with a classical potential  plus a quantum deformation.  
%as in  supersymmetric  or QES systems.  
%We study  exact WKB  method for such  quantum  deformed polynomial potentials,  and focus on   double- and triple- well.  
Using exact WKB,  we derive  exact quantization condition 
and its median resummation. The solution of median resummed form gives physical 
Borel-Ecalle resummed results, as we show explicitly in quantum deformed double- and triple- well potentials. 
 Despite the fact that instantons have finite actions,   for  
 generic  quantum deformation,   they  do not contribute to the  energy spectrum 
 at leading order in  semi-classics.   For certain quantized quantum deformations, 
 where the alignment of levels to all order in perturbation theory occurs, 
 instantons 
 contribute  to  the spectrum. If deformation parameter is not properly quantized, 
 their effect  disappears, but higher order effects in semi-classics survive. 
 In this sense, we classify saddle contributions as fading  and  robust. 
 %depending on  quantization of deformation parameter.
 % under quantum deformation. 
% Solution of exact quantization condition also clarifies patterns of convergent/divergent expansion for states below a critical level and divergent expansion above that.  
 Finally, for 
quantum deformed  triple-well potential, we  demonstrate the P-NP  
relation, by computing  period integrals and Mellin transform.
%between perturbation theory around the two types of perturbative vacua and  the non-perturbative 
%bion configurations, by computing  period integrals and Mellin transform.

\end{abstract}

\maketitle

\tableofcontents

\newpage

%%%%%%%%%%%%%%%%%%%%%%%%%%%%%%%%%%%%%%%%%%%%%
\section{Introduction and background}
%%%%%%%%%%%%%%%%%%%%%%%%%%%%%%%%%%%%%%%%%%%%%

%Exact WKB method is an invaluable, yet not sufficiently appreciated  tool in quantum mechanics \cite{  BPV, Voros1983,Silverstone,  DDP2,DP1, Takei1,Takei2, Takei3,Kawai1,AKT1, Schafke1,  Iwaki1}.  
WKB method underwent a silent transformation, starting around 80s, turning it from  an
approximation with  drawbacks  concerning connection problem  \cite{Berry, Silverstone} to a rigorous formalism  for both mathematicians and physicists alike \cite{  BPV, Voros1983,Silverstone,  DDP2,DP1, 
Takei1,Takei2, Takei3,Kawai1,AKT1, Schafke1,  Iwaki1, 
Alvarez3, Zinn-Justin:2004vcw, Zinn-Justin:2004qzw,  Dunne:2013ada, Dunne:2014bca}. 
The new version, called  
exact WKB (sometimes  called resurgent WKB)   is an indispensable tool, not only in quantum mechanics, but  also in diverse parts of theoretical physics and mathematics.  
The method has  deep connections with 4d  ${\cal N}=2$ gauge theories in the contexts of 
$\Omega$-deformation \cite{Nekrasov:2009rc,Mironov:2009uv} and wall-crossing phenomena  \cite{Gaiotto:2012rg}, as well as with  ODE/IM correspondence \cite{Dorey:2001uw, Dorey:2007zx},  
topological  string theory,   see e.g.  \cite{
  Grassi:2014cla,Grassi:2014zfa,  Kashani-Poor:2015pca,  Kashani-Poor:2016edc,Ashok:2016yxz,Ito:2018eon,Hollands:2019wbr,Ashok:2019gee,Ito:2019jio,Imaizumi:2020fxf,Coman:2020qgf,Allegretti:2020dyt,Kuwagaki:2020pry, Emery:2020qqu,Enomoto:2020xlf,Taya:2020dco,Yan:2020kkb,Duan:2018dvj}. The common tread of all these different examples is the quantization of an underlying classical curve. 

The exact WKB method is combination of  two parts, topological   and asymptotic analysis.  
The construction  starts with the  complexification of  the coordinate space and then uses the classical energy conservation relation $p^2 = 2(E-V(x))$ to turn the  $x \in \mathbb C$  plane  into a Stokes graph.   
The exact quantization condition is  determined  by the topology of the  Stokes graph, and 
 remains invariant  under variations 
of the parameters and energy,  as long as one does not  encounter a topology change in Stokes graph. 
%The construction  starts with the  complexification of  the coordinate space and then  uses the classical potential data  to turn the complex  $x \in \mathbb C$  plane  into a Stokes graph.  
%Stokes graph is a network of Airy or Weber type building blocks, and cuts.  The  exact quantization only depends on the topological aspects of the Stokes graph, and remains invariant  under variations of of the parameters and energy,  as long as one does not  encounter a Stokes line. 
%\footnote{ The exact quantization condition can change only if the topology of the Stokes graph changes.  Considering Stokes graph as a function of energy, it is piece-wise smooth. It only changes abruptly exactly at Stokes lines. }
The computational part of the problem  starts with  the determination of  the WKB wavefunction, which is an asymptotic series, obtained recursively from the  non-linear Riccati equation. 
%which in turn is  obtained from Schrodinger equation.  
From this asymptotic series, one can obtain   the  expressions for 
 perturbative/non-perturbative cycles  (Voros multipliers), which are also   asymptotic.  For a streamlined   introduction, see \cite{Kawai1}, for  explicit construction and calculation  in a physical problem, see \cite{DDP2,Sueishi:2020rug,Sueishi:2021xti}. 
% The analytic  part of the story is the   calculations of the Voros symbols by using all orders  asymptotic expansion of WKB wave-functions. 
 The calculations  can become tedious, however, at the end of the day, 
 they carry  tremendous  merit which justifies  the  hard work.  The final result of the formalism captures all perturbative, non-perturbative phenomena for all energy levels,  large-order/low order resurgence,  low-order-low-order constructive form of the resurgence and fairly 
 non-trivial  aspects of the spectrum at once.

In our earlier works, we considered various  quantum mechanical systems with  classical polynomial and periodic  potentials, constructed exact quantization conditions,  and  
provided a dictionary between the  Airy-type and  degenerate Weber-type building blocks 
of Stokes graphs, which is important for explicit computations \cite{Sueishi:2020rug,Sueishi:2021xti}.  
In this work, we consider systems where  potential is modified by an $O(\hbar)$ quantum deformation. These systems are not rare, and they are physically interesting.    
Consider quantum mechanics of systems with  a bosonic field $x(\tau)$ and $N_f$ Grassmann valued field $\psi_i(\tau)$ (generalization of  supersymmetric QM   \cite{Witten:1982df}) , or  the systems in which   bosonic field $x(\tau)$ is coupled to Wess-Zumino term with  a spin-$S$ coupling \cite{Stone:1988fu}. 
Such generalization appeared in \cite{Balitsky:1985in, Aoyama:2001ca, Behtash:2015loa}.
For  $N_f =2,3, \ldots $ and general half-integer and integer $S$,  these    are related to quasi-exactly solvable systems \cite{Turbiner:1987nw}.  
In the first case, by diagonalizing fermion number, and in the second case, by diagonalizing spin coupling, we reach to a collection of graded Hamiltonians. For example, for supersymmetric theory,  we end up with two Hamiltionians, in the fermion number zero and one sectors, and in $N_f$-flavor theory, a collection of $2^{N_f}$ Hamiltonians.   
One can think of each Hamiltonian in   this decomposition in terms of a potential, 
\be
V(x,\hbar) = \underbrace{ V_0(x)}_{\rm classical} + \underbrace{\hbar p  V_1(x)}_{\rm quantum \;  deformation} \equiv  
\frac{1}{2} W'(x)^2 + \frac{\hbar p}{2}  W''(x), % \qquad p \in {\mathbb R}.
\label{cq} 
\ee
where the first term  $V_0(x)$ is {\it classical} and the second term $\hbar V_1(x)$  is {\it quantum} induced. The quantum potentials that arise from $N_f$-Grassmann field theories, $p$ takes values in a  quantized  discrete set.  We will also  analytically continue the coefficient  $p$ into ${\mathbb R}$.  There are many interesting phenomena that arise from this continuation. It is our goal to investigate this interesting class of quantum deformed potentials via exact-WKB method.\footnote{We assume $p$ is fixed in the   $\hbar \rightarrow 0$ limit, so that the quantum deformation always remains $O(\hbar)$ relative to classical potential. If one wishes to take a scaling limit  $p \hbar \sim O(\hbar^0)$, then $\hbar p  V_1(x)$  becomes classical. 
Even the basic Stokes graphs have different geometry.   In this work, we do not consider this scaling  limit.}

These systems provide useful prototypes to semi-classically calculable  instanton analysis in QFTs with a bosonic 
and multiple fermionic fields  in diverse dimensions, if instanton size moduli is kept under control in some way 
\cite{Unsal:2007vu,Unsal:2007jx,Shifman:2008ja,Poppitz:2009uq,Anber:2011de,Poppitz:2012sw,Misumi:2014raa,Fujimori:2019skd,Misumi:2019upg,Fujimori:2020zka, Dorigoni:2017smz, 
Demulder:2016mja, Schepers:2020ehn}.
In the asymptotically free theories such as 2d sigma models and 4d gauge theories, taking control of the   instanton size moduli by  circle compactification, twisted boundary conditions or Higgsing,  the semi-classical  analysis  become similar to quantum mechanical systems.   
Furthermore, we now understand that there are compactification of QFTs with 't Hooft fluxes  to QM systems which preserve various anomalies and non-perturbative saddle points, see eg. \cite{Unsal:2020yeh}.  In these theories, integrating out fermions in a weak coupling calculable regime induces an $O(\hbar)$ deformation to classical action, similar to our quantum deformed system.  As such,  deeper understanding of  the quantum deformed potentials     via exact WKB should be useful more broadly. 
%Also, the exact WKB is extremely useful in the context 

From the exact WKB analysis, we construct exact quantization conditions at energies  below  the  barrier.
These  remain  invariant under  DDP formula \cite{DDP2, DP1}, which is same as reality of resurgent transseries \cite{Aniceto:2013fka}.  Next, we point out that the 
 median resummation of exact quantization condition is immensely useful to extract physical results.  In the latter form, all resurgent cancellations takes place at the level of median resummation of the quantization condition.   This idea is  used by Pham et.al.\cite{DDP2, DP1} to show the reality of energy spectrum, but not to calculate the spectrum.  We take their analysis one step beyond, and solve for the non-perturbative contributions, which are automatically in physical form, free of all  ambiguities. This is one of the main technical achievements of our analysis. 

The solutions of median resummed   quantization condition reveals a number of nice and  intriguing physical results. 
The systems we study have finite action instantons and  bions\footnote{
%\textcolor{red}{
In quantum deformed potentials, bions appears as exact Euclidian solutions. Even for the undeformed potential, the partition function include bion contributions because of the periodic boundary condition. This is why we have to consider bion contributions in the quantum theories with multiple classical vacua.
%}
}
\cite{Unsal:2007vu,Unsal:2007jx,Shifman:2008ja,Poppitz:2009uq,Anber:2011de,Poppitz:2012sw,Misumi:2014raa,Fujimori:2019skd,Misumi:2019upg,Fujimori:2020zka}. 
Despite that, for certain quantized deformation parameters,  part of the spectrum does not receive any  non-perturbative contribution at all and perturbation theory converges and captures all (e.g. algebraically solvable states in QES and SUSY systems).  In certain cases, there is a convergent/divergent alternation in the perturbative  spectrum of low lying states.    
If the deformation parameter is not properly quantized, instantons end up not contributing to energy levels at leading order in semi-classics $e^{-S_{\rm I}/\hbar}$, and leading non-perturbative contributions are 
due to bions $e^{-S_{\rm B}/\hbar} \sim e^{-2S_{\rm I}/\hbar}$ and they are robust.  
The emergence of the traditional low-order/high order resurgence relations 
and all orders resurgent cancellations via the DDP formula \cite{DDP2, DP1} is also captured  by the formalism. 
More profoundly,  we used asymptotic expansion of  WKB wave functions,  quantum integrals over perturbative and non-perturbative  cycles, and 
Mellin transform  to prove that 
 perturbation theory around perturbative vacuum for generic level number $k$ and generic $p$ determines  all perturbative and non-perturbative information around non-perturbative bion saddles. This generalize  P-NP relation between perturbation theory and instanton sectors 
 \cite{Alvarez1, Alvarez2, Alvarez3, Dunne:2013ada,Dunne:2014bca, Dunne:2016qix, Gahramanov:2015yxk, Raman:2020sgw}, and  perturbation theory and bion sectors \cite{Kozcaz:2016wvy,Dunne:2016jsr} to quantum deformed potentials. 
 This is a constructive low-order/low-order relation between different saddles in the problem and  a generalization of the P-NP  relations for  these systems.   
The  exact WKB formalism is capable to  address all these  phenomena  in a unified fashion.   

%Finally, let us point out that throughout this paper, we consider  exact WKB and  exact quantization conditions at energies below the barrier, such that  all turning points are real.   At the top of barrier, real turning points merge and above the barrier, they  become complex conjugate pairs. In this process,  the topology of the Stokes curves and the form of the exact quantization condition changes abruptly. We hope to return to this problem in future work, and describe the exact quantization conditions everywhere in the spectrum. 

%%%%%%%%%%%%%%%%%%%%%%%%%%%%%%%%%%%%%%%%%%%%%
\section{Goals and Structure of this paper}
%%%%%%%%%%%%%%%%%%%%%%%%%%%%%%%%%%%%%%%%%%%%%
\label{sec:GS}

In this section, we clarify the goal and the structure of this paper.
The goals of this paper includes

\begin{enumerate}

\item
to analyze the quantum mechanical systems with quantum deformed double- and triple-well potentials
to obtain exact quantization conditions, understand the resurgent structure, and drive the median-resumed (ambiguity-cancelled) physical quantities,

\item
to show the striking fact that the instantons with real and finite actions do not contribute to the energy spectrum in some cases depending on the deformation parameter,

\item
to obtain a perturbative/non-perturbative (P-NP) relation for the quantum deformed triple-well quantum mechanics.

\end{enumerate}

Since the Hamiltonians we focus on originate in the quantum mechanics including both bosonic and fermionic degrees of freedom,
the above goals have significant implications for many fields as long as quantum mechanics is involved in them.
For example, once we obtain an exact quantization condition for a certain quantum mechanical system, we can derive an exact energy spectrum and an exact partition function. 
We will show that one can obtain the exact quantization condition of quantum-deformed double- and triple-well systems by the exact-WKB analysis.
We note that the eaact quantization condition for the triple-well system will be derived for the first time.
Although we do not perform the numerical calculation to confirm our results, we check that these results are consistent with those obtained in the literature.

Now, let us introduce the structure of this paper:
In Sec.III, we consider QM with Grassmann field and QM with topological Wess-Zumino term and explain how these theories relate to QM with the quantum deformed $N$-tuple well potential.
In Sec.IV, in order to solve the puzzling of the energy spectra described above and their $p$-dependence, we introduce the exact WKB analysis and obtain the quantization condition by the cycle (Voros multiplier) expression.
In this paper, in particular, we consider the cases of double-well and triple-well potentials as examples.
In Sec.V, we solve the quantization condition and investigate the structure of the energy spectra from the aspect of Borel resummability.
We also derive the P-NP relation and describe the connection to the path integral expression.
In Sec.VI, we consider the median resummation of the quantization condition through the perturbative-non-perturbative relationship, i.e. the DDP formula, and derive the asymptotic solution of energy spectra without imaginary ambiguities.
The results provide the solution of the puzzling.
Sec.VII is devoted to the conclusions and prospects.

%%%%%%%%%%%%%%%%%%%%%%%%%%%%%%%%%%%%%%%%%%%%%
\section{QM  with multiple  Grassmann fields  vs.  Wess-Zumino terms}
%%%%%%%%%%%%%%%%%%%%%%%%%%%%%%%%%%%%%%%%%%%%%

\label{sec:Grassmann-WZ}

We consider  two types of quantum mechanical systems.  
\begin{itemize} 
\item A bosonic field $x(t)$ and $N_f$ Grassmann valued field.  
\item A bosonic field $x(t)$ coupled to topological  spin-$S$    Wess-Zumino term.  
\end{itemize}
For $N_f=1$  and Wess-Zumino with $S=\half$ \cite{Stone:1988fu, Altland:2006si},   the systems we are considering is 
${\cal N}=1 $ supersymmetric quantum mechanics.   For  $N_f = 2, 3, \ldots$ and $S=1, \frac{3}{2}, \ldots,$  they are related to quasi-exactly solvable systems \cite{Turbiner:1987nw}. 
We construct equivalent physical systems by using these two descriptions, and map them 
  to quantum modified potentials  \cite{Behtash:2015loa}.

%%%%%%%%%%%%%%%%%%%%%%%%%%%%%%%%%%%%%%%%%%%%%%%%%%%%%%%%%%%%%%%%%%%%%%%%%%%
\subsection{QM with  $N_f$-Grassmann field}
%%%%%%%%%%%%%%%%%%%%%%%%%%%%%%%%%%%%%%%%%%%%%%%%%%%%%%%%%%%%%%%%%%%%%%%%%%
The  Euclidean action for the first formulation is given by \cite{Behtash:2015loa} 
 \begin{align}
  S=    \frac{1}{g}    \int d\tau  \Big( \half \dot x^2  +  \half (W')^2 
          +   \bar \psi_i (\partial_\tau  + W'') \psi_i  \Big )   \qquad  i=1, \ldots, N_f   
\label{useful}
\end{align} 
Below, we follow \cite{Behtash:2015loa, Aoyama:2001ca} to turn this system into a collection of 
quantum deformed potentials.\footnote{Another way to obtain quantum deformed potentials is 
to modify the Yukawa interaction of the $N_f =1$ flavor theory as  $W'' \bar \psi  \psi  
\rightarrow  p W'' \bar \psi   \psi$ \cite{Balitsky:1985in,Verbaarschot:1990ga}.}  
 Let us consider either the thermal or   graded partition functions: 
 \begin{align}
\label{eq:tpf}
Z(\beta) = {\rm Tr}_{\cal H}   e^{- \beta H}, \qquad  \qquad 
\tilde {Z}(L) \equiv& {\rm Tr}_{\cal H}  (-1)^F e^{-L H}    . 
\end{align}
For the  thermal partition function in path integral formulation, one imposes 
 anti-periodic boundary conditions for Grassmann valued fields.     
 The insertion of  $ (-1)^F $ converts the anti-periodic boundary conditions to periodic boundary conditions, hence, for graded partition function, one 
 needs to use periodic boundary conditions. The integrations over the fermions can be done exactly in these QM systems:
 \begin{align}
&  Z_{\mp} \equiv \int Dx \prod_{i=1}^{N_f}   D\bar \psi_i D\psi_i   \; 
        e^{ -\int d\tau \left( \half \dot x^2  +  \half (W')^2 
          +   \bar \psi_i (\partial_\tau  + W'') \psi_i  \right)}  \cr
  &= \int Dx   \; e^{ -\int d\tau {\cal L}_{\rm bos} }  \;  
          \Big[{\rm det}_{\mp} (\partial_\tau  + W'')\Big]^{N_f}   \, .
\label{useful2}
\end{align} 
The fermionic determinant can be calculated exactly \cite{Gozzi:1983qxk}, and  
for anti-periodic and    periodic  boundary conditions, it gives 
\begin{align}	
[{\rm det}_{\mp} (\partial_\tau + W'')] =  \left\{ \begin{array}{cc}
  \Big[ 2 \cosh \Big( \half \int W'' d\tau \Big)\Big]  & \qquad  {\rm apbc} \cr \cr
  \Big[ 2 \sinh \Big( \half \int W'' d\tau \Big)\Big]  & \qquad  {\rm pbc}
  \end{array} \right.
\end{align}  
Therefore, we can express the  thermal   partition function  in the following useful  form: 
\begin{align}	
 {Z}(\beta)
 &= \int Dx \;   e^{ -	 \int d\tau   {\cal L}_{\rm bos} }   \; 
       \left[ 2 \cosh \half    \int d\tau  W''  \right]^{N_f}  \cr
 &= \int Dx \;   e^{ -	 \int d\tau   {\cal L}_{\rm bos} }  
      \sum_{k=0}^{N_f} {N_f  \choose k}  
          \left[e^{ \int d\tau  \frac{W''}{2} } \right]^k   
         \left[ e^{ -\int d\tau  \frac{W''}{2} } \right]^{N_f-k}   \cr
 &=  \sum_{k=0}^{N_f} {N_f  \choose k} 
    \int Dx \;   e^{ - \frac{1}{g} \int d\tau \left( \half \dot x^2  +  \half (W')^2    
     + (2k-N_f)  g \frac{W''}{2}  \right)}  
 \equiv  \sum_{k=0}^{N_f} {N_f  \choose k}  Z_k   
\label{decom}
\end{align}	
and  graded one as: 
\begin{align}	
 \widetilde {Z}(L) = \sum_{k=0}^{N_f} {N_f  \choose k} (-1)^{N_f -k}  Z_k
\label{decom2}
\end{align}	
Here, 
%\begin{align}	
%\tilde {Z}(L) 
% &= \int Dx \;   e^{ -	 \int d\tau   {\cal L}_{\rm bos} }   \; 
%       \left[ 2 \sinh \half    \int d\tau  W''  \right]^{N_f}  \cr
% &= \int Dx \;   e^{ -	 \int d\tau   {\cal L}_{\rm bos} }  
%      \sum_{k=0}^{N_f} {N_f  \choose k} (-1)^{N_f -k}   
%          \left[e^{ \int d\tau  \frac{W''}{2} } \right]^k   
%         \left[ e^{ -\int d\tau  \frac{W''}{2} } \right]^{N_f-k}   \cr
% &=  \sum_{k=0}^{N_f} {N_f  \choose k} (-1)^{N_f -k}   
%    \int Dx \;   e^{ - \frac{1}{g} \int d\tau \left( \half \dot x^2  +  \half (W')^2    
%     + (2k-N_f)  g \frac{W''}{2}  \right)} \cr 
% &= \sum_{k=0}^{N_f} {N_f  \choose k} (-1)^{N_f -k}  Z_k  
%\label{decom}
%\end{align}	
 $Z_k$ is the partition function for a  bosonic system with  quantum modified potentials 
\begin{align}
V(x)=  \half (W')^2    
     + (2k-N_f)  g \frac{W''}{2} 
    \label{c+q}
\end{align}
which involves the classical part $O(g^0)$ and a quantum induced part $O(g^1)$. 
 The decomposition \eqref{decom} implies that the 
Hamiltonian  ${\hat H}$  and the Hilbert space  ${\cal H}$ of the 
$N_f$ flavor theory  split as:
\begin{align} 
{\hat H} & = \bigoplus_{k=0}^{N_f}  {\rm deg} ( {\cal H}_{(N_f, k)})   
                  {\hat H}_{(N_f,k)}, \qquad    
{\rm deg} ( {\cal H}_{(N_f, k)}) = \textstyle{ {N_f \choose k}} \cr
{\cal H}&= \bigoplus_{k=0}^{N_f}  {\rm deg} ( {\cal H}_{(N_f, k)})   
            {\cal H}_{(N_f,k)},  
\label{directsum-0}
\end{align}
where  ${\rm deg} ( {\cal H}_{(N_f, k)}) $ is the degeneracy of a 
given sector, and $k$ acquires an interpretation as fermion number. 
 Under  fermion number modulo two, these sectors exhibit an alternating Bose-Fermi structure, similar to supersymmetric theory
\cite{Witten:1982df} 
\begin{align} 
   (-1)^F         {\cal H}_{(N_f,k)} = (-1)^k   {\cal H}_{(N_f,k)} 
\label{Bose-Fermi}
\end{align}

%%%%%%%%%%%%%%%%%%%%%%%%%%%%%%%%%%%%%%%%%%%%%%%%%%%%%%%%%%%%%%%%%%%%%%%%%%%
\subsection{QM with topological Wess-Zumino term}
\label{sec-WZ}
%%%%%%%%%%%%%%%%%%%%%%%%%%%%%%%%%%%%%%%%%%%%%%%%%%%%%%%%%%%%%%%%%%%%%%%%%%%
	%\vspace{0.3cm}	 
	
Let us now consider the following bosonic QM system coupled to WZ term following \cite{Stone:1988fu}. 
 \begin{align}
S=  \int d\tau  \frac{1}{g} \Big( \half \dot x^2  +  \half (W')^2   \Big)  +  \int  d\tau  \; W'' \;  S   \cos \theta  +  i S   \int  d\tau  \; (1- \cos \theta ) 
    \partial_\tau \phi  
\end{align}
where the last term is topological    Wess-Zumino  term (also called  Berry phase action),
see \cite{Altland:2006si} for an introduction. 
The partition function for this QM system is the one of quantum mechanical particle with spin-$S$ coupled to ``magnetic field" $W''$:
 \begin{align}
\label{eq:spf-2}
{Z}_S(\beta) = \int  Dx D(\cos \theta) D \phi  \; 
   e^{ - \frac{1}{g} \int  \big ( \half \dot x^2  +  \half (W')^2   \big)
    +  \int  S   W''  \cos \theta   +  i  S (1- \cos \theta ) \partial_\tau \phi } \;. 
\end{align}
Here, $\tau \in S^1_\beta$ parametrize the base space, and  $(\theta, \phi) \in S^2$  are target space coordinates corresponding to path integral representation of spin.

The path integral over $ \int D(\cos \theta) D \phi $  can be done exactly and provides a nice example of localization.  Consider 
 \begin{align}
\label{eq:spf-3}
{I}_S(\beta) = \int   D(\cos \theta) D \phi  \; 
   e^{   \int  S   W''  \cos \theta   +  i  S (1- \cos \theta ) \partial_\tau \phi } \;.  
\end{align}
Decomposing $\phi(\tau)$  to periodic part and  winding number valued in integers, we can write 
\begin{align} 
	\phi(\tau) = \underbrace{ {\widetilde \phi}(\tau) }_{\rm periodic} + \frac{2 \pi k}{\beta} \tau,  \qquad k \in \Z
\end{align}
Therefore, the path integral  \eqref{eq:spf-3} can be written as a sum over the winding sectors, and path integral over periodic fields:
\begin{align}
\label{eq:spf-4}
{I}_S(\beta) = \sum_{k \in \Z}   \int   D(\cos \theta) D  \widetilde   \phi  \; 
   e^{   \int  S   W''  \cos \theta   +  i  S  \int   \widetilde   \phi  \;  \partial_{\tau}  (1- \cos \theta )  + i S   \int \frac{2 \pi}{\beta}k  (1- \cos \theta )  } \;. 
\end{align}
In the middle term, we used an integration by parts. Now, we can perform  $ D  \widetilde   \phi $  integration exactly.  It gives a Dirac-delta constraint over the $\theta (\tau)$ field of the form $\delta(  S \partial_{\tau}  (1- \cos \theta )  ) \propto \delta(  \partial_{\tau}  \cos \theta   )  $.  This forces $\theta(\tau) =  \theta_0 $  to be time  independent.   Therefore,  the path integration \eqref{eq:spf-3} reduces to an ordinary integration:
\begin{align}
\label{eq:spf-4-ord}
{I}_S(\beta) = \sum_{k \in \Z}   \int   d(\cos \theta_0)    \;\;   e^{     S    \cos \theta_0  \int W''    + i    2 \pi S  k  (1- \cos \theta_0 )  } \;. 
\end{align}
Using Poisson resummation, we can convert the sum over winding number to something we momentarily  identify as  a spectral sum 
% (or sector sum over the Hilbert space): 
\begin{align}
\label{eq:Poisson}
 \sum_{k \in \Z}   e^{   i    2 \pi S  k  (1- \cos \theta_0 )  }  = \sum_{\widetilde m \in \Z}  \delta(S(1-\cos \theta_0) - \widetilde m)
\end{align}
Therefore, \eqref{eq:spf-4} can be written as: 
\begin{align}
\label{eq:spf-5}
{I}_S(\beta) = \sum_{\widetilde m \in \Z}   \int   d(\cos \theta_0)    \;\;   e^{    S  \cos \theta_0 \int W'' }   \delta(S(1-\cos \theta_0) - \widetilde m)  \;. 
\end{align}
Dirac delta function gives a non-vanishing contribution if and only if $ \cos \theta_0  =1 - \frac{ \widetilde m}{S}$.  Since  $ |\cos \theta_0 | \leq 1$, 
the number of  $\widetilde m $ contributing to the sum is  constrained and  finite.  In fact, there are only $(2S+1)$ terms  with  $ 0 \leq \widetilde m  \leq 2S$  contributing to the integral.  Defining  $m= \widetilde m  -S$, we can write \eqref{eq:spf-5} as 
 \begin{align}
\label{eq:spf-6}
{I}_S(\beta) = \sum_{ m =-S}^{+S}    e^{   m \int    W''  }   \;. 
\end{align}
and the partition function in the presence of the Wess-Zumino term takes the form: 
 \begin{align}
\label{eq:spf-7}
{Z}_S(\beta) =  \sum_{ m =-S}^{+S}   \;   \int  Dx   \; 
   e^{ - \frac{1}{g} \int  \big ( \half \dot x^2  +  \half (W')^2    +  m g  W''   \big) }
\end{align}

Clearly, $S=\half$ system is equivalent to $N_f=1$ theory \cite{Stone:1988fu}, which is   ${\cal N}=1$  supersymmetric quantum mechanics \cite{Witten:1982df}.  
For  
$N_f=2$, we have a  spin  $\half \otimes \half= 1\oplus 0$ particle,  
a spin-1, and a spin-0 particle, therefore, we need to consider $S=1$ and $S=0$ systems together. 	
In general,   the mapping between QM with $N_f$ Grassmann fields and QM with WZ terms is following.   
\begin{align}
 \underbrace{ \half \otimes \ldots \otimes \half}_{N_f}    =  \bigoplus_{S=S_{\rm min}}^{S_{\rm max}}  
   {\rm mult}(S) \; S \, ,
\label{spin}
\end{align} 
where  
\begin{align}  
S_{\rm max} = \frac{N_f}{2}, \qquad S_{\rm min} 
          =  \left\{ \begin{array} {ll} 0  \qquad  
 &  N_f{\rm \; even}\, ,  \cr
\half  \qquad  &  N_f{\rm \;odd}\, ,  \cr 
\end{array}\right. 
\end{align} 	 
and   the  multiplicity of the spin-$S$ sub-sector   is given by  
 \begin{align}
S = \frac{N_f}{2} -k, \qquad 	{\rm mult}(S) 
  =  \left\{ \begin{array} {ll}
    1  \qquad  &  k =0 \, , \cr
   {N_f \choose k} - {N_f \choose k-1}  \qquad  &  1 \leq k \leq  
                \lfloor \frac{N_f}{2} \rfloor \, . \cr 
    \end{array}\right.
\end{align} 
The relation between the  partition function graded according to fermion 
number and spin representations is given by 
\begin{align}
\label{eq:spin-ghs}
Z(\beta) \equiv  \sum_{k=0}^{N_f} {\rm deg}(  {\cal H}_k)  
        Z_k  
 = \sum_{S=S_{\rm min}}^{S_{\rm max}}    
    {\rm mult}(S) Z_S  \, . 
\end{align}

Our main point is simple:  The quantum mechanical systems with potential  of the form 
\begin{align}
V(x, g) = \half (W' (x))^2    + g  m   W'' (x)
\label{c+q2}
\end{align}
arise naturally by the  integration over Grassmannian fields $\psi_i (\tau)$ or integrating the spin fields.   The leading term is classical potential  and sub-leading one is quantum deformation. 
%These types of potentials are examples of quasi-exactly solvable systems. For QES systems,    a definite number of lowest lying states are algebraically  solvable provided $\exp( \pm W)$ is normalizable. 
 Since quantum deformed potentials are fairly general,  we generalize exact WKB
  analysis for systems with such potentials. 
%  analysis both  for the lowest lying states as well as higher states of such systems. 
 
\subsection{General remarks on SUSY, QES and in between} 
The quantum mechanical systems obtained above, either by coupling the system to  Grassmann valued fields or to a Wess-Zumino term, decompose to  systems with quantum deformed  potentials given in \eqref{cq}.  For certain discrete subsets of $p$, these systems are related to either ${\cal N}=1$ supersymmetric  (SUSY) \cite{Witten:1982df} or quasi-exactly solvable  (QES) quantum mechanics \cite{Turbiner:1987nw, Kozcaz:2016wvy}. 

Let us restrict our attention to a  properly quantized $p$, to a potential of the form \eqref{cq}.
The SUSY and QES  has the common feature that a subset of states 
are solvable to all orders in perturbation theory (denoted as  ${\cal H}_0$ in \eqref{split}), and a further subset of ${\cal H}_0$ is solvable algebraically.  
%such as ground state in quantum mechanics.  
In these systems, the Hilbert space split up as 
\begin{align}
    {\cal H}=  \underbrace{{\cal H}_0}_{\rm finite} \oplus \underbrace{{\cal H}_1}_{\rm infinite}
    \label{split}
\end{align}
 In general, ${\cal H}_0$ is composed of  finitely many states (only one state in SUSY case and $2,3, \ldots, M$ 
 states in QES  where 
$M$ is determined by $p$), and ${\cal H}_1$ is composed of infinitely many states. 
Which states in ${\cal H}_0$ are algebraically solvable depends on a criterion explained below. 
The states in 
${\cal H}_1$ are not solvable algebraically. All the states in   ${\cal H}$  can be treated in the framework of WKB, and asymptotic series for WKB wave-functions can be obtained as explained in the next section. 

The states in ${\cal H}_0$  may or may not be exactly solvable non-perturbatively.  The well-known criterion in SUSY QM which determines whether supersymmetry is broken or not   generalize to QES, and indicates whether a state is   algebraically solvable or not.  
Let us recall briefly the standard description in SUSY and then, state how it generalize to QES systems. 

\noindent
{\bf Unbroken vs. broken  SUSY =  QES vs. Pseudo-QES:} The ${\cal N}=1$ supersymmetric QM  is described by the supersymmetry  algebra 
\begin{align}
\{Q , Q^{\dagger}\} =2 H, \qquad \{Q, Q\} =0, \qquad [Q, H]= 0,
\end{align}
where $H$ is Hamiltonian and $Q$ is supercharge. 
A representation for supersymmetry generators is 
\begin{align}
Q= (i p + W'(x)) \psi_{-}, \qquad Q^{\dagger} = (-i p + W' (x)) \psi_{+}.
\end{align}
Supersymmetry  algebra implies  that the spectrum of supersymmetric theory is positive semidefinite,  
Spec$(H) \geq 0$, and every $E>0$ state  is Bose-Fermi paired.  
To all orders in perturbation theory,  there exists an $E=0$ state. 
However, there may or may not exists an $E=0$ state non-perturbatively 
depending on the nature superpotential $W(x)$ \cite{Witten:1982df}.  
 
Now, assume $E=0$ state  exists.  Supersymmetry algebra  and the saturation of lower bound imply 
$Q^{\dagger}    | \Psi_0 \rangle=  Q  | \Psi_0 \rangle =0$.  
 If ground state is of the form $|0\rangle \otimes |0\rangle_F$, the action of $Q$ 
is automatically zero because $\psi_{-} |0\rangle_F = 0$ and the action of $Q^{\dagger}$ can be zero iff   $(-i p + W' )\Psi_0=0$, which implies 
$ \Psi_0 (x) = e^{+ \frac{1}{\hbar} W(x)}$.    If ground state is of the form $|0\rangle \otimes |1\rangle_F$, then we find 
$ \Psi_0 (x) = e^{- \frac{1}{\hbar} W(x)}$.

For  polynomial $W(x)$, there are three  possibilities  concerning $e^{\pm \frac{1}{\hbar} W(x)}$, which can be used to show if a supersymmetric ground state exists. Exactly the same possibilities   will  be crucial  in QES as well in determination of  exact algebraic solvability of low lying states.  These are:
\begin{itemize} 
\item[(1)]  If $W(x)  \rightarrow   + \infty$ as $|x| \rightarrow \infty$, then $e^{- W(x)/\hbar }$  is normalizable,  $e^{+W(x)/\hbar }$ is not.  
\item[(2)]  If $W(x)  \rightarrow   - \infty$ as $|x| \rightarrow \infty$, then $e^{+W(x)/\hbar  }$  is normalizable,  $e^{-W(x)/\hbar}$ is not.   

{\bf Unbroken SUSY or QES:} If at least one of the  $e^{\pm W(x)/\hbar } \in L^2(\mathbb R)$, SUSY is unbroken 
and for QES, a subset of states in ${\cal H}_0$ are algebraically  solvable. These states can be written as  $P_i(x) e^{- W(x)/\hbar }, \;  i=1, \ldots, M$  where $P_i(x)$ are polynomials.

\item [(3)]  If $ \lim_{x \rightarrow + \infty} W(x)  =  - \lim_{x \rightarrow - \infty} W(x)   $, then neither one of the $e^{\pm W(x)/\hbar }$ is normalizable.  

{\bf Broken SUSY or pseudo-QES:} If    $e^{\pm W(x)/\hbar } \notin L^2(\mathbb R)$, SUSY is broken
and none of the  states in ${\cal H}_0$ are exactly solvable.   In both SUSY and QES, the states in 
${\cal H}_0$ are solvable perturbatively, but these are not normalizable, and there is no  non-perturbative algebraic solutions. This latter case is called pseudo-QES.  

\end{itemize} 

  For polynomial $W(x)$, the leading large $|x|$ behaviour determines the nature of states in 
  ${\cal H}_0$ and this is given by 
 \begin{align}
   W(x) =  \left\{ \begin{array}{ll}
      x^{2m} + \ldots,       &   |x| \rightarrow \infty, \qquad \rm unbroken \;  SUSY \; or \;  QES  \\
       x^{2m+1} + \ldots,      & |x| \rightarrow \infty, \qquad \rm broken \;  SUSY \; or \; pseudoQES
   \end{array}  \right\}
 \end{align}
Therefore,  
the double-well potential $ W(x) \sim  x^3+ \ldots $ is an example of second category and triple-well potential  
$ W(x) \sim  x^4 + \ldots$ is an example of first category.

There are a set of  puzzling facts about these systems that the exact WKB analysis  address successfully. 
In the  unbroken SUSY or   QES case, 
 algebraically solvable states in ${\cal H}_0$ do not receive instanton 
 $e^{-S_{\rm I}/\hbar}$  or bion  $e^{-S_{\rm B}/\hbar}$ 
contribution to their energies, despite the fact that both instantons and bions are finite action configurations   
that are present in the theory. 
  Yet, states in ${\cal H}_1$ receive either $e^{-S_{\rm I}/\hbar}$  and/or  $e^{-S_{\rm B}/\hbar}$ type contributions.  
In the broken SUSY and pseudo-QES cases, the states in ${\cal H}_0$ do not receive instanton $e^{-S_{\rm I}/\hbar}$,
but receive bion  $e^{-S_{\rm B}/\hbar}$ contribution to their energies. 
We explain these  phenomena in terms of 
destructive/constructive interference associated with the  hidden topological angles \cite{Behtash:2015kna}. Higher states living in ${\cal H}_1$ receive either $e^{-S_{\rm I}/\hbar}$  and/or  $e^{-S_{\rm B}/\hbar}$ type contributions.  

If $p$ is generic real number, than these system are not part of SUSY or QES descriptions, but they are continuously connected to them.  There are no algebraically solvable states in the spectrum.   In such cases,  there is no instanton contribution anywhere in the spectrum and all NP contributions 
are due to bions.  The solutions of the exact quantization conditions explains all these phenomena at once.

%%%%%%%%%%%%%%%%%%%%%%%%%%%%%%%%%%%%%%%%%%%%%
\section{Exact-WKB analysis for quantum deformed potentials} \label{sec:EWKB_analysis}
%%%%%%%%%%%%%%%%%%%%%%%%%%%%%%%%%%%%%%%%%%%%%
As described in the previous section,  quantum mechanical systems with extra Grassmann fields or WZ terms naturally decompose into potential problems in which the potential is a combination of classical and quantum induced terms, see \eqref{c+q} and \eqref{c+q2}. Since $\hbar$ and $g$ are formally on the same footing, we can view these systems as Schr\"{o}dinger equation with a potential $V(x,\hbar)=V_0(x)+ \hbar V_1(x)$:\footnote{In the analysis below, we take $\hbar$ as the expansion parameter.% instead of $g$.
% The potentials (\ref{eq:def_W_double})(\ref{eq:def_W_triple} do not have the $g$-dependence.
 }
%In this section, we would review exact-WKB analysis. %based on the degenerate Weber (DW)-type Stokes graph.
%In our analysis, we start with the following Schr\"{o}dinger equation with a potential $V(x,\hbar)=V_0(x)+V_1(x)\hbar$, expressed as 
\be
\label{Sch1}
\left[ - \hbar^2 \frac{\pd^2}{\pd x^2} + Q(x,\hbar) \right] \psi(x,\hbar) = 0,
\ee
where $Q(x,\hbar)=\sum_{n=0}  Q_n(x)\hbar^n$  now includes both the classical  as well as a quantum induced terms in potential.  $Q(x,\hbar)$ is   given by
\be
Q_0(x) = 2(V_0(x)-E), \qquad Q_1(x) = 2V_1(x), \qquad Q_{n \ge 2}(x) = 0,
\ee
with the energy $E$. The fact that we only included $Q_1$ is motivated by the two physical applications 
described in \S. \ref{sec:Grassmann-WZ}.  
%\in {\mathbb R}_{> 0}$.
Here, the factor $2$ comes from $\hbar^2/2$ in the first term as a convention.  Note that for this application, we only need $\hbar Q_1$ as quantum deformation to exact-WKB analysis.  
In this section, we would like to describe   exact-WKB analysis for such  
quantum deformed potentials. \footnote{In Ref.\cite{Gaiotto:2021tsq}, the  exact  WKB analysis is called quantum deformed WKB.  The nature of exact WKB is already an asymptotic expansion in $\hbar$, and modern treatments of WKB analysis (that started in 80s) already takes into account the fact that WKB wave-functions, exponentials, periods are all asymptotic expansions in $\hbar$  \cite{DDP2,DP1, 
Silverstone, Voros1983, BPV, 
Takei1,Takei2, Takei3,Kawai1, Alvarez3, Zinn-Justin:2004vcw, Zinn-Justin:2004qzw,  Dunne:2013ada, Dunne:2014bca}.    We reserve quantum deformation for an addition of an  $O(\hbar)$  term to the  classical potential. As described above, such quantum deformed  potentials  appear naturally with the addition  of Grassmann fields or  Wess-Zumino terms to the classical Lagrangians.} 
\textcolor{black}{The below part for exact-WKB analysis is written based on \cite{Kawai1,Iwaki1,Sueishi:2021xti,DDP2,Takei3}.
See those Refs. for detail.
}

Consider  the ansatz for the WKB-wave function: 
\be
&& \psi_a (x,\hbar) = \sigma(\hbar) \exp \left[\int^x_a dx^\prime \, S(x^\prime,\hbar) \right], \qquad S(x,\hbar) = \sum_{n=-1}^\infty S_n(x) \hbar^n, \label{eq:wave_ansatz}
\ee
Substituting to \eqref{Sch1} gives the nonlinear Riccati equation, which can be mapped to a recursive equation for $S_n(x)$.  
\be
&& S(x,\hbar)^2 + \frac{\pd S(x,\hbar)}{\pd x} = \hbar^{-2}  Q(x) \label{eq:Scho_Q_S} \\
&\Rightarrow \quad& 
\begin{cases}
  S_{-1}(x) = \pm \sqrt{Q_0(x)} &   \\
  2 S_{-1}(x) S_{j}(x) + \sum_{k,\ell = 0}^{k+\ell=j-1} S_{k}(x) S_{\ell}(x) + \frac{d S_{j-1}(x)}{d x} = Q_{j+1}(x) & \mbox{for } \  j \ge 0
\end{cases}. \label{eq:Sg_Qgen}
\ee
The formal  power series solution of the wave function can ultimately be   obtained  from the $S_n(x)$.
$\sigma(\hbar)$ in Eq.(\ref{eq:wave_ansatz}) is a normalization factor, which is a  function independent of $x$.
Since the Schr\"{o}dinger equation is an second order equation, it gives two independent solution associated with the sign in  Eq.(\ref{eq:Sg_Qgen}).
Let us denote them $S^{(\pm)}(x,\hbar)=\sum_{n=-1}^{\infty}S^{(\pm)}_{n}(x)\hbar^n$ with $S^{(\pm)}_{-1}=\pm \sqrt{Q_0(x)}$.
Define $S_{\rm odd/even}(x,\hbar)$ as the formal expansion with $S_{\rm odd/even}(x,\hbar) = \sum_{n=-1}^\infty S_{{\rm odd/even},n}(x)\hbar^n$ given by 
%\footnote{``odd/even'' does not mean that $S_{\rm odd/even}(x,\hbar)$ has only odd/even powers of $\hbar$ when $Q_{n >0}(x) \ne 0$.}
\be
&& S_{\rm odd}(x,\hbar) := \frac{S^{(+)}(x,\hbar)-S^{(-)}(x,\hbar)}{2}, \qquad  S_{\rm even}(x,\hbar) := \frac{S^{(+)}(x,\hbar)+S^{(-)}(x,\hbar)}{2}.
\ee
Note that odd/even is the name for the way linear combination is formed. It does not imply that  
$S_{\rm odd/even}(x,\hbar)$ has only odd/even powers of $\hbar$ when $Q_{n >0}(x) \ne 0$, unlike the case of classical potentials for which only $Q_{0}(x) \ne 0$.  
From Eq.(\ref{eq:Scho_Q_S}), one can directly show that
\be
S_{\rm even}(x,\hbar)  = - \frac{1}{2}  \frac{\pd \log S_{\rm odd}(x,\hbar)}{\pd x},
\ee
and thus, the formal wavefunction $\psi_{a}(x,\hbar) = (\psi^+_a(x,\hbar),\psi^{-}_a(x,\hbar))^\top$ is obtained as
\be
\label{WKBWF}
\psi_a^{\pm}(x,\hbar) = \frac{\sigma_{\pm}(\hbar)}{\sqrt{S_{\rm odd}(x,\hbar)}} \exp \left[ \pm \int^x_a dx^\prime \, S_{\rm odd}(x^\prime,\hbar) \right].
\ee
The pre-factor 
$\sigma_{\pm}(\hbar)$ is chosen such that $\sigma_{\pm}(\hbar)=1$ when $a \in {\rm TP}$ where ${\rm TP}$ is a set of turning points determined by 
$Q_0(x)=0$. Note that the quantum deformation of the potential does not alter the Stokes graph or   $S_{-1}(x)$ which are dictated by classical data, 
but it feeds into all the higher order terms in  $S(x,\hbar)$, which is obtained by recursively solving \eqref{eq:Sg_Qgen}.

\begin{figure}[t]
  \begin{center}
    \begin{tabular}{cc}
      \begin{minipage}{0.4\hsize}
        \begin{center}
          \includegraphics[clip, width=55mm]{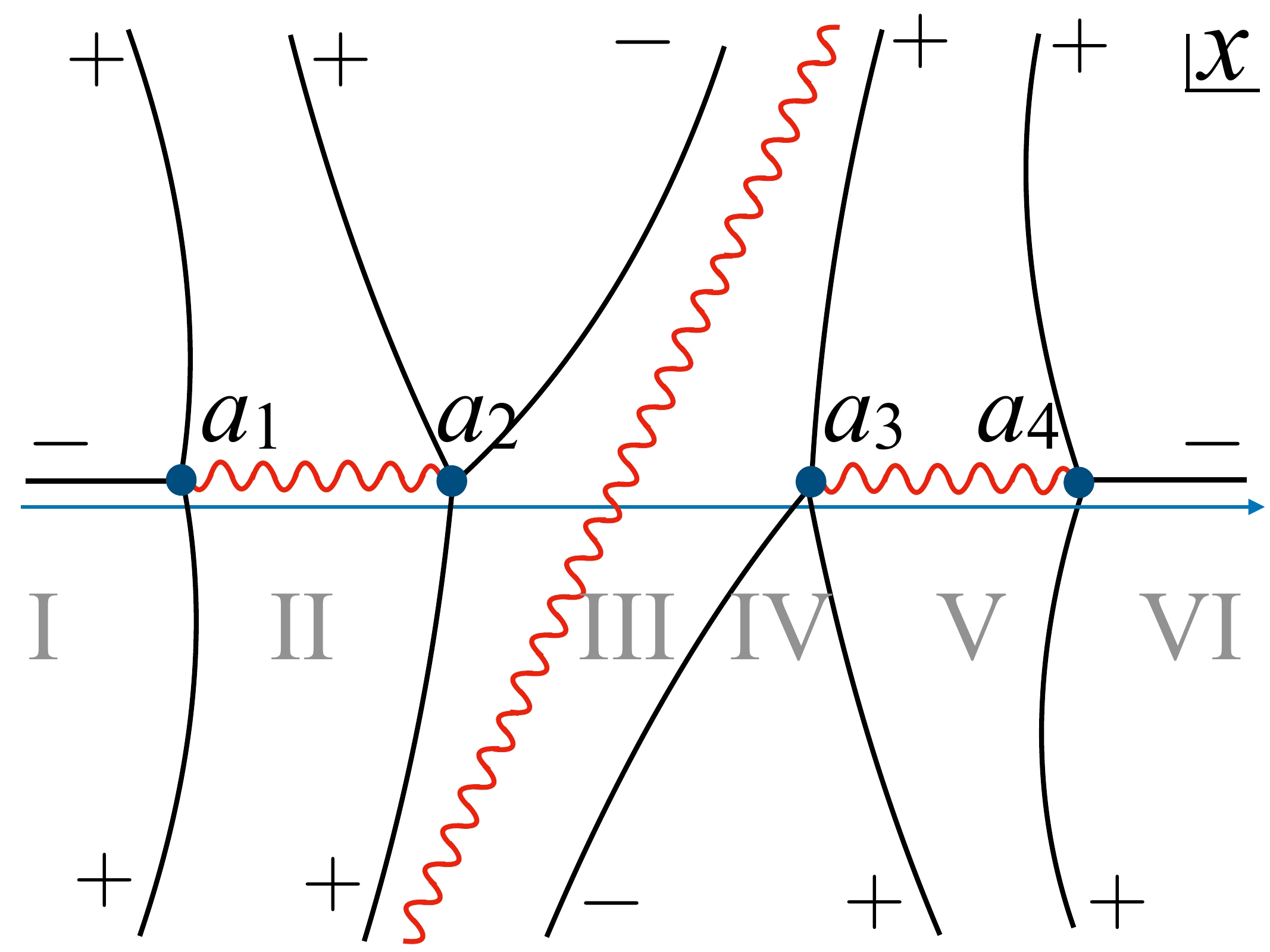}
          \hspace{1.6cm} 
        \end{center}
        \subcaption{Double-well (${\rm Im}\,\hbar>0$)}
      \end{minipage}   \quad \quad
      \begin{minipage}{0.4\hsize}
        \begin{center}
          \includegraphics[clip, width=55mm]{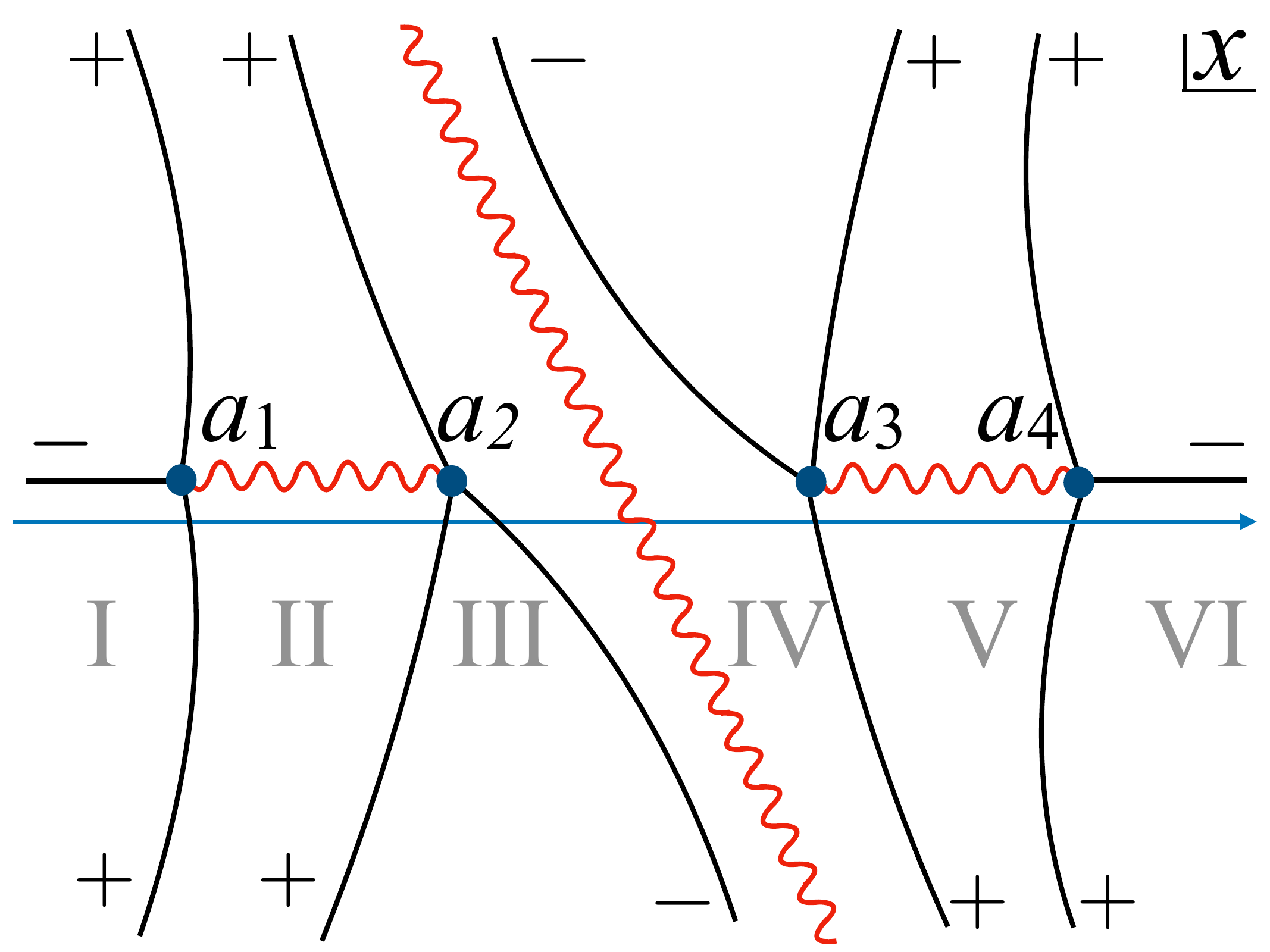}
          \hspace{1.6cm} 
        \end{center}
        \subcaption{Double-well (${\rm Im}\,\hbar<0$)}
      \end{minipage} \vspace{5.5mm} \\  
        \begin{minipage}{0.4\hsize}
        \begin{center}
          \includegraphics[clip, width=65mm]{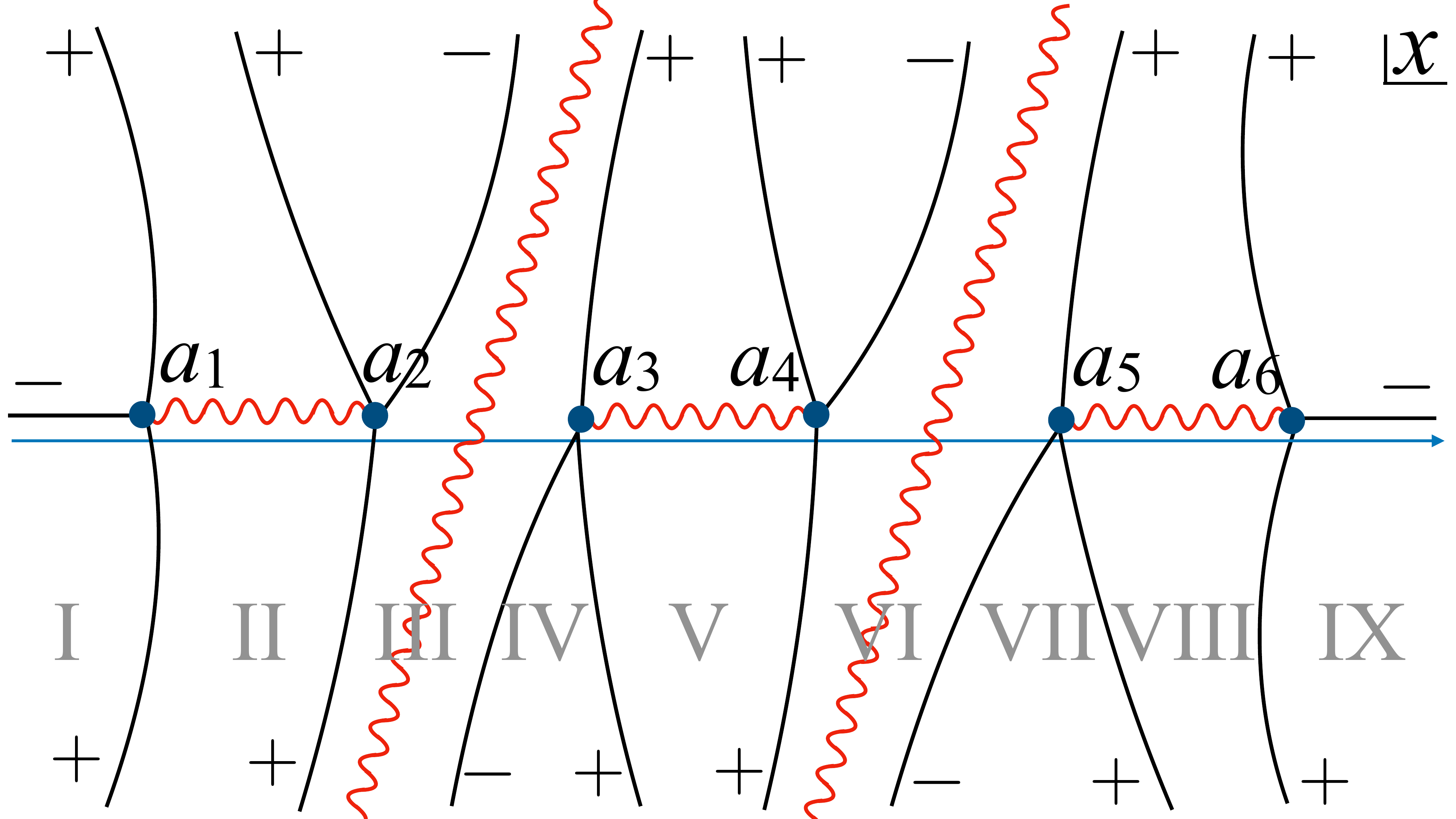}
          \hspace{1.6cm} 
        \end{center}
        \subcaption{Triple-well (${\rm Im}\,\hbar>0$)}
      \end{minipage}   \quad \quad
      \begin{minipage}{0.4\hsize}
        \begin{center}
          \includegraphics[clip, width=65mm]{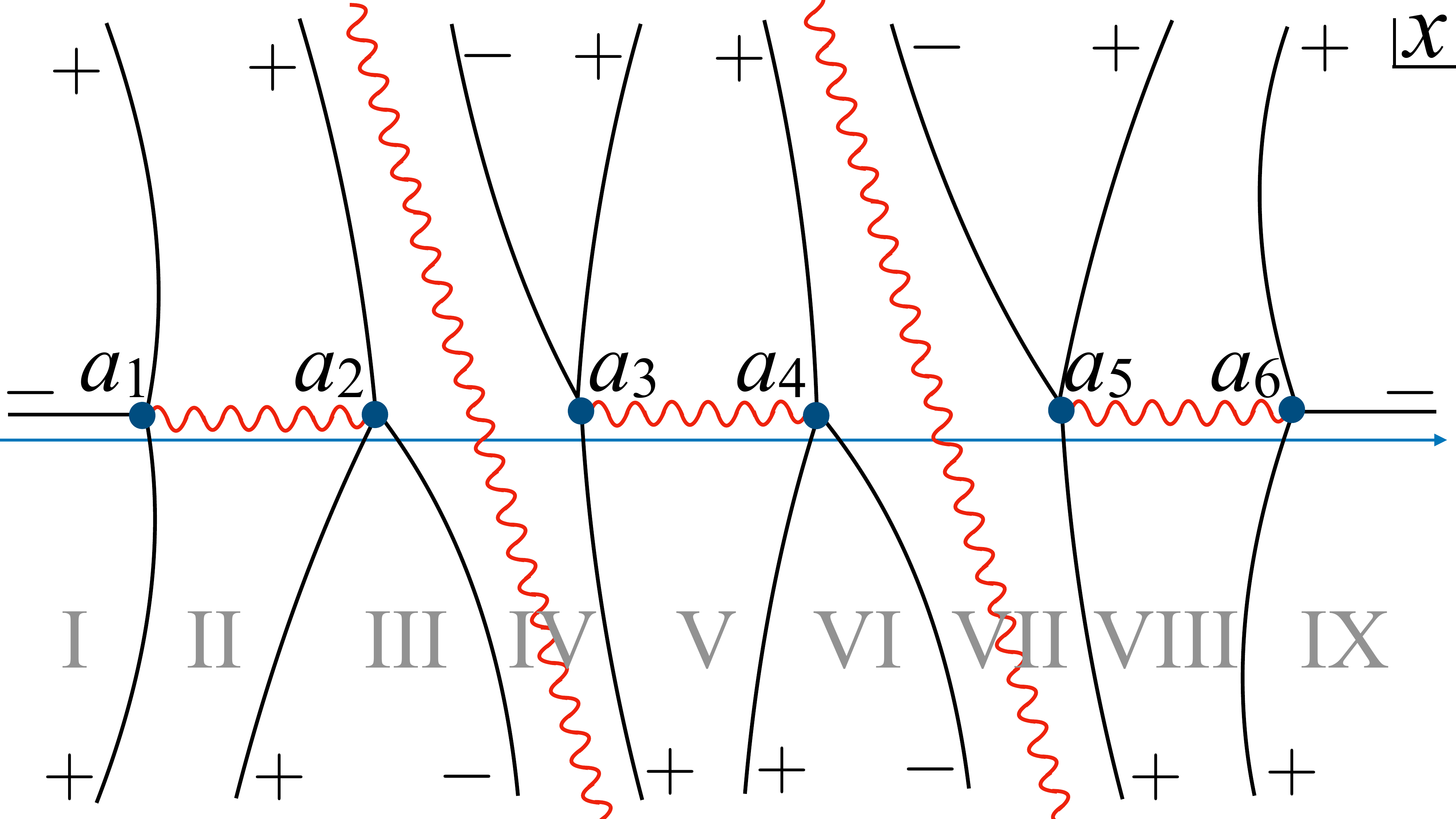}
          \hspace{1.6cm} 
        \end{center}
        \subcaption{Triple-well (${\rm Im}\,\hbar<0$)}
      \end{minipage}      
    \end{tabular} 
    \caption{Stoke graphs for the double-well and triple-well potentials. We 
    consider energy levels 
      less than the top of the potential barrier, so all turning points are real. 
    \textcolor{black}{
    The blue points labeled by $a_1, a_2, \cdots$ are turning points, and the three Stokes lines, denoted by solid black lines, emerge from each of them.
%    The symbols $\pm$ attached to the tip of each Stokes line denote the signs of asymptotic behavior of $S_{{\rm odd},-1}$ when taking the limit toward this direction for $x$.
%    Since this graph is defined as the first Riemann sheet, $\psi_+(\psi_-$) is divergent when taking the limit to the region labeled by $+(-)$ but vanishes when reaching to $-(+)$.
    The red wave lines indicate branch-cuts, where the roles of $\psi^{+}$ and $\psi^{-}$ exchange.
    The quantization condition is computed going along the blue colored trajectory drawn slightly below the real axis from the left to the right as taking into account of the Stokes lines crossing the trajectory. 
%    For the construction of the quantization condition ${\frak T}$, we go along the blue line taken from $x = -\infty$ to $x=+\infty$ but slightly below the real axis as taking into account of monodromy matrices associated with each Stokes line $M_\pm$, normalization matrices $N_{a_{n_1},a_{n_2}}$, and branch-cut matrices $T$.
}     
}
    \label{fig:Stokes_graph}
  \end{center}
\end{figure}

Whether the WKB wave function is Borel summable or not is dictated by the classical 
data of the problem, namely the classical potential  $V_0(x)$ or  $S_{\rm odd,-1}(x)$.  
So it is helpful to draw the Stokes graph for classical potential, see 
Fig.\ref{fig:Stokes_graph}. Note that quantum deformation does not enter to the determination of the Stokes graph. 
When $E=E(\hbar):=\sum_{n=0}^\infty E_{n} \hbar^n$ with finite $E_0$ in the $\hbar \rightarrow 0$ limit, the graph defined by $V_0(x)=W^\prime(x)^2/2$ locally constitutes of the Airy-type Stokes graph.
\textcolor{black}{
We define a set of turning points as 
\be
{\rm TP} := \{ x \in {\mathbb C} \ | \  Q_0(x) = 0 \},
\ee
and the Stokes curve associated with $a \in {\rm TP}$ is defined as
\begin{align}
	\Im \frac{1}{\hbar}\int_a^x \sqrt{Q_0(x)} dx=0 \,.
\end{align}
}
Clock-wise crossing a Stokes line emerging from a turning point $a$ can be expressed by $\psi_{{\rm I},a}(x,\hbar)=M_{\pm} \psi_{{\rm II},a}(x,\hbar)$, where 
\be
M_{+} = 
\begin{pmatrix}
  1 && i \\
  0 && 1
\end{pmatrix}, \qquad M_{-} = 
\begin{pmatrix}
  1 && 0 \\
  i && 1
\end{pmatrix}. \label{eq:Airy_conn}
\ee
Since the connection formula Eq.(\ref{eq:Airy_conn}) is locally defined around each turning point, one must change a normalization point before crossing a line emerging from other turning points.
This  is carried out by $\psi_{a_1}(x,\hbar)=N_{a_1,a_2}\psi_{a_2}(x,\hbar)$ with a normalization matrix $N_{a_1,a_2}$, where 
\be
N_{a_1,a_2} = 
\begin{pmatrix}
  e^{+\int_{a_1}^{a_2} dx \, S_{\rm odd}(x,\hbar)} && 0 \\
 0 &&    e^{-\int_{a_1}^{a_2} dx \, S_{\rm odd}(x,\hbar)} 
\end{pmatrix}, \qquad a_{1,2} \in {\rm TP}.
\ee
Additionally, branch-cuts also exist the Stokes graph.
These cuts are nothing but exchange of $\psi^{+}_a(x,\hbar)$ with  $\psi^{-}_a(x,\hbar)$, i.e. $+S_{\rm odd}(x,\hbar)$ with $-S_{\rm odd}(x,\hbar)$, hence, it can be realized by the branch-cut matrix $T$ given by
\be
T=
\begin{pmatrix}
 0 && - i \\
 -i && 0
\end{pmatrix}.
\ee

The quantization condition ${\frak D}(E,\hbar)$ is obtained from the analytic continuation from $x=-\infty$ to $x=+\infty$ and  the normalizability of  the  wavefunction.   Starting  
with a  WKB wavefunction that vanishes  
at $x=-\infty$, and using the rules of the passages through the Stokes lines, cuts, change of turning points as one proceeds to $x=+\infty$,  one obtains a linear combination of the two WKB wavefunctions \eqref{WKBWF}, an exponentially decaying and increasing wave function at $x=+\infty$. 
Demanding that wave function remains 
normalizable amounts to quantization condition. This amounts to setting the coefficient of the diverging WKB exponent at  $x=+\infty$ to zero.  

For ${\rm Arg}(\hbar)=0$, the WKB wave function is non-Borel summable since the theory is on a Stokes line.  
In order to  go around this obstacle, (resolve the Borel nonsummability for the wavefunction), we introduce an infinitesimally small complex phase for $\hbar$ as $\hbar = |\hbar| e^{i \theta}$ with $0<|\theta| \ll 1$.
Since the topology of the Stokes graph between positive and negative $\theta$ is different from each other, the difference in ${\frak D}(E,\hbar)$ by the positive and negative $\theta$ corresponds to the source of the imaginary ambiguity cancellations.
%\footnote{In general, some of  nonperturvative contributions entangled in the perturbative expansion by avoidance of singular points on the Borel plane does not always give a gap.[Costin, Asymptotics and Borel summability]}

For an $N$-tuple well potential with an energy $E$ less than any convexes of the potential,  
there are $2N$ real turning points.  We denote them as $a_{1} < \cdots < a_{2N} \in {\rm TP} \subset {\mathbb R}$.  Emanating from each turning point, there is a local Airy type building block, and Stokes graph can be viewed as  a network of these building blocks as shown in  Fig.\ref{fig:Stokes_graph} for $N=2, 3$.  
\textcolor{black}{
%The Stokes graph consists of three objects: turnninig points, Stokes lines, and branch-cuts.
The symbols $\pm$ attached to the tip of each Stokes line denote the signs of asymptotic behavior of $S_{{\rm odd},-1}$ when taking the limit toward this direction for $x$.
Since this graph is defined as the first Riemann sheet, $\psi_+(\psi_-$) is divergent when taking the limit to the region labeled by $+(-)$ but vanishes when reaching to $-(+)$.
For the construction of the quantization condition ${\frak D}$, we go along the blue line taken from $x = -\infty$ to $x=+\infty$ but slightly below the real axis as taking into account of monodromy matrices associated with each Stokes line $M_\pm$, normalization matrices $N_{a_{n_1},a_{n_2}}$, and branch-cut matrices $T$.
}
For both $\Im (\hbar)= 0^{\pm} $, there are $3N$ Stokes regions, where the WKB wave-functions are written as 
$\psi_{{\rm I},a_1}(x,\hbar), \ldots,  \ldots, \psi_{{\rm 3N},a_1}(x,\hbar)$. 
The monodromy matrix of the $N$-tuple well potential  connecting the WKB-wave functions can be expressed as 
\be
\label{minustoplus}
&& \psi_{{\rm I},a_1}(x,\hbar) = {\cal M} \psi_{{\rm 3N},a_1}(x,\hbar), \nl
&&{\cal M} = 
\begin{cases}
  M_+ N_{a_1,a_2} \left[ \prod_{n=1}^{N-1} M_+ N_{a_{2n},a_{2n+1}} T M_- M_+ N_{a_{2n+1},a_{2n+2}} \right] M_+ N_{a_{2N},a_1} =: {\cal M}^+ & \mbox{for $\theta>0$} \\
  M_+ N_{a_1,a_2} \left[ \prod_{n=1}^{N-1} M_+ M_- N_{a_{2n},a_{2n+1}} T M_+ N_{a_{2n+1},a_{2n+2}} \right] M_+ N_{a_{2N},a_1}  =: {\cal M}^- & \mbox{for $\theta<0$} \\
\end{cases}. \nl
\ee
The Stokes graph for the double-well and triple-well potentials are shown in Fig.\ref{fig:Stokes_graph}\footnote{In this figure, we introduced a branch cut between $a_{2n}$ and $a_{2n+1}$ due to a technical reason.
By using this way, one can keep the same asymptotic behavior of the Stokes lines from $a_{n}$ for odd (even) $n$.
This form of the graph can be directly applied  to  the dictionary proposed in Ref.\cite{Sueishi:2021xti} to reduce  to the degenerate Weber-type graphs.}. 
%\textcolor{red}{
For example, the monodromy matrix for $N=2$ is obtained as
\be
{\cal M}^+ &=& 
\begin{pmatrix}
  1 + B^{-1} + A_1^{-1} B^{-1} & i  (1 + A_1)(1 + A_2^{-1}) + i A_1 A_2^{-1}  B   \\
 - i A_1^{-1} B^{-1}& 1 + A_2^{-1} 
\end{pmatrix}, \\
{\cal M}^- &=& 
\begin{pmatrix}
  B^{-1} (1 + A_1^{-1}) & i  (1 + A_1)(1 + A_2^{-1}) + i B  \\
 - i A_1^{-1} B^{-1}& 1 + A_2^{-1} + B 
\end{pmatrix},
\ee
where we defined $A$-(perturbative) and $B$-(non-perturbative) cycles in terms of all orders $S_{\rm odd}(x,E,\hbar)$, which dictates the WKB wave function:
\be
\label{ABcycles}
&& A_\ell(E,\hbar) = \exp \left[ \frac{1}{\hbar}  \oint^{a_{2\ell}}_{a_{2\ell-1}} dx \, S_{\rm odd}(x,E,\hbar)\right], \qquad B_\ell(E,\hbar) = \exp \left[ \frac{1}{\hbar}  \oint^{a_{2\ell+1}}_{a_{2\ell}} dx \, S_{\rm odd}(x,E,\hbar)\right].
\ee
$A_\ell$ and $B_\ell$ give, respectively,   oscillation in a locally bounded potential and tunneling effect between neighboring vacua,  because $\oint^{a_{2\ell}}_{a_{2\ell-1}} dx \, S_{{\rm odd},-1}(x,E) \in i {\mathbb R}$ and $\oint^{a_{2\ell+1}}_{a_{2\ell}} dx \, S_{{\rm odd},-1}(x,E) \in {\mathbb R}_{<0}$.
This monodromy matrix provides us the quantization condition 
by using  the fact that  normalizable  WKB-wave function must vanish at infinity. 
Now, the upper and lower components of the wavefunction asymptotically behave $e^{-c/\hbar}$ and $e^{+ c/\hbar}$ with a real positive $c$ at $|x|=\infty$, respectively, so that the normalizability of the analytically-continued function, $\psi_{{\rm I}a_1}$, requires the condition that the 1-2 component of ${\cal M}$ must be zero.
%}
Hence, one can determine  the quantization condition as: 
\begin{align}
 \lim_{|x| \rightarrow \infty} \psi_{{\rm I}_{a_1}}(x,|\hbar|e^{i 0^{\pm}} ) =0 \Longrightarrow  \;\;  {\frak D}^{\pm} (E, \hbar) :={\cal M}^{\pm}_{12} =0.
 \label{eq:EQC}
\end{align}
%\textcolor{red}{we should refer the definition of ${\cal M}^{\pm}_{12}$ (the meaning of indices).}
For $N=2,3$, $ {\frak D}^{\pm} (E, \hbar) $  is given by:
\be
\mbox{Double-well} &:& \nl
   {\frak D}^{\pm}(E,\hbar)
   &=& \left( 1 + {A}_1 \right) \left( 1 + {A}_2^{-1} \right) + \begin{cases}
     {A_1}{B} & \mbox{for $+$} \\
     {A_2}^{-1}{B} & \mbox{for $-$} 
   \end{cases}, \label{eq:DpmDW_Ai} \\
\mbox{Triple-well} &:& \nl
 {\frak D}^{\pm}(E,\hbar) &=& 
 (1+{A}_1)( 1+{A}_2^{-1}) (1+{A}_3)  \nl
&& + \begin{cases}
   {A}_1 (1+{A}_3) {B}_1 + (1+{A_1}){A}_3 {B}_2+  {A}_1 {A}_3 {B}_1 {B}_2 & \quad \mbox{for $+$} \\
   {A}_2^{-1}(1+{A}_3) {B}_1  + (1+{A}_1) {A}_2^{-1} {B}_2 + {A}_2^{-1} {B}_1{B}_2  & \quad \mbox{for $-$}
 \end{cases}.  \label{eq:DpmTW_Ai}
\ee

%%%%%%%%%%%%%%%%%%%%%%%%%%%%%%%%%%%%%%%%%%%%%
\section{Analysis via the quantization condition}% (temporary)}
%%%%%%%%%%%%%%%%%%%%%%%%%%%%%%%%%%%%%%%%%%%%%
In this section, we derive the quantization condition expressed by perturbative and non-perturbative cycles and consider energy spectra. 
We also derive a nontrivial relation, which is called  P-NP  relation, among a perturbative cycle, 
a nonperturbarive cycle, and energy for quantum deformed double and triple well potentials.  This is an exact and constructive  version of resurgence which relates perturbative expansion around the perturbative vacuum to the expansion around instanton or bion configurations. It is an early term/early term relation, unlike traditional resurgence which is late term/early term relation.  For example, given the perturbative expansion of 
energy as a function of level number $N$ and coupling $g$ at some order $K$,  we can derive the  perturbative expansion around the instanton or bion  at order  $K-1$.\footnote{
 In our analysis of the quantum mechanics performing below, ``bion" is charactorized by the energy, $\exp (-S_B/\hbar)$. As we can see later, in the double-well, it is an instanton-antiinstanton pair only. In the triple-well, it comes from three types of contributions, instanton-antiinstanton, instanton-instanton, and antiinstanton-antiinstanton pairs. This issue would be argued in Sec V.E.
 }

In this paper, we  consider quantum deformed  potentials 
$V(x,\hbar) = \frac{1}{2} W'(x)^2 \pm \frac{\hbar p}{2}  W''(x) $ where 
\be
\mbox{Double-well} &:& \quad W'(x)=x^2-\frac{1}{4}, \label{eq:def_W_double} \\
\mbox{Triple-well} &:& \quad W'(x)=\frac{1}{2}x(x^2-1), \label{eq:def_W_triple}
\ee
 $O(\hbar^0)$ part in the potential is classical and   $O(\hbar)$ part    is quantum deformation.
 We also also argue the extension of our quantization conditions  to generic superpotential $W(x)$.
 
%Notice  that if one takes a  scaling (large-$p$) limit  where $\hbar p \sim O(1)$, then, the deformation becomes classical and the Stokes curves associated with the new classical systems is  different from the one that arises from the order $O(\hbar^0)$ part in \eqref{cq}.  For example, in the above case, with  generic quantum deformation of double-well where $p$ is a finite real number, perturbation theory for  the low lying states   is non-Borel summable, yet if we take the scaling limit $\hbar p \sim O(1)$, they become Borel summable. These  facts can be traced to the topology of the Stokes graphs. It is an interesting question to understand in detail how  an asymptotic non-Borel summable  perturbation theory for  some level $N$ and finite  $p$ given by  $E(N, p, \hbar) $ transmutes to a Borel summable series in the $\hbar p \sim O(1)$ limit. 

%%%%%%%%%%%%%%%%%%%%%%%%%%%%%%%%%%%%%%%%%%%%%
\subsection{Voros multipliers and quantization condition} \label{sec:Voros_quant}
%%%%%%%%%%%%%%%%%%%%%%%%%%%%%%%%%%%%%%%%%%%%%
The exact-WKB method is a combination of  two parts, topological construction and analytic computation.  
The analysis that leads to exact quantization condition \eqref{eq:EQC} 
  is topological, and   is dictated by the  Stokes graph data.  
  This is, however,   half of  the story.   The  important   points provided by the exact-WKB analysis is that the quantization condition can be expressed by \textit{Voros multipliers} or  \textit{cycles}. The analytic  part of the story is the calculation of the Voros symbols by using all orders  asymptotic expansion of \eqref{eq:wave_ansatz}. 
Since the Stokes automorphism can be simply written for each cycle \cite{DDP2}, the resurgence relation for objects derived from the quantization condition can be explicitly written. 
One can define two type of cycles,  perturbative cycles ($A$-cycles) and non-perturbative cycles ($B$-cycles), given in \eqref{ABcycles}.
Roughly speaking, $A$-cycle  express the perturbative fluctuations around a perturbative vacuum and $B$-cycle express the  tunneling effect between two vacua including   perturbative fluctuation around it. 
 The practical side of the exact-WKB  analysis involves explicit computation of these Voros symbols. 

%comment
%we can remove As discussed in our earlier work...
%
\textcolor{black}{
In \S.~\ref{sec:EWKB_analysis}, we have explained the exact-WKB method using the Airy-type, i.e., each of Stokes curve emerges from a simple turning point.
Now, we introduce the degenerate Weber(DW)-type exact-WKB method.
In this method, we firstly rescale the energy $E$ in the Schr\"{o}dinger equation as $E \hbar $, and then draw the Stokes graph.
This is just a redefinition of the variables, so it is physically equivalent to the Airy-type. But, since the definition of the Stokes curve is determined by the lowest order of $\hbar$ in the WKB expansion, it changes the condition giving the stokes curve as
\begin{align}
    \Im\frac{1}{\hbar}\int_a^\infty \sqrt{2(V_0-E)}dx = 0 \quad \to \quad \Im\frac{1}{\hbar}\int_a^\infty \sqrt{2V_0}dx = 0,
\end{align}
where $a$ is a turning point. Also, if we set $E=O(\hbar)$ for a classical vacuum (minimum point), two simple turning points collide to each other and become a double turning point.
Because of these differences, the monodromy matrix of the connection formula also changes significantly. (See (60) in \cite{Sueishi:2021xti}.)
%As a result, computation of the monodromy matrix also becomes complicated from the Airy-type which is relatively easy to compute. 
Although computation of the DW-type is more complicated than that of the Airy-type, there are two reasons to introduce the DW-type method; \\ \\
\noindent
(I) The $\hbar$ expansion in the path integral calculation corresponds to the $\hbar$ expansion of the DW-type: 
In the path integral, the $\hbar$ expansion is performed by rescaling the quantum fluctuation as $x=x_{cl}+\sqrt{\hbar}\tilde{x}$ so that the gaussian part of $\tilde{x}$ is the zero-th order of $\hbar$ (regarding $\hbar$ expansion as coupling expansion), as follows:
\begin{align}
    Z &=\int \mathcal{D}x\; e^{-\frac{S[x]}{\hbar}} = e^{-\frac{S[x_{cl}]}{\hbar}}\int \mathcal{D}\tilde{x}\; e^{-\tilde{x} M \tilde{x}+O(\sqrt{\hbar})}.
\end{align}
Let us consider this procedure from the viewpoint of Schr\"{o}dinger equation. In the case of $V(x)=\displaystyle{\sum_{n=2} a_n x^n}$, the Schr\"{o}dinger equation is expressed as
\begin{align}
    \qty( -\frac{\hbar^2}{2}\frac{\partial^2}{\partial x^2}+a_2x^2+a_3x^3+...) \psi=E\psi.
\end{align}
In order to regard $\hbar$ as a coupling constant, we rescale $x=\sqrt{\hbar}\tilde{x}$ and $E= \tilde{E} \hbar$.
Then, the Schr\"{o}dinger equation becomes
\begin{align}
    \qty( -\frac{1}{2}\frac{\partial^2}{\partial\tilde{x}^2}+a_2\tilde{x}^2+O(\sqrt{\hbar}) )\psi = \tilde{E}\psi.
\end{align}
%Therefore, we have to consider the rescaled energy $\tilde{E}$. 
Notice that we rescaled not only $x$ but also $E$. 
This procedure should not change physics because it is just a redefinition of variables,
%Although there is nothing special in physics because it is just a redefinition of variables. 
but the difference arises when using exact-WKB to find the exact $\hbar$ asymptotic expansion of the partition function and energy.
For example, terms related to $\log \hbar$ and the fluctuation determinant ($\det M$) cannot be naturally derived without using the DW-type.
 \\ \\
(II) To obtain the hidden topological angle analytically: 
The hidden topological angle (HTA) is a phase produced by a quantum deformation part in the potential.
As studied in \cite{Behtash:2015loa,Behtash:2015kna, Alireza2018}, this can also be interpreted as a phase associated with the action evaluated by a complex classical solution.
In order to incorporate this contribution analytically in the form of Voros symbols, it is necessary to use the DW-type instead of the Airy-type.
This is because even if the Voros symbol cannot be calculated analytically in the Airy-type, the DW-type can produce the HTA by factorization as $\mathfrak{A}=e^{ip\pi}\bar{\mathfrak{A}}$, where ${\frak A}$ is a Voros symbol of the DW-type.
\\ 
\noindent
}

There is a precise dictionary connecting the Airy-type and the DW-type, and one needs to rewrite the exact quantization conditions \eqref{eq:DpmDW_Ai} and \eqref{eq:DpmTW_Ai}  based on the Airy-type building blocks in terms of DW type building blocks.
This amounts to 
\begin{align}
  {\rm Airy:}\;\; (A,B) \longrightarrow {\rm Degenerate \;  Weber:}\;\;   ({\frak A},{\frak B}) \label{eq:dicAB}
\end{align}
and using the cycle expressions 
\be
   {\frak A}_\ell (E,\hbar) &=& e^{2 \pi i F_{\ell}(E,\hbar)}, \label{eq:A-cy} \\ %=: e^{(-1)^{\ell} \pi i p} \bar{\frak A}(E,\hbar) \nl
   %&=&  e^{-2 \pi i [E +(-1)^{\ell+1} p/2]} + O(\hbar), \\
   {\frak B}_{\ell}(E,\hbar) &=& 2 \pi {\frak B}_0 \prod_{\ell^\prime=0}^1 \frac{C_{\ell+\ell^\prime -}(E,\hbar)}{C_{{\ell+\ell^\prime} +}(E,\hbar)} \frac{e^{(-1)^{\ell^\prime}\pi i F_{\ell+\ell^\prime}(E,\hbar)}\hbar^{F_{\ell+\ell^\prime}(E,\hbar)}}{\Gamma(1/2-F_{\ell+\ell^\prime}(E,\hbar))}, \label{eq:B-cy}  %=: \bar{\frak B}(E,\hbar) \nl
   %&=&  \frac{2 \pi {\frak B}_0}{\Gamma(E + \frac{1+p}{2}) \Gamma(E + \frac{1-p}{2})} \left( \frac{\hbar}{2}\right)^{-2 E} + O(\hbar),
\ee
where ${\frak A}_{\ell}$ and ${\frak B}_{\ell}$ denote $A$-cycle and $B$-cycle, respectively, and $\ell$ is a label of each vacuum.
In addition, ${\frak B}_0$ the (classical) bion action, i.e. ${\frak B}_0 = e^{-S_{\rm B}/\hbar}$.
Since a $B$-cycle attaches two $A$-cycles on the left and right sides, it includes the information of two $F_{\ell}(E,\hbar)$s. 
$F_\ell(E,\hbar)$ is easily calculable by performing residue integration around the $\ell$-th vacuum: 
\be
F_{\ell}(E,\hbar) = -{\rm Res}_{a=a_{\ell}} S_{\rm odd}(E,\hbar). \label{eq:Fell_A}
\ee
$C_{\ell \mp}(E,\hbar)$ in Eq.(\ref{eq:B-cy}) can be obtained from  
the 
transformation connecting  local and global coordinates \cite{Sueishi:2021xti}.
\textcolor{black}{
Calculation for the quantization condition using the DW-type requires reconstruction of the connection formula.
However, without the explicit connection formula of the DW-type, we can easily perform the transformation of (\ref{eq:dicAB}) by taking the following steps:
\begin{enumerate}
    \item  Find the Fredholm determinant $D$ in terms of the Voros symbols using the connection formula of Airy-type. (e.g. $D=1+A+AB$)
    \item Replace the symbol via the Airy-Weber dictionary. (See Table I in \cite{Sueishi:2021xti}) (e.g. $D=1+\mathfrak{A}+\mathfrak{AB}$)
\end{enumerate}
Therefore, using the results by the Airy-type, which are easy to calculate, we can produce the ones by DW-type. 
%We calculate the partition function and the $\hbar$-dependence of the asymptotic expansion of energy using this procedure.
}
\textcolor{black}{
We emphasize that the DDP(Dellabaere-Dillinger-Pham) formula linking each Voros symbol with the Stokes phenomenon is also symbolically equivalent (e.g. $({\cal S}_+-{\cal S}_-)[A]= {\cal S}_+[AB]\to({\cal S}_+-{\cal S}_-)[\mathfrak{A}]={\cal S}_+[\mathfrak{AB}]$), so that we can identify the resurgent structure directly\cite{Sueishi:2021xti}.
}

One of the nice properties of the quantum-deformed systems \eqref{cq}  is that 
%the $\log {\frak A}_\ell$ and $\log {\frak B}$ for each order are even functions of $p$ except $O(\hbar^0)$, and 
in the asymptotic  expansion of $\log {\frak A}_\ell$ and $\log {\frak B}$  in  $\hbar$, 
the $O(\hbar^0)$ term plays a special role, intimately related to hidden topological angle which is the phase of the bions as we  describe  later. 
It takes the form  $ \pi i s p + ({\rm energy})$ where $s \in \{-1,0,+1\}$ depending on a given potential.
Thus, %if $\log {\frak A}_\ell$ and $\log {\frak B}$ include $p$ in $O(\hbar^0)$, 
when $s = \pm 1$, it is convenient to split ${\frak A}_{\ell}(E,\hbar)$ and/or ${\frak B}_{\ell}(E,\hbar)$ into the  phase factor  and the other part for the later analysis:
\be
{\frak A}_{\ell}(E,\hbar) =: e^{\pi i s p }\bar{\frak A}_{\ell}(E,\hbar),  \qquad {\frak B}_\ell(E,\hbar) =: e^{\pi i s p }\bar{\frak B}_{\ell}(E,\hbar),
\ee
where $\bar{\frak A}_{\ell}(E,\hbar)$ and $\bar{\frak B}(E,\hbar)$ are even functions of $p$.
Similarly,
\be
F_{\ell}(E,\hbar)=\frac{\log {\frak A}_{\ell}}{2 \pi i } =: \frac{sp}{2}  + \bar{F}_{\ell}(E,\hbar),
\ee
where $F_\ell(E,\hbar)$ is given in \eqref{eq:Fell_A}. Here, $\bar{F}_{\ell}(E,\hbar)  \equiv \sum_{n=0}^{\infty}\bar{F}_{\ell,n} \hbar^n $ is a formal power series dictated by  WKB-wave function data $S(E, \hbar)$ \eqref{eq:wave_ansatz}, which is itself a formal series obtained by recursively solving Riccati equation \eqref{eq:Sg_Qgen}. 

For the double-well potential (\ref{eq:def_W_double}), the cycles  \eqref{eq:A-cy} and 
\eqref{eq:B-cy}  including only their leading order structure in $\hbar$ are given by:
\be
&& \mbox{Double-well}: \nl
   &&\qquad {\frak A}_\ell (E,\hbar) %&=& e^{2 \pi i F_{\ell}(E,\hbar)} 
   = e^{(-1)^{\ell} \pi i p} \bar{\frak A}(E,\hbar)
   =  e^{-2 \pi i [E +(-1)^{\ell+1} p/2]} + O(\hbar),  \label{harmonicDW}  \\ 
 && \qquad {\frak B}(E,\hbar) %&=& 2 \pi {\frak B}_0 \prod_{\ell=1}^2 \frac{C_{\ell-}(E,\hbar)}{C_{\ell+}(E,\hbar)} \frac{e^{(-1)^{\ell+1}\pi i F_\ell(E,\hbar)}\hbar^{F_\ell(E,\hbar)}}{\Gamma(1/2-F_\ell(E,\hbar))} 
   = \bar{\frak B}(E,\hbar) 
   =  \frac{2 \pi {\frak B}_0}{\Gamma(E + \frac{1+p}{2}) \Gamma(E + \frac{1-p}{2})} \left( \frac{\hbar}{2}\right)^{-2 E} \left( 1 + O(\hbar) \right),  \\
   && \qquad {\frak B}_{0} = e^{-\frac{1}{3\hbar}},   
\ee
%where
%\be
%&& {\frak B}_{0}=e^{-\frac{1}{3\hbar}}, \\
%&& F_{\ell} (E,\hbar) =  - E + (-1)^{\ell} \frac{p}{2} + O(\hbar), \\
%&& \prod_{\ell=1}^{2} \frac{C_{\ell-}(E,\hbar)}{C_{\ell+}(E,\hbar)} = 2^{2E} e^{+ \pi i p}  + O(\hbar),
%\ee
%and $O(\hbar)$ is an even function of $p$.
%Notice that the extra factor $e^{+ \pi i p}$ showed up from $\prod_{\ell=1}^{2}\frac{C_{\ell-}}{C_{\ell+}}$.

%The cycles are given by
%\be
%   {\frak A}_\ell (E,\hbar) &=& e^{2 \pi i F_{\ell}(E,\hbar)}, \\ 
%   {\frak B}_{\ell}(E,\hbar) &=& 2 \pi {\frak B}_0 \prod_{\ell^\prime=0}^1 \frac{C_{\ell+\ell^\prime-}(E,\hbar)}{C_{\ell+\ell^\prime+}(E,\hbar)} \frac{e^{(-1)^{\ell^\prime}\pi i F_{\ell+\ell^\prime}(E,\hbar)}\hbar^{F_{\ell+\ell^\prime}(E,\hbar)}}{\Gamma(1/2-F_{\ell+\ell^\prime}(E,\hbar))},
%\ee
%where
%\be
%&& F_{1,3}(E,\hbar) = \bar{F}(E,\hbar) + \frac{p}{2}, \qquad F_{2}(E,\hbar) = 2 \bar{F}(E,\hbar) - \frac{p}{2}.
%\ee
%By using the following symbols
\noindent
Similarly, for the triple-well potential (\ref{eq:def_W_triple}), the cycles  
take the form: 
\be 
&& \mbox{Triple-well}: \nl
&& \qquad {\frak A}_{1,3}(E,\hbar) = e^{+\pi i p}\bar{\frak A}(E,\hbar) =  e^{-2 \pi i (E - p/2)} + O(\hbar),  \label{harmonicTW} \\
&& \qquad {\frak A}_{2}(E,\hbar) = e^{-\pi i p}\bar{\frak A}^{2}(E,\hbar) =  e^{-2 \pi i (2E + p/2)} + O(\hbar),  \\
&& \qquad {\frak B}_{\ell}(E,\hbar) = \bar{\frak B}(E,\hbar) = \frac{2 \pi {\frak B}_0 }{\Gamma(E+\frac{1-p}{2})\Gamma(2E+\frac{1+p}{2})} \cdot \frac{\hbar^{-3 E}}{2^{E-\frac{p}{2}}} \left( 1 + O(\hbar) \right), \\ %\nl
%&& \bar{\frak A}(E,\hbar) = e^{2 \pi i \bar{F}(E,\hbar)}, \qquad \bar{F} (E,\hbar) =  - E  + O(\hbar), \\
%&& \bar{\frak B}(E,\hbar) = \left| \prod_{\ell=1}^2 \frac{C_{\ell-}(E,\hbar)}{C_{\ell+}(E,\hbar)} \right| \frac{2 \pi {\frak B}_0 \hbar^{3\bar{F}(E,\hbar)}}{\Gamma(\frac{1-p}{2}-\bar{F}(E,\hbar))\Gamma(\frac{1+p}{2}-2\bar{F}(E,\hbar))}, \\
&& \qquad {\frak B}_{0}=e^{-\frac{1}{4\hbar}}, %\qquad  \prod_{\ell^\prime=0}^{1} \frac{C_{\ell+\ell^\prime-}(E,\hbar)}{C_{\ell+\ell^\prime+}(E,\hbar)} = %2^{-3E} e^{(-1)^{\ell} \pi i(E+p)} 2^{-\left(E-\frac{p}{2}\right)} e^{(-1)^{\ell}\pi i (E +p)} + O(\hbar),
\ee
%where $O(\hbar)$ is an even function of $p$,  Eq.(\ref{eq:QcondTW}) can be expressed by
Note that in the symmetric triple-well system (including its quantum deformation), 
 the three  $A$-cycles and two $B$-cycles can be expressed only by $\bar{\frak A}(E,\hbar)$ and $\bar{\frak B}(E,\hbar)$, respectively, meaning that it  is in fact a  genus-1 system, just like 
 double-well potential.  
 If one considers an  asymmetric triple-well potential, it is  no longer a  genus-1 classical potential.

For the double-well and triple-well potentials and their quantum deformations, the quantization condition  in terms of the DW-type cycles 
building blocks can be written as:
\be
\mbox{Double-well} &:& \nl
   {\frak D}^{\pm}
   &=& \left( 1 + {\frak A}_1 \right) \left( 1 + {\frak A}_2^{-1} \right) + \begin{cases}
     {\frak A_1}{\frak B} & \mbox{for $+$} \\
     {\frak A_2}^{-1}{\frak B} & \mbox{for $-$} 
   \end{cases} \nl
   &=& \left( 1 + e^{-\pi i p} \bar{\frak A}  \right) \left( 1 + e^{-\pi i p} \bar{\frak A}^{-1} \right) + e^{- \pi i p} \bar{\frak A}^{\pm 1} \bar{\frak B}, \label{eq:DpmDW2} \\
%   &=& \left( 1 + e^{2 \pi i(-E - p/2)}  \right) \left( 1 + e^{2\pi i (E-p/2)}  \right) \nl
%   && + \frac{2 \pi {\frak B}_0e^{2 \pi i \left( \mp  E -  \frac{p}{2} \right)} }{\Gamma(E + \frac{1+p}{2}) \Gamma(E + \frac{1-p}{2})} \left( \frac{\hbar}{2}\right)^{-2 E} 
%   + O(\hbar), \label{eq:DpmDW}
\mbox{Triple-well} &:& \nl
 {\frak D}^{\pm} &=& 
 (1+{\frak A}_1)( 1+{\frak A}_2^{-1}) (1+{\frak A}_3)  \nl
&& + \begin{cases}
   {\frak A}_1 (1+{\frak A}_3) {\frak B}_1 + (1+{\frak A_1}){\frak A}_3 {\frak B}_2+  {\frak A}_1 {\frak A}_3 {\frak B}_1 {\frak B}_2 & \quad \mbox{for $+$} \\
   {\frak A}_2^{-1}(1+{\frak A}_3) {\frak B}_1  + (1+{\frak A}_1) {\frak A}_2^{-1} {\frak B}_2 + {\frak A}_2^{-1} {\frak B}_1{\frak B}_2  & \quad \mbox{for $-$}
 \end{cases} \nl %\label{eq:QcondTW}
 &=& 
 (1+e^{\pi i p} \bar{\frak A})^2( 1+ e^{\pi i p}\bar{\frak A}^{-2})\nl
 && + \begin{cases}
   2e^{\pi i p} \bar{\frak A} (1+e^{\pi i p} \bar{\frak A}) \bar{\frak B}  +  e^{2 \pi i p}\bar{\frak A}^{2} \bar{\frak B}^2 & \qquad \mbox{for $+$} \\
   2e^{\pi i p} \bar{\frak A}^{-2} (1+e^{\pi i p} \bar{\frak A}) \bar{\frak B}  +  e^{\pi i p} \bar{\frak A}^{-2} \bar{\frak B}^2 & \qquad \mbox{for $-$}
 \end{cases}  \nl
 %&\propto&    e^{\pm \pi i p}\bar{\frak A}^{\mp 2} ( 1+ e^{\mp \pi i p}\bar{\frak A}^{\mp1})^2 +   \left[ (1+e^{\mp \pi i p} \bar{\frak A}^{\mp 1}) + \bar{\frak B} \right]^2 \nl
 &\propto&  \prod_{\varepsilon \in  \{-1,+1 \}} \left[ (1+e^{\mp \pi i p} \bar{\frak A}^{\mp 1}) (1  + i \varepsilon  e^{\pm \pi i \frac{p}{2}}\bar{\frak A}^{\mp 1}) + \bar{\frak B}   \right], \label{eq:QcondTW2}
\ee
where $\pm$ is the sign of the phase of complexified $\hbar=e^{i \theta}|\hbar|$, $\theta=0^{\pm}$. 

These quantization conditions are invariant under the DDP formula\cite{DP1,DDP2}:
\be
{\cal S}_{+}[\bar{\frak A}] = {\cal S}_{-}[\bar{\frak A}](1+{\cal S}[\bar{\frak B}])^{-1} &\quad \Rightarrow \quad & {\cal S}_+[{\frak D}_+] = {\cal S}_-[{\frak D}_-]. \label{eq:bar_DDP}
\ee
which relates the  left/right Borel resummation
of the perturbative $A$-cycle  to  Borel resummation of the non-perturbative 
$B$-cycle.  As a result, the Fredholm determinant remains invariant under Stokes automorphism, 
as $\arg(\hbar) $ moves from $0^{-}$ to $0^{+}$. This implies that all resurgent cancellations to all orders are already built-in the quantization condition.  
%Therefore, the resurgence relation of all of expression such as the Gutzwiller trace formula derived by the quantization condition can be traced from Eq.(\ref{eq:bar_DDP}).

We also should keep in our mind that the complex phase $e^{\pm i p \pi}$ attached to $\bar{\frak A}(E,\hbar)$ is known as the Maslov index in the Gutzwiller trace formula\cite{Gutzwiller,Alireza2018,Sueishi:2020rug} and is related to a hidden topological angle in the path integral formalism, which will be discussed in Sec.~\ref{sec:path-integral}.

%%%%%%%%%%%%%%%%%%%%%%%%%%%%%%%%%%%%%%%%%%%%%
\subsection{Nonpertubative effects in the energy spectrum, SUSY, QES, and in between}
%%%%%%%%%%%%%%%%%%%%%%%%%%%%%%%%%%%%%%%%%%%%%

%%%%%%%%%%%%%%%%%%%%%%%%%%%%%%%%%%%%%%%%%%%%%
\subsubsection{non-perturbative effect to the energy spectra} 
%%%%%%%%%%%%%%%%%%%%%%%%%%%%%%%%%%%%%%%%%%%%%
\label{sec:nettes}
We now start to discuss   the non-perturbative effect to the energy spectra  of the quantum deformed theories, see \cite{Kozcaz:2016wvy, Dunne:2016jsr,Dunne:2020gtk, Sato:2001ac}  for an earlier work from path integral perspective.  
One may naively be tempted to think that 
it is just the usual instanton analysis. However, the story is far more interesting, and captures many interesting non-perturbative phenomena. There are indeed some cases where instantons provide leading non-perturbative contributions, but also cases in which instantons do not contribute despite the fact that they are finite action solutions.  There are also cases in which bions contribute at leading order, and other ones  in which different bions  have desctructive/constructive interference depending on hidden topological angle. When such destructive interference happens,  perturbation theory is convergent, and the corresponding states turn out to be algebraically solvable. There is also a phenomenon called Cheshire cat resurgence taking place in these systems. When the quantum deformation parameter $p$ is analytically  continued away from its quantized  values, the convergence and destructive interference effects are replaced with the asymptotic nature of perturbative series and ambiguity of bion events which cancel each other out according to resurgence. 
In this sense, the non-perturbative phenomena we describe  in these systems are quite rich.

The beauty of exact WKB is that 
all the interesting perturbative and non-perturbative phenomena can be extracted 
from the exact quantization condition obtained in Sec.~\ref{sec:Voros_quant}, at once,  by replacing the cycles with their explicit forms. We will do this analysis in detail, starting with the simple aspects. 
The energy can be generally decomposed into two parts, that is a perturbative  and a non-perturbative part. % which is coupled with the bion contribution.
In order to obtain the energy spectrum, we first extract the  perturbative part labeled the energy level denoted $k$ below, and then non-perturbative part by setting the boundary condition for the perturbative part with $\hbar \rightarrow 0_+$. The solution of exact quantization condition 
generates  a transseries with all non-perturbative factors. 
%It is also the necessary process to obtain the transmonomials in the energy spectra.
%Notice that in the SUSY deformed model with $N$ vacua, there generally exists $\lfloor (N+1)/2 \rfloor$ solutions.

Let us demonstrate the process explicitly using the double-well and the triple-well potential. 
Let us first demonstrate how the simple harmonic energy levels emerge in the 
quantum-deformed double and triple well potentials. 
The quantization condition are given by Eqs.(\ref{eq:DpmDW2}) and (\ref{eq:QcondTW2}).  Replacing $\bar{\frak A}(E,\hbar)$ with the explicit form \eqref{harmonicDW}  and \eqref{harmonicTW}, and ignoring the non-perturbative contribution at first, we find:
\be
 \mbox{Double-well} &:& \qquad \left( 1 + e^{2 \pi i(-E - p/2)}  \right) \left( 1 + e^{2\pi i (E-p/2)}  \right) = 0, \\
 \mbox{Triple-well} &:& \qquad \left( 1 + e^{-2 \pi i(E - p/2)}  \right)^2 \left( 1 + e^{2\pi i (2E+p/2)}  \right) = 0.
\ee
Here, we picked up only the dominant parts in terms of $\hbar$.
This provides harmonic energy spectrum for the quantum deformed potential.
\be
 \mbox{Double-well} &:& \qquad E_{\rm pt} = k + \frac{1 \mp p}{2}, \\
 \mbox{Triple-well} &:& \qquad E_{\rm pt} = 
 \begin{cases}
\frac{1}{2} \left(k + \frac{1-p}{2} \right) & \mbox{for the inner-vacuum} \\
  k + \frac{1+p}{2} & \mbox{for the outer-vacua}   
 \end{cases},
 \label{harmonic}
\ee
with $k \in {\mathbb N}_0$ is the harmonic level number.   $p = \pm 1$ correspond to supersymmetric 
pairs, with one zero energy level, and exhibiting the degeneracy (at the harmonic level) for higher states.  For double-well, natural frequency is $\omega=1$ and for triple-well, it is $\omega_{\rm in}=1/2, \omega_{\rm out}=1$. 

To obtain the leading  non-perturbative contribution, we add $\delta(\hbar)$ to the energy as $E(\hbar)=E_{\rm pt} + \delta (\hbar)$, and keep non-perturbative term. 
Substituting $E(\hbar)$ into Eqs.(\ref{eq:DpmDW2}) and (\ref{eq:QcondTW2}) gives
\be
&&  \mbox{Double-well} : \nl
 && \quad 4 \sin  (\pi \delta(\hbar)) \sin [\pi  (\delta(\hbar)-p)] \nl
 && \quad \quad - \frac{2 \pi {\frak B}_0e^{ \pm \pi i  (p - 2\delta(\hbar))} }{\Gamma(1+k+ \delta(\hbar)) \Gamma(1+k-p + \delta(\hbar))} \left( \frac{\hbar}{2}\right)^{- (1+2k -p + 2\delta(\hbar))} = 0,  \\ \nl
&&  \mbox{Triple-well (inner-vacuum)} : \nl
&& \quad 4 \cos \left[ \pi \left( \frac{k}{2} + \frac{1-3p}{4} + \delta (\hbar) \right)\right] \cos \left[ \pi \left( \frac{k}{2} + \frac{1+\varepsilon}{4} + \delta (\hbar) \right)\right]  \nl
&& \quad \quad + (-1)^{k} \frac{2 \pi {\frak B}_0 e^{\mp \pi i \left( \frac{1}{2}  -\frac{3p-\varepsilon}{4} + 2\delta (\hbar) \right) } 2^{-\left( \frac{1}{4}+\frac{k}{2} -\frac{3p}{4} + \delta (\hbar) \right)} \hbar^{-3 \left( \frac{1}{4}+\frac{k}{2} - \frac{p}{4} + \delta (\hbar) \right)}}{\Gamma(\frac{3}{4}+\frac{k}{2} - \frac{3p}{4} + \delta (\hbar) ) \Gamma(1 + k + 2\delta (\hbar))}=0, \nl \\ 
&&  \mbox{Triple-well (outer-vacua)} : \nl
&&  \quad  4 \sin \left[ \pi \delta (\hbar) \right] \sin \left[ \pi \left( \frac{3p+\varepsilon}{4} + \delta (\hbar) \right)\right]   \nl
&& \quad \quad - \frac{2 \pi {\frak B}_0 e^{\mp \pi i\left(\frac{3p+\varepsilon}{4} + 2 \delta (\hbar) \right)}  2^{-\left(\frac{1}{2}+k+\delta (\hbar) \right)} \hbar^{-3 \left(\frac{1}{2} + k +\frac{p}{2}+ \delta (\hbar) \right)}} {\Gamma(1  + k + \delta (\hbar) ) \Gamma( \frac{3}{2} + 2k + \frac{3p}{2}+ 2\delta (\hbar) )} =0,
\ee
where $\varepsilon \in \{- 1, +1\}$ and we used $E_{\rm pt}=k + \frac{1-p}{2}$ for the double-well potential without the loss of generality.
By setting the boundary condition $\lim_{\hbar \rightarrow 0_+}\delta(\hbar) = 0$, the solutions are obtained as
\be
  \mbox{Double-well} &:& \nl
 \delta_{p \in {\mathbb Z}}(\hbar) &=& 
 {\cal P} \sqrt{\frac{{\frak B}_0}{\pi \hbar \Gamma (1+k) \Gamma (1+k-p)}} \left( \frac{\hbar}{2} \right)^{-k+\frac{p}{2}} \nl
&& - \frac{{\frak B}_0}{\pi \hbar \Gamma (1+k) \Gamma (1+k-p)} \left( \frac{\hbar}{2} \right)^{-2k+p}  \Phi_{\pm 1}(k,p)+ O({\frak B}_0^{3/2}), \label{eq:deltaDW_pint}  \\ \nl
\delta_{p \notin {\mathbb Z}}(\hbar) &=& (-1)^{1+k} \frac{{\frak B}_0e^{\pm \pi i p}\Gamma(-k+p)}{\pi \hbar \Gamma(1+k)}  \left( \frac{\hbar}{2} \right)^{-2k+p}  \nl
&& - \frac{2{\frak B}_0^2 e^{\pm 2 \pi i p} \Gamma(-k+p)^2}{\pi^2 \hbar^2 \Gamma(1+k)^2} \left( \frac{\hbar}{2}\right)^{-4k+2p} \Phi_{\pm 1}(k,p) + O({\frak B}_0^3),  \label{eq:deltaDW_pnonint} 
\ee
where ${\cal P} \in \{-1,+1\}$ is the parity, and\footnote{
$\Phi_n(k,p)$ and $\Psi^{(1,2)}_n(k,p)$ defined by (\ref{eq:Phin}) and (\ref{eq:Psin1}) (\ref{eq:Psin2}), respectively, include only $\log \hbar$ as a transmonomial.
Hence, $\delta_p(\hbar)$ is expanded by $e^{-S_{\rm B}/\hbar}$, $\hbar^{-1}$, and $\log h$.
}
\be
\Phi_{n}(k,p):=  \frac{\psi^{(0)}(1+k) + \psi^{(0)}(1+k-p)}{2} + \log \frac{\hbar}{2} + \pi n i, \label{eq:Phin}
\ee
with the polygamma function $\psi^{(n)}(x)$. Probably, the most striking feature of the solution 
\eqref{eq:deltaDW_pint} and \eqref{eq:deltaDW_pnonint} is that the instanton may or may not be  the leading non-perturbative contribution depending on details. In fact, in generic case where 
$p$ is not quantized, the instanton contribution does not appear at all. 
%Notice that ${\cal P}$ can be regarded as the parity and causes the energy splitting.It is also important to mention that the polygamma function for the double-well is divergent when $\psi^{(0)}(x)$ with $x \in {\mathbb Z}_{0-}$, but the divergence is cancelled by the gamma function in the denominator.Therefore, the bion contribution remains.
%and is relevant to the SUSY breaking when $|p|=1$.
Multiple  remarks about this result are in order:%  starting with the  $p \in {\mathbb Z}^{>}$ case. 

     \begin{figure}[t]
    \centering 
  \hspace{2cm}
    \includegraphics[width=15cm]{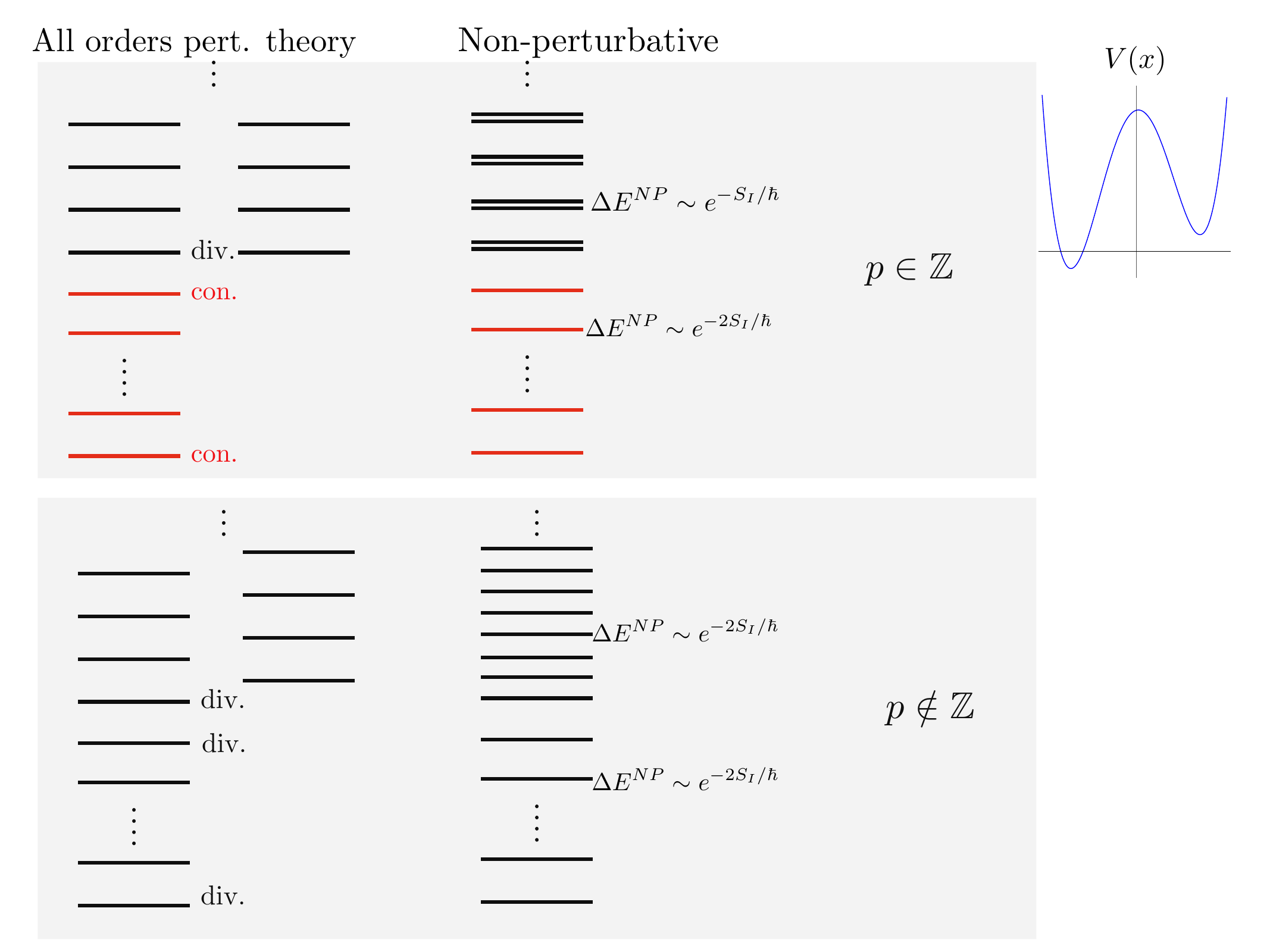}
      \vspace{-0.1cm}
    \caption{Perturbative vs non-perturbative spectrum of the quantum deformed 
    double-well potential. For  $p \in \mathbb Z$ (upper), lowest $p$ states have convergent , and higher ones have divergent perturbation theory.  Non-perturbatively, the shift of the lowest $p$ states are of order $e^{-2S_{\rm I}/\hbar}$, and level splitting for higher states is $e^{-S_{\rm I}/\hbar}$. 
    For $p \notin \mathbb Z$ (lower), all states have divergent perturbation theory. 
    All perturbative energy levels are non-perturbatively shifted (up or down, depending on level number and $p$) by 
    $e^{-2S_{\rm I}/\hbar}$ factor. Order $e^{-S_{\rm I}/\hbar}$ effects disappear from the spectrum. 
    }
     \label{fig:DWPNP}
\end{figure}

\begin{itemize}
    \item For the  $p \in {\mathbb Z}$ case,   consider the term proportional to instanton factor, $\sqrt{{\frak B}_0} \sim e^{-S_{\rm I}/\hbar}$ in \eqref{eq:deltaDW_pint}.
This term is proportional to $\frac{1}{\Gamma(1+k - p)}$.  Therefor, for level number $k=0, 1, \ldots, p-1$, the argument of the gamma function is a negative integer,  which corresponds to  a pole of   $\Gamma(x)$. 
As such, the instanton contribution vanishes for these states. The leading non-perturbative contribution arises from the 
bions for the lowest $p$ states. For $p=0$ (textbook case), of course, instanton contribution is present and it leads to level splitting between parity even and odd states. Note that for 
unquantized $p$ \eqref{eq:deltaDW_pnonint},  there is no contribution to the spectrum at the one-instanton level, despite the fact that instantons are solutions of classical equations.  This comes from the fact that harmonic states are not aligned for unquantized $p$. 

\item   
 Since  $\frac{\psi^{(0)}(z) }{\Gamma(z)}$ is an entire function, 
 there is a non-perturbative contribution that arise from $\frac{\psi^{(0)}(1+k-p) }{\Gamma(1+k - p)}$ factor even when $k=0, 1, 2,  p-1.$ 
  In particular, the ambiguity of the bion contribution vanishes in these cases.  Relatedly, perturbation series is convergent for these levels.  However, for double-well in contrast with triple well, these levels are not exactly solvable mainly because $e^{\pm W(x)}$ are non-normalizable, 
while  $e^{- W(x)}$  is normalizable for the  triple-well.

 \item  The bion contribution   to the level $k$ of the $p$ deformed theory is of the form: 
 \be
 \Delta E_{p, k} =   -  \frac{1}{2 \pi  \Gamma (1+k) } \left( \frac{\hbar}{2} \right)^{p-2k-1}
 \Gamma(p-k)e^{i \pi (p-k)} e^{-S_{\rm B}/\hbar} 
 \ee
%{ \color{red} A factor of 2 disagreement with the Cheshire cat paper... But it is great that everything else is an exact match}
  For supersymmetric theory, $p=1, k=0$ and the contribution is positive consistent with supersymmetry algebra. 
 Note that the contribution of this bion to ground state ($k=0$) alternate as a function of $p$, increases the ground state energy for $p$ odd, and decreases it for $p$ even.  
 For fixed $p$, bion  contribution to higher states also alternates as a function of level number $k$. These two types of alternation in sign and the phase of the bion event 
 is a consequence of what is called  {\it hidden topological angle}. It also  arise from the phase associated with Lefschetz thimble in path integral realization of this problem. 
 
 \item For  quantized $p$,  
  states with level number $k \geq p-1$  on the lower harmonic well have degenerate pairs on the other well, and the degeneracy holds to all    orders in perturbation theory.  There is both instanton and bion contribution to the energy levels, leading to level splitting and shifting of the center-of-energy for the given level. 
 
 \item For generic unquantized values of   $p$, since there is no alignment of levels between the two wells, instanton contribution completely disappear for any level number $k$. The leading non-perturbative effects are due to bions as it arises naturally from exact-WKB 
  \eqref{eq:deltaDW_pnonint}.
 
\end{itemize}
These non-trivial features  are  consistent with  supersymmetry and  quasi-exact solvability 
 structure in special cases. 
 %\textcolor{red}{
 To convince ourselves, we can also compare the results with the calculations given in the literature. It is indeed in agreement with the results obtained by using a combination of path integral and Bender-Wu method for these problems \cite{Kozcaz:2016wvy}.
 %}
 This does not yet complete our discussion of the quantum-deformed double-well potential.  We will also determine
 order by order perturbation theory by using WKB-wave function, and verify explicitly P-NP relation between perturbation theory between perturbative and non-perturbative saddles.

For the triple-well, we set $p=(2q+1)/3$. 
The non-perturbative contribution is obtained as
\be
&& \mbox{Triple-well} \ \mbox{(inner-vacuum)}: \nl \nl
&& \qquad \delta^{\varepsilon=(-1)^{q}}_{k+q\in 2{\mathbb Z}+1}(\hbar) = \mp i \frac{ {\frak B}_0 2^{-\frac{1}{2}(2+k-q)} \hbar^{-\frac{1}{2} (1+3 k-q)}}{\Gamma (1+k) \Gamma(\frac{1}{2}+\frac{k}{2}-\frac{q}{2})} \nl
&& \qquad \qquad \qquad \qquad  \quad  + \frac{{\frak B}_0^2 2^{-(2+k-q)} \hbar ^{-(1+3k-q)}}{\Gamma (1+k)^2 \Gamma (\frac{1}{2}+\frac{k}{2}-\frac{q}{2})^2} \Psi^{(1)}_{\pm 2}\left(k,\frac{2q+1}{3} \right) + O({\frak B}_0^3) \label{eq:TW_delta1_in_Bborel}, \\ \nl
&& \qquad \delta^{\varepsilon=(-1)^{q+1}}_{k+q\in 2{\mathbb Z}+1}(\hbar) = {\cal P} \sqrt{\frac{{\frak B}_0}{\pi \Gamma(1+k) \Gamma(\frac{1}{2}+\frac{k}{2}-\frac{q}{2})}} 2^{-\frac{1}{4}(2+k-q)} \hbar^{-\frac{1}{4}(1+3k-q)} \nl
&& \qquad \qquad \qquad \qquad  \quad - \frac{{\frak B}_0 2^{-\frac{1}{2}(2+k-q)} \hbar^{-\frac{1}{2}(1+3k-q)}}{2\pi \Gamma(1+k) \Gamma(\frac{1}{2}+\frac{k}{2}-\frac{q}{2})} \Psi^{(1)}_{\pm 2}\left(k,\frac{2q+1}{3} \right) + O({\frak B}_{0}^{3/2}) \label{eq:TW_delta2_in_Bborel}, \\ \nl 
&& \qquad \delta_{k+q \notin 2{\mathbb Z}+1} (\hbar) =  \mp i \frac{{\frak B}_0 2^{- \frac{1}{2}\left(2 + k - q \right)} e^{\mp \frac{\pi i}{2}(k-q)} \hbar ^{- \frac{1}{2} \left(1+3k- q \right)} \Gamma  (\frac{1}{2}-\frac{k}{2}+ \frac{q}{2})}{\pi \Gamma (1+k)} \nl
&& \qquad \qquad \qquad \qquad  \quad   +  \frac{{\frak B}_0^2 2^{- \left(2+k-q \right)} e^{\mp \pi i (k-q)}  \hbar ^{- \left(1+3k-q \right)} \Gamma  (\frac{1}{2}-\frac{k}{2}+ \frac{q}{2})^2}{\pi^2 \Gamma (1+k)^2} \Psi^{(1)}_{\pm 2}\left(k,\frac{2q+1}{3} \right) + O({\frak B}_0^3), \nl \label{eq:TW_delta3_in_Bborel}
\ee
\be
&& \mbox{Triple-well} \ \mbox{(outer-vacua)}: \nl \nl
&& \qquad \delta^{\varepsilon=(-1)^q}_{q \in {\mathbb Z}}(\hbar) = \mp i \frac{{\frak B}_0 2^{-\frac{1}{2}(3+2k)} \hbar^{-(2+3k+q)}}{\Gamma (1+k) \Gamma (2+2 k+q)} \nl
&& \qquad \qquad \quad \quad \quad \quad \, + \frac{{\frak B}_0^2 2^{-(3+2k)} \hbar^{-2(2+3 k+ q)}}{\Gamma (1+k)^2 \Gamma (2+2 k+q)^2} \Psi^{(2)}_{\pm 2}\left(k,\frac{2q+1}{3}\right)  +O({\frak B}_0^3), \label{eq:TW_delta1_out_Bborel} \\ \nl
&& \qquad \delta^{\varepsilon=(-1)^{q+1}}_{q \in {\mathbb Z}}(\hbar) = {\cal P} \sqrt{\frac{{\frak B}_0}{\pi\Gamma (1+k) \Gamma (2+2 k+q)}} 2^{-\frac{1}{4}(3+2k)} \hbar^{-\frac{1}{2}(2+3k+q)} \nl
&& \qquad \qquad \quad \quad \quad \quad \ \, \, - \frac{{\frak B}_0 2^{-\frac{1}{2}(3+2k)} \hbar^{-(2+3 k+ q)}}{\pi \Gamma (1+k) \Gamma (2+2 k+q)} \Psi^{(2)}_{\pm 2}\left(k,\frac{2q+1}{3}\right) + O({\frak B}_0^{3/2}),  \label{eq:TW_delta2_out_Bborel} \\ \nl
&& \qquad \delta_{q \notin {\mathbb Z}} (\hbar) =  - \frac{{\frak B}_0 2^{-\frac{1}{2}(3+2k)} \hbar^{- ( 2 +3 k+ q)} \Gamma (-1-2 k-q)}{ \pi \Gamma (1+k) } \left( \varepsilon - e^{\mp \pi i q} \right) \nl
&& \qquad \qquad \qquad \ \, -\frac{{\frak B}_0^2 2^{-(3+2k)} \hbar^{- 2(2+3 k+q)} \Gamma (-1-2 k-q)^2}{\pi^2 \Gamma (1+k)^2} \left(\varepsilon -  e^{\mp \pi i q}   \right)^2 \Psi^{(2)}_{\pm 2}\left(k,\frac{2q+1}{3}\right)  + O({\frak B}_0^3),  \label{eq:TW_delta3_out_Bborel} \nl 
\ee
where ${\cal P} \in \{-1,+1\} $, and
\be
&& \Psi^{(1)}_{n}(k,p) := 2\psi ^{(0)}(1+k) + \psi ^{(0)}\left(\frac{3}{4} + \frac{k}{2} -\frac{3p}{4}\right)+ \log 2 + 3 \log \hbar + \pi n i,  \label{eq:Psin1} \\
&& \Psi^{(2)}_{n}(k,p) := \psi ^{(0)}(1+k)+2   \psi ^{(0)}\left(\frac{3}{2} + 2k +\frac{3p}{2}\right)+ \log 2 + 3 \log \hbar + \pi n i, \label{eq:Psin2}
\ee
with  $\psi^{(n)}(x)$ is the polygamma function.

These equations are solutions of the exact quantization conditions for quantum deformed triple-well potential.    In order not to be repetitive, we will provide  the implications of these equations after describing 
the exact resurgent cancellations between Borel-resummation of perturbation theory and bion amplitudes in \S. \ref{sec:quantized}  and  \ref{sec:unquantized}

\begin{comment}
We can comment on their implications as we did for quantum deformed double-well.  
Multiple remarks are in order about these results. (COME BACK TO THIS AGAIN.) 
\begin{itemize}
    \item First, consider the term proportional to instanton factor, $\sqrt{{\frak B}_0} \sim e^{-S_{\rm I}/\hbar}$
This term is proportional to $\frac{1}{\Gamma((1+k - q)/2)}$.  Therefor, for  level numbers $k$ for which $(1+k - q)/2$ is a negative integer, the instanton contribution vanishes.   Then, outer and inner well states are not aligned and the leading contribution is from bion. 

\end{itemize}
\end{comment}

%%%%%%%%%%%%%%%%%%%%%%%%%%%%%%%%%%%%%%%%%%%%%
\subsubsection{Quasi-solvable cases}
%%%%%%%%%%%%%%%%%%%%%%%%%%%%%%%%%%%%%%%%%%%%%
When the perturbative part of the energy spectra is convergent by tuning $(k,p)$, the energy is said to be quasi-solvable.
In this case, the discontinuity disappears from the energy spectra.
The condition can be obtained from $\delta(\hbar)$ in Eqs.(\ref{eq:deltaDW_pint})-(\ref{eq:TW_delta1_in_Bborel})  as
\be
 \mbox{Double-well} &:& \quad \mbox{$p \in {\mathbb Z}$ \ and \ $k-|p|<0$}, \nl
 \mbox{Triple-well} \ \mbox{(inner-vacuum)}&:& \quad \mbox{$q \in {\mathbb N}$ \ and \  $k - q \in 2{\mathbb Z}_{\le 0}-1$} = \{-1,-3, \ldots\},  \label{eq:cond_QES} \\
 \mbox{Triple-well} \ \mbox{(outer-vacua)}&:& \quad \mbox{$q \in {\mathbb Z}_{<0} -1$ \ and \ $2k + q \in {\mathbb Z}_{\le 0}-2$} = \{-2,-4, \ldots\}. \nn
\ee
{\bf SUSY case:} This corresponds to  $q=1$ or $q=-2$ which implies $p=\pm1$.  For $q=1$,  $k-q$ is in the special  set  \eqref{eq:cond_QES}  only for $k=0$, and for $q=-2$,  $2k+q$ is in the special  set 
only for $k=0$. This signifies that in the SUSY case, these two types of states  are solvable to all orders in perturbation theory. However, only one of $e^{\pm W(x)}$ is normalizable, $L^2(\mathbb R)$ function, leading to only one harmonic state. The other cases will be discussed in detail in \S. \ref{sec:quantized}.

%%%%%%%%%%%%%%%%%%%%%%%%%%%%%%%%%%%%%%%%%%%%%
\subsection{Perturbation theory as a function of level number and deformation parameter}
%%%%%%%%%%%%%%%%%%%%%%%%%%%%%%%%%%%%%%%%%%%%%

We can  obtain the (perturbative part of) energy spectra by explicitly calculating $F_{\ell}(E,\hbar)$ which constitutes an $A$-cycle by Eq.(\ref{eq:Fell_A}). The data that enters there is $\oint^{a_{2\ell}}_{a_{2\ell-1}} dx \, S_{\rm odd}(x,E,\hbar)$, 
which appears in the exponent of the   the WKB-wave function, and  it  is  a formal asymptotic series. 
After obtaining $F_{\ell}(E,\hbar)$,  we would like to determine the perturbative expansion of energy as a function of level 
number. 
Since all of $A$-cycles can be expressed by $F_{\ell}(E,\hbar)$, the perturbative energy spectra can be obtained  by solving
\be
{\frak D}_{\rm pt}(F_{\ell}(E(\hbar),\hbar)) = 0,
\ee
%under the condition given by Eq.(\ref{eq:cond_QES})
%{\color{red} I do not think we  need this condition. Perturbation theory also holds for other states in the theory. Is the description of  rationale in this part  correct?}
%\textcolor{blue}{SK: If you do not care of QES, you do not need it, as you said.
%I commented it out.}, where ${\frak D}_{\rm pt}(F_{\ell}(E,\hbar))$ denotes  
which is the perturbative part of the quantization condition obtained by setting 
 the   $B$-cycle contribution is set to zero.   Inverting this relation, we obtain perturbative expansion of the energy 
\be
E \rightarrow E(k, p, \hbar) = \sum_{n=0}^{\infty} E_n (k, p) \hbar^n.
\ee
Explicitly, this gives an all order  perturbative expansion for the energy
eigenvalues as a function of coupling $\hbar$, level number $k$, and deformation 
parameter $p$. 
\begin{align}
 \mbox{Double-well}:  \nl
  E_0 =&k+\frac{1-p}{2}, \nl
 E_1 =&\frac{1}{2} \left(-6 k^2+6 k p-6 k-p^2+3 p-2\right), \nl
 E_2 =&  -\frac{1}{4} (2 k-p+1) \left(34 k^2-34 k p+34 k+4 p^2-17 p+18\right), \nl
  E_3 =& \frac{1}{4} \left(-750 k^4+1500 k^3 p-1500 k^3-996 k^2 p^2+2250 k^2 p -1584 k^2+246 k p^3 \right. \\
 &\left. -996 k p^2+1584 k p-834 k-16 p^4+123 p^3-346 p^2+417 p-178\right), \nl
 \qquad E_4 =& -\frac{1}{16} (2 k-p+1) \left(21378 k^4-42756 k^3 p+42756 k^3+27612 k^2 p^2-64134 k^2 p \right. \nl
 &\left. +55472 k^2 -6234 k p^3+27612 k p^2-55472 k p+34094 k+336 p^4-3117 p^3 \right. \nl
 & \left. +10907 p^2 -17047 p+10026\right), \nn
\end{align}
It is reassuring  to see   that the perturbative expansion obtained from  
all orders Bohr-Sommerfeld quantization condition 
in which  all orders WKB wave function is used  agrees precisely with the result of 
 Bender-Wu analysis \cite{Kozcaz:2016wvy}.

The same analysis for the triple-well potential yields perturbative expansion for the inner-well energy levels (recall $\omega_{\rm in} = \frac{1}{2}$ and $\omega_{\rm out} = 1$ in our convention)
\begin{align}
\mbox{Triple-well} \ &\mbox{(inner-vacuum)}:  \nl
 E_{\rm in, 0} =& \frac{1}{2}\left(k + \frac{1}{2} -p \right), \nl
 E_{\rm in, 1} =& -\frac{3 k^2}{2}+\frac{3 k p}{2}-\frac{3 k}{2}+\frac{3 p}{4}-\frac{3}{4},\nl
E_{\rm in, 2} =& -6 k^3+9 k^2 p-9 k^2-\frac{9 k p^2}{4}+9 k p-\frac{39 k}{4}-\frac{9 p^2}{8}+\frac{9 p}{2}-\frac{27}{8},\nl
E_{\rm in, 3} =& -\frac{105 k^4}{2}+105 k^3 p-105 k^3-54 k^2 p^2+\frac{315 k^2 p}{2}-\frac{363 k^2}{2}+\frac{27 k p^3}{4} -54 k p^2 \label{pt-inner} \\  
 &+\frac{705 k p}{4} -129 k+\frac{27 p^3}{8}-27 p^2+\frac{495 p}{8}-\frac{153}{4}, \nl
 E_{\rm in, 4} =& -603 k^5+\frac{3015 k^4 p}{2}-\frac{3015 k^4}{2}-1170 k^3 p^2+3015 k^3 p-3645 k^3+324 k^2 p^3 \nl
 &-1755 k^2 p^2+5391 k^2 p-3960 k^2-\frac{405 k p^4}{16}  +324 k p^3-\frac{15885 k p^2}{8}+\frac{7767 k p}{2} \nl
 &-\frac{39897 k}{16}-\frac{405 p^4}{32}+162 p^3-\frac{11205 p^2}{16}+1188 p-\frac{20385}{32}. \nn 
\end{align}
and outer-well energy levels 
\begin{align}
\mbox{Triple-well} \ &\mbox{(outer-vacua)}:  \nl
E_{\rm out, 0} =& k+\frac{p}{2}+\frac{1}{2},\nl
E_{\rm out, 1} =& -6 k^2-6 k p-6 k-\frac{9 p^2}{8}-3 p-\frac{15}{8}, \nl
E_{\rm out, 2} =& -48 k^3-72 k^2 p-72 k^2-\frac{63 k p^2}{2}-72 k p-\frac{93 k}{2}-\frac{27 p^3}{8}-\frac{63 p^2}{4}-\frac{189 p}{8}-\frac{45}{4},\nl
E_{\rm out, 3} =& -840 k^4-1680 k^3 p-1680 k^3-1161 k^2 p^2-2520 k^2 p-1671 k^2-\frac{621 k p^3}{2} \label{pt-outer} \\
 &-1161 k p^2-\frac{3363 k p}{2} -831 k-\frac{729 p^4}{32}-\frac{621 p^3}{4}-\frac{6219 p^2}{16}-\frac{1683 p}{4}-\frac{5265}{32},\nl
 E_{\rm out, 4} =& -19296 k^5-48240 k^4 p-48240 k^4-45540 k^3 p^2-96480 k^3 p-65340 k^3-19764 k^2 p^3 \nl
  &-68310 k^2 p^2  -98316 k^2 p-49770 k^2-3726 k p^4-19764 k p^3-46449 k p^2-50076 k p \nl
 & -20259 k -\frac{6561 p^5}{32}-1863 p^4-\frac{107109 p^3}{16}-\frac{23679 p^2}{2}-\frac{327861 p}{32}-\frac{6885}{2}. \nn
\end{align}
Remarkably, when the inner-well levels   coincide with the outer-well levels at the harmonic level, 
this agreement continues to {\it all orders} in perturbation theory.se
%{\color{red} Is there a simple proof of this using perturbative exact quantization condition? It feels like it. }
%\textcolor{blue}{
%SK: 
By expressing $E_{{\rm in/out},n}=E_{{\rm in/out},n}(k,p)$, it can be made sure that $E_{{\rm in},n}(2k+q+1,\frac{2q+1}{3})=E_{{\rm out},n}(k,\frac{2q+1}{3})$ for any $n \in {\mathbb N}_0$.
%}
%This is true for half of the states   in the central well. living in the common parts of the well. 
This  phenomena happens only for certain quantized values of $p$. Defining $p= \frac{2q+1}{3}$, the alignment of outer and inner levels happen for $q \in \mathbb Z$, see Fig. \ref{fig:Pspec1} and \ref{fig:Pspec2}.
For classical triple-well potential $p=0$, this alignment never happens. For supersymmetric  and QES cases where $ q \in \mathbb Z$ is quantized, 
exact alignment of outer and inner well  takes place.  For $ q \notin \mathbb Z$, this does not  happen. As a result, when alignment happens, instantons will also contribute, while for the generic case, despite the fact that instantons are present as a solution to the first order BPS equation, they do not contribute to spectrum at leading order, and only  their correlated events, bions, do.

%%%%%%%%%%%%%%%%%%%%%%%%%%%%%%%%%%%%%%%%%%%%%
\subsection{Low order P-NP relation for bions}
%%%%%%%%%%%%%%%%%%%%%%%%%%%%%%%%%%%%%%%%%%%%%
The  P-NP relation  is a  low-order/low order constructive relation connecting perturbation theory around perturbative vacuum and perturbation theory around a non-perturbative saddle, such as instanton or bion.  In the WKB language, it connects P and NP cycles,  and it  is expressed by a partial differential equation depending on energy and coupling 
\cite{Alvarez1, Alvarez2, Alvarez3, Dunne:2013ada,Dunne:2014bca, Dunne:2016qix, Gahramanov:2015yxk, Kozcaz:2016wvy,Dunne:2016jsr} 
There are two known versions of this relation, one is connecting to perturbative  fluctuations around instantons, and the other is the one around bions. 
In \cite{Dunne:2013ada,Dunne:2014bca, Dunne:2016qix}, P-NP  relation  is formulated for  genus-1 classical systems. 

For quantum deformed potentials
an earlier suggestive work  is Ref.~\cite{Kozcaz:2016wvy}, which treats  quantum deformed sine-Gordon and double-well potentials,  and  provides  convincing numerical evidence that P-NP relation   holds for  bions.  Ref.~\cite{Kozcaz:2016wvy},  using Bender-Wu method   \cite{Bender:1973rz} automatizied in the BW package \cite{Sulejmanpasic:2016fwr} calculates 
the large order growth of perturbation theory as $\frac {n!}{(S_{\rm B}/\hbar)^n} \left ( 1+  \frac{ S_{\rm B}/\hbar}{n} \tilde b_1(k, p)
+ \frac{ (S_{\rm B}/\hbar)^2}{n(n-1)} \tilde b_2( k, p)  + \ldots \right)$ for various $k$ and $p$ including  the sub-leading corrections,  and  found the coefficients  $\tilde b_n (k, p)$ by using asymptotic analysis numerically. 
 Then, 
assuming that P-NP holds, it calculated perturbation theory  at low orders around bions, 
from which one obtains $ \sim e^{-S_{\rm B}/\hbar} (1+b_1(k, p) \hbar  + b_2(k, p)\hbar^2 + \ldots)$. And remarkably,  the numerical results $\tilde b_n (k, p)$ agree with the polynomials  $b_n(k, p)$  obtained from P-NP relation within an error of order $10^{-6}$.  In this section, 
 by starting with definition of  $A$ and $B$ cycles given by Eqs.(\ref{eq:A-cy}) and (\ref{eq:B-cy}) and using the Mellin transform, we analytically demonstrate  
 the  P-NP relation for   the quantum deformed double- and triple-well potentials.

In order to obtain the P-NP  relation, we redefine the $B$-cycles in terms of
$G_{\ell}(E,\hbar)$ as
\be
   %{\frak B}_{\ell}(E,\hbar) &=& 2 \pi {\frak B}_0 \prod_{\ell=1}^2 \frac{C_{\ell-}(E,\hbar)}{C_{\ell+}(E,\hbar)} \frac{e^{(-1)^{\ell+1}\pi i F_\ell(E,\hbar)}\hbar^{F_\ell(E,\hbar)}}{\Gamma(1/2-F_\ell(E,\hbar))} 
      %&=&  e^{-2 \pi i [E +(-1)^{\ell+1} p/2]} + O(\hbar), \\
   {\frak B}_{\ell}(E,\hbar) &=& 2 \pi {\frak B}_0 \prod_{\ell^\prime=0}^1 \frac{C_{\ell+\ell^\prime -}(E,\hbar)}{C_{{\ell+\ell^\prime} +}(E,\hbar)} \frac{e^{(-1)^{\ell^\prime}\pi i F_{\ell+\ell^\prime}(E,\hbar)}\hbar^{F_{\ell+\ell^\prime}(E,\hbar)}}{\Gamma(1/2-F_{\ell+\ell^\prime}(E,\hbar))} %=: \bar{\frak B}(E,\hbar) \nl
   \nl
   %&\rightarrow& 2 \pi e^{-{G}_\ell(E,\hbar)} \prod_{\ell^\prime=0}^1 \frac{C_{\ell+\ell^\prime -,0}(E)}{C_{\ell+\ell^\prime+,0}(E)} \frac{e^{(-1)^{\ell^\prime}\pi i F_{\ell+\ell^\prime,0}(E)}\hbar^{F_{\ell+\ell^\prime}(E,\hbar)}}{\Gamma(1/2-F_{\ell+\ell^\prime}(E,\hbar))}, 
   &\rightarrow& 2 \pi e^{-{G}_\ell(E,\hbar)} {\frak C}_{\ell}(E,\hbar) \prod_{\ell^\prime=0}^1  \frac{\hbar^{F_{\ell+\ell^\prime}(E,\hbar)}}{\Gamma(1/2-F_{\ell+\ell^\prime}(E,\hbar))}, \label{eq:B-cy_G} \\
   {\frak C}_{\ell}(E,\hbar) &:=&  \prod_{\ell^\prime=0}^1 \left[ \frac{C_{\ell+\ell^\prime -,0}(E(F_{\ell+\ell^\prime,0}))}{C_{\ell+\ell^\prime+,0}(E(F_{\ell+\ell^\prime,0}))} e^{(-1)^{\ell^\prime}\pi i F_{\ell+\ell^\prime,0}} \right]_{F_{\ell+\ell^\prime,0} \rightarrow F_{\ell+\ell^\prime}(E,\hbar)},
\ee
Here $E(F_{\ell,0})$ is the inverse function of $F_{\ell,0}(E)$, $F_{\ell,0}(E)$ and $C_{\ell \pm,0}(E)$ are the coefficients of $O(\hbar^0)$ terms in the expansion of  $F_{\ell}(E,\hbar)$ and $C_{\ell \pm}(E,\hbar)$, respectively.
We assume that $G_{\ell}(E,\hbar)$ has the expanded form in terms of $\hbar$ as 
\be
\label{eq:NP-function}
G_{\ell}(E,\hbar) = \sum_{n=-1}^\infty G_{\ell,n}(E) \hbar^n.
\ee
where $G_{\ell,-1}(E) \equiv S_{\rm B}$ is the bion action, and other terms are related to the perturbative fluctuations around bions in a  simple way. %{\bf I know this is easy, but do we need to explain this a bit more...}

In the double and symmetric triple-well cases,  the $A$-cycles and $B$-cycles can be expressed  by a single  $\bar{\frak A}(E,\hbar)$ and single $\bar{\frak B}(E,\hbar)$, respectively. For more generic polynomial potentials, we have more than one perturbative and non-perturbative cycle. Note that even for  the symmetric triple well, there are  actually two perturbative $A$ cycles, but they are related in a precise way \eqref{inout}. 

The $A$-cycles can be written through $\bar{\frak A}(E,\hbar)$, 
\be
&& \qquad \qquad \bar{\frak A}(E,\hbar) =e^{2 \pi i \bar{F}(E,\hbar)}, 
\label{eq:Abar}
\ee
whereas $B$-cycles are expressed by
\be
&& \mbox{Double-well}: \nl
&& \qquad \qquad  \bar{\frak B}(E,\hbar) = \frac{2 \pi e^{-\bar{G}(E,\hbar)}}{\Gamma(\frac{1+p}{2}-\bar{F}(E,\hbar)) \Gamma(\frac{1-p}{2}-\bar{F}(E,\hbar))} \left( \frac{\hbar}{2} \right)^{2 \bar{F}(E,\hbar)}, \label{eq:B_DW} \\ \nl
&& \mbox{Triple-well}: \nl
&& \qquad \qquad \bar{\frak B}(E,\hbar) = \frac{2 \pi e^{-\bar{G}(E,\hbar)}}{\Gamma(\frac{1-p}{2}-\bar{F}(E,\hbar)) \Gamma(\frac{1+p}{2}-2\bar{F}(E,\hbar))} 2^{\bar{F}(E,\hbar)+p/2}\hbar^{3 \bar{F}(E,\hbar)}, \label{eq:B_TDW}
\ee
where $\bar{F}(E,\hbar)= \sum_{n=0}^{\infty} \bar{F}_n(E)\hbar^{n}$ which depends on the potential. 
For the systems we have, we use slight abbreviations: 
\be
&& \mbox{Double-well}: \nl
&& \qquad \bar{F}_0(E)=F_{\ell,0}(E) - (-1)^\ell \frac{p}{2} = -E, \qquad \bar{F}_{n \in {\mathbb N}}(E) = F_{\ell,n \in {\mathbb N}}(E),\label{eq:F_DW} \\ \nl
&& \mbox{Triple-well}: \nl
&& \qquad \bar{F}_0(E) =F_{\ell=1(3),0}(E) - \frac{p}{2} %= F_{3,0}(E) - \frac{p}{2} 
= \frac{F_{\ell=2,0}(E)}{2} + \frac{p}{4} = -E, \qquad \bar{F}_{n \in {\mathbb N}}(E) = F_{\ell,n \in {\mathbb N}}(E). \label{eq:F_TDW}
\ee
$\bar{G}(E,\hbar)$ can be evaluated by using the Mellin transform 
explained in  Appendix~\ref{sec:Mellin} in detail.
By obtaining $\bar{F}(E,\hbar)$ and $\bar{G}(E,\hbar)$ using the residue integral around the turning point and the Mellin transform, respectively, we  convert 
\begin{align}
\label{eq:convertEF}
    (\bar{F}(E,\hbar),\bar{G}(E,\hbar)) \longrightarrow (E(\bar{F},\hbar),\bar{G}(\bar{F},\hbar))
\end{align}
where $\bar{F}$  appears as the argument of these function.  This form is the natural one to find 
a  relationship between $E(\bar{F},\hbar)$ and $\bar{G}(\bar{F},\hbar)$.

Let us now derive the P-NP  relation.
The procedure to obtain the functional form of $\bar{G}(E,\hbar)$ for the double-well and triple-well are  explicitly explained in Appendix~\ref{sec:derivation_G_DW} and \ref{sec:derivation_G_TW}, respectively. 
The first few expansion coefficients of the non-perturbative function $\bar{G}(E)$ given in  \eqref{eq:NPfunc2} and \eqref{eq:NPfunc3} are of the form:
\begin{align}
\mbox{Double-well}: \nl
\bar{G}_{-1}(E) =& S_{\rm B} = \frac{1}{3}, \nl
\bar{G}_{0}(E) =& 0, \nl
\bar{G}_1(E) =& 17 E^2-\frac{3 p^2}{4}+\frac{19}{12}, \\
\bar{G}_2(E) =& 227 E^3-\frac{77 E p^2}{4}+\frac{187 E}{4}, \nl
\bar{G}_3(E) =& \frac{47431 E^4}{12}-\frac{3717 E^2 p^2}{8}+\frac{34121 E^2}{24}+\frac{341 p^4}{64}-\frac{1281 p^2}{32}+\frac{28829}{576}, \nl
\bar{G}_4(E) =& \frac{317629 E^5}{4}-\frac{35560 E^3 p^2}{3}+\frac{264725 E^3}{6}+\frac{19215 E p^4}{64}-\frac{253045 E p^2}{96}+\frac{842909 E}{192}, \nn
\end{align}
where the leading term is the bion action, and  the other terms play a role in determination of perturbation theory around bion.  Similarly, for the triple well, it is: 
\begin{align}
\mbox{Triple-well}: \nl
\bar{G}_{-1}(E) =& S_{\rm B} = \frac{1}{4}, \nl
\bar{G}_{0}(E) =& 0, \nl
\bar{G}_{1}(E) =& 36 E^2-\frac{9 p^2}{8}+\frac{21}{8}, \nl
\bar{G}_{2}(E) =& 852 E^3-\frac{207 E p^2}{4}+\frac{519 E}{4}-\frac{21 p^3}{16}+\frac{21 p}{16}, \\
\bar{G}_{3}(E) =& 25536 E^4-2106 E^2 p^2+6606 E^2-78 E p^3+78 E p+\frac{531 p^4}{32}-\frac{2043 p^2}{16}+\frac{5139}{32}, \nl
\bar{G}_{4}(E) =& 875304 E^5-90603 E^3 p^2+344475 E^3-\frac{15237 E^2 p^3}{4}+\frac{15237 E^2 p}{4}+\frac{49653 E p^4}{32}\nl
 &-\frac{222669 E p^2}{16}+\frac{750501 E}{32}+\frac{6633 p^5}{128}-\frac{22599 p^3}{64}+\frac{38565 p}{128}. \nn
\end{align}
\if0
\be
&& \mbox{Triple-well}: \nl
&& \qquad \bar{G}_{-1}(E) = S_{\rm B} = \frac{1}{4}, \nl
&& \qquad \bar{G}_{0}(E) = 0, \nl
&& \qquad \bar{G}_{1}(E) = 36 E^2-\frac{9 p^2}{8}+\frac{21}{8}, \nl
&& \qquad \bar{G}_{2}(E) = 852 E^3-\frac{207 E p^2}{4}+\frac{519 E}{4}-\frac{21 p^3}{16}+\frac{21 p}{16}, \nl
&& \qquad \bar{G}_{3}(E) = 25536 E^4-2106 E^2 p^2+6606 E^2-78 E p^3+78 E p+\frac{531 p^4}{32}-\frac{2043 p^2}{16}+\frac{5139}{32}, \nl
&& \qquad \bar{G}_{4}(E) = 875304 E^5-90603 E^3 p^2+344475 E^3-\frac{15237 E^2 p^3}{4}+\frac{15237 E^2 p}{4}+\frac{49653 E p^4}{32}\nl
&& \qquad \qquad \qquad \ -\frac{222669 E p^2}{16}+\frac{750501 E}{32}+\frac{6633 p^5}{128}-\frac{22599 p^3}{64}+\frac{38565 p}{128}. \nn
\ee
\fi
Then, we change the independent variable from $E$ to $\bar{F}$ in the formal series expansion 
following  \eqref{eq:convertEF} and write 
$\bar{G}(\bar{F},\hbar)$ and $E(\bar{F}, \hbar )$ series expansions as:
\be 
&& \mbox{Double-well}: \nl
&& \qquad \bar{G}_{-1}(\bar{F}) = \frac{1}{3}, \nl
&& \qquad \bar{G}_{0}(\bar{F}) = 0, \nl
&& \qquad \bar{G}_{1}(\bar{F}) = 17 \bar{F}^2-\frac{3 p^2}{4}+\frac{19}{12},  \nl
&& \qquad \bar{G}_{2}(\bar{F}) = -125 \bar{F}^3+\frac{43 \bar{F} p^2}{4}-\frac{153 \bar{F}}{4},\\
&& \qquad \bar{G}_{3}(\bar{F}) = \frac{17815 \bar{F}^4}{12}-\frac{1485 \bar{F}^2 p^2}{8}+\frac{23405 \bar{F}^2}{24}+\frac{101 p^4}{64}-\frac{821 p^2}{32}+\frac{22709}{576}, \nl
&& \qquad \bar{G}_{4}(\bar{F}) = -\frac{87549 \bar{F}^5}{4}+\frac{7105 \bar{F}^3 p^2}{2}-\frac{50715 \bar{F}^3}{2}-\frac{4775 \bar{F} p^4}{64}+\frac{47675 \bar{F} p^2}{32}-\frac{217663 \bar{F}}{64}, \nl \nl
&& \qquad E_{0}(\bar{F}) = -\bar{F}, \nl
&& \qquad E_{1}(\bar{F}) = -3 \bar{F}^2+\frac{p^2}{4}-\frac{1}{4}, \nl
&& \qquad E_{2}(\bar{F}) = 17 \bar{F}^3-\frac{9 \bar{F} p^2}{4}+\frac{19 \bar{F}}{4}, \\
&& \qquad E_{3}(\bar{F}) = -\frac{375 \bar{F}^4}{2}+\frac{129 \bar{F}^2 p^2}{4}-\frac{459 \bar{F}^2}{4}-\frac{11 p^4}{32}+\frac{71 p^2}{16}-\frac{131}{32}, \nl
&& \qquad E_{4}(\bar{F}) = \frac{10689 \bar{F}^5}{4}-\frac{4455 \bar{F}^3 p^2}{8}+\frac{23405 \bar{F}^3}{8}+\frac{909 \bar{F} p^4}{64}-\frac{7389 \bar{F} p^2}{32}+\frac{22709 \bar{F}}{64}. \nn
\ee
For triple-well system, we obtain:
\be
&& \mbox{Triple-well}: \nl
&& \qquad \bar{G}_{-1}(\bar{F}) = \frac{1}{4}, \nl
&& \qquad \bar{G}_{0}(\bar{F}) = 0, \nl
&& \qquad \bar{G}_{1}(\bar{F}) = 36 \bar{F}^2-\frac{9 p^2}{8}+\frac{21}{8}, \nl
&& \qquad  \bar{G}_{2}(\bar{F}) = -420 \bar{F}^3+\frac{99 \bar{F} p^2}{4}-\frac{411 \bar{F}}{4}-\frac{21 p^3}{16}+\frac{21 p}{16}, \\
&& \qquad \bar{G}_{3}(\bar{F}) = 8040 \bar{F}^4-675 \bar{F}^2 p^2+4275 \bar{F}^2+51 \bar{F} p^3-51 \bar{F} p+\frac{9 p^4}{4}-\frac{279 p^2}{4}+117, \nl
&& \qquad \bar{G}_{4}(\bar{F}) = -192024 \bar{F}^5+20655 \bar{F}^3 p^2-181575 \bar{F}^3-\frac{7155 \bar{F}^2 p^3}{4}+\frac{7155 \bar{F}^2 p}{4} \nl
&& \qquad \qquad \qquad \ -\frac{6399 \bar{F} p^4}{32} +\frac{107055 \bar{F} p^2}{16}-\frac{547551 \bar{F}}{32}+\frac{1701 p^5}{128}-\frac{15795 p^3}{64}+\frac{29889 p}{128}. \nl \nl
&& \qquad E_{{\rm in},0}(\bar{F}) = -\frac{\bar{F}}{2}, \nl
&& \qquad E_{{\rm in},1}(\bar{F}) = -\frac{3 \bar{F}^2}{2}+\frac{3 p^2}{8}-\frac{3}{8}, \nl
&& \qquad E_{{\rm in},2}(\bar{F}) = 6 \bar{F}^3-\frac{9 \bar{F} p^2}{4}+\frac{21 \bar{F}}{4}+\frac{3 p^3}{8}-\frac{3 p}{8}, \\
&& \qquad E_{{\rm in},3}(\bar{F}) = -\frac{105 \bar{F}^4}{2}+\frac{99 \bar{F}^2 p^2}{4}-\frac{411 \bar{F}^2}{4}-\frac{21 \bar{F} p^3}{4}+\frac{21 \bar{F} p}{4}-\frac{9 p^4}{32}+\frac{153 p^2}{16}-\frac{297}{32}, \nl
&& \qquad E_{{\rm in},4}(\bar{F}) = 603 \bar{F}^5-\frac{675 \bar{F}^3 p^2}{2}+\frac{4275 \bar{F}^3}{2}+\frac{153 \bar{F}^2 p^3}{2}-\frac{153 \bar{F}^2 p}{2}+\frac{27 \bar{F} p^4}{2}-\frac{837 \bar{F} p^2}{2} \nl
&& \qquad \qquad \quad \quad \ \ +702 \bar{F}-\frac{81 p^5}{32}+\frac{675 p^3}{16}-\frac{1269 p}{32}. \nl \nl
&& \qquad E_{{\rm out},0}(\bar{F}) = -\bar{F}, \nl 
&& \qquad E_{{\rm out},1}(\bar{F}) = -6 \bar{F}^2+\frac{3 p^2}{8}-\frac{3}{8}, \nl
&& \qquad E_{{\rm out},2}(\bar{F}) = 48 \bar{F}^3-\frac{9 \bar{F} p^2}{2}+\frac{21 \bar{F}}{2}+\frac{3 p^3}{8}-\frac{3 p}{8}, \\
&& \qquad E_{{\rm out},3}(\bar{F}) = -840 \bar{F}^4+99 \bar{F}^2 p^2-411 \bar{F}^2-\frac{21 \bar{F} p^3}{2}+\frac{21 \bar{F} p}{2}-\frac{9 p^4}{32}+\frac{153 p^2}{16}-\frac{297}{32}, \nl
&& \qquad E_{{\rm out},4}(\bar{F}) = 19296 \bar{F}^5-2700 \bar{F}^3 p^2+17100 \bar{F}^3+306 \bar{F}^2 p^3-306 \bar{F}^2 p+27 \bar{F} p^4 \nl
&& \qquad \qquad \qquad \quad \ -837 \bar{F} p^2+1404 \bar{F}-\frac{81 p^5}{32}+\frac{675 p^3}{16}-\frac{1269 p}{32}, \nn
\ee
Notice that the triple-well has two independent perturbative vacua, meaning that two energy spectra can be defined.  These are $E_{\rm in}$ and $E_{\rm out}$. Note that the perturbative expansion in these two harmonic vacua are related in a precise way as: 
\be
\label{inout}
&& E_{\rm in}(2\bar{F},\hbar) = E_{\rm out}(\bar{F},\hbar).
\ee
At this stage, we calculated perturbative information around perturbative saddle and non-perturbative bion saddles as a  function of  $(\bar{F},\hbar)$ where $\bar{F} $ can be viewed as continuation of level number and  $\hbar$ is the coupling. 
 By inspection, we observe that  the formal perturbative 
$E(\bar{F},\hbar)$ 
and non-perturbative series $\bar{G}(\bar{F},\hbar)$ are related to each other in a precise and simple way\footnote{
\textcolor{black}{
The results are observations from the above computations up to $O(\hbar^4)$.
}
}
. 
\be
&& \mbox{Double-well}: \nl
&& \qquad \qquad \frac{\pd E(\bar{F},\hbar)}{\pd \bar{F}} = \frac{\hbar}{S_{\rm B}} \left( -2 \bar{F} + \hbar \frac{\pd \bar{G}(\bar{F},\hbar)}{\pd \hbar} \right), \\ \nl
&& \mbox{Triple-well}: \nl
&& \qquad \qquad \frac{\pd E_{\rm in}(\bar{F},\hbar)}{\pd \bar{F}} = \frac{\hbar}{2S_{\rm B}} \left(-\frac{3}{2} \bar{F} +\hbar \frac{\pd \bar{G}(\bar{F}/2,\hbar)}{\pd \hbar} \right), \label{DU2}\\
&& \qquad \qquad \frac{\pd E_{\rm out}(\bar{F},\hbar)}{\pd \bar{F}} = \frac{\hbar}{S_{\rm B}} \left(-3 \bar{F} +\hbar \frac{\pd \bar{G}(\bar{F},\hbar)}{\pd \hbar} \right).
\label{DU3}
\ee
\textcolor{black}{Notice that the difference of the factor $1/2$ multiplying $\bar{F}$ between Eqs.(\ref{DU2})and (\ref{DU3}) comes from the curvature around each vacuum, i.e. $\omega_{\rm in}=\omega_{\rm out}/2$.}
There are the P-NP  relations for the quantum deformed double-and triple well potentials between perturbative vacuum and bion configurations. 
These relations tells us that the information around non-perturbative saddles can be extracted completely from perturbative expansion in a simple constructive way.  
Since the solution of the exact quantization condition is a transseries and it is  encoded into these two functions, this implies that 
the perturbative expansion encodes all non-perturbative  information in the problem. 
Note that in our problem, we assumed that we are only considering energies below the potential barrier. It is interesting to investigate this problem above the barrier, where the topology of the Stokes graph and the form of exact quantization condition changes.

%%%%%%%%%%%%%%%%%%%%%%%%%%%%%%%%%%%%%%%%%%%%%
\subsection{Path integral expression} \label{sec:path-integral}
%%%%%%%%%%%%%%%%%%%%%%%%%%%%%%%%%%%%%%%%%%%%%
The relation between partition function   $Z^\pm(\beta)$ (evaluatet at $\arg(\hbar)=0^{\pm}$)  in path integral formulation and Fredholm determinant ${\frak D}^\pm$ 
that appears in   exact quantization condition    can be written as
\be
Z^\pm(\beta) &=& -\frac{1}{2 \pi i} \int^{\epsilon + i \infty}_{\epsilon - i \infty} \frac{\pd \log {\frak D}^\pm}{\pd E} e^{-\beta E} dE \nl
&=& -\frac{\beta}{2 \pi i} \int^{\epsilon + i \infty}_{\epsilon - i \infty} \log {\frak D}^\pm  e^{-\beta E} dE. \label{eq:Zpm}
\ee
%Since $\arg(\hbar)=0$ is Stokes line,  both sides of the equation are naturally analytically continued to  $\arg(\hbar)=0^{\pm}$.  
In order to obtain the path-integral, ${\frak D}^{\pm}$ can be decomposed into the perturbative and the non-perturbative parts in \eqref{eq:DpmDW2}   and \eqref{eq:QcondTW2}  as
\be
\label{DTFactor1}
&& \mbox{Double-well}: \nl
&& \qquad \qquad  {\frak D}^{\pm} = {\frak D}^{+}_{\bar{\frak A},-p} {\frak D}^{-}_{\bar{\frak A},-p} \left[ 1 + \frac{\bar{\frak B}}{{\frak D}^{\mp}_{\bar{\frak A},-p} {\frak D}^{\mp}_{\bar{\frak A},+p}} \right], \\
&& \mbox{Triple-well}: \nl
&& \qquad \qquad {\frak D}^{\pm}  =  \left( {\frak D}^{+}_{\bar{\frak A},+p} \right)^2 {\frak D}^{-}_{\bar{\frak A}^2,+p} \left[1+   \frac{2\bar{\frak B}}{{\frak D}^{\mp}_{\bar{\frak A},\mp p} {\frak D}^{\mp}_{\bar{\frak A}^2,\pm p}}  +  \frac{\bar{\frak B}^2}{\left( {\frak D}^{\mp}_{\bar{\frak A},\mp p} \right)^2 {\frak D}^{\mp}_{\bar{\frak A}^2,\pm p}} \right],
\label{DTFactor2}
\ee
where the factors
\be
   {\frak D}^{\pm}_{\bar{\frak A},+p} = 1 + e^{+\pi i p} \bar{\frak A}^{\pm 1}.
\ee
The product that factorize  in \eqref{DTFactor1} and \eqref{DTFactor2} has all orders perturbation theory data for all levels below the barrier,  and inside the parenthesis includes the non-perturbative cycles. This decomposition will help us to express the exact quantization as a sum over non-perturbative saddles in the path integral representation. Since this construction involves all orders (below the potential barrier), it should help tremendously to  decode information about arbitrary levels from the path integral, not only ground state. 
The  $\log {\frak D}^\pm$ is given by
\be
\mbox{Double-well}:&& \nl
\log {\frak D}^{\pm}
&=& \log {\frak D}^{+}_{\bar{\frak A},-p} + \log {\frak D}^{-}_{\bar{\frak A},-p} + \log \left[ 1 + \frac{\bar{\frak B}}{{\frak D}^{\mp}_{\bar{\frak A},-p} {\frak D}^{\mp}_{\bar{\frak A},+p}} \right] \nl
   &=& -\sum_{n=1}^{\infty} \frac{(- e^{-\pi i p})^n}{n} \left( \bar{\frak A}^{+n} + \bar{\frak A}^{-n} \right) - \sum_{n=1}^{\infty} \frac{(-1)^n}{n} \left[ \sum_{\ell=0}^{\infty} \sum_{m=0}^\ell \bar{\frak B} (-1)^{\ell} e^{-\pi i p (\ell-2m)}  \bar{\frak A}^{\mp \ell} \right]^n,  \nl
%   &=& -\sum_{n=1}^{\infty} \frac{(- e^{-\pi i p})^n}{n} \left( \bar{\frak A}^{+n} + \bar{\frak A}^{-n} \right) - \sum_{n=1}^{\infty} \frac{(-1)^n}{n} \left[ \sum_{\ell=0}^{\infty}  \bar{\frak B} (-1)^{\ell}{P}_\ell  \bar{\frak A}^{\mp \ell} \right]^n, 
\label{eq:logD_DW}
\ee
\be
\mbox{Triple-well}:&& \nl
\log {\frak D}^{\pm}
&=& \log  \left( {\frak D}^{+}_{\bar{\frak A},+p} \right)^2  + \log {\frak D}^{-}_{\bar{\frak A}^2,+p} + \log \left[1+   \frac{2\bar{\frak B}}{{\frak D}^{\mp}_{\bar{\frak A},\mp p} {\frak D}^{\mp}_{\bar{\frak A}^2,\pm p}}  +  \frac{\bar{\frak B}^2}{\left( {\frak D}^{\mp}_{\bar{\frak A},\mp p} \right)^2 {\frak D}^{\mp}_{\bar{\frak A}^2,\pm p}} \right] \nl
%&=& 2\log  {\frak D}^{+}_{\bar{\frak A},+p}  + \log {\frak D}^{-}_{\bar{\frak A}^2,+p} - \sum_{n=1}^\infty \sum_{k=0}^{\lfloor \frac{n}{2} \rfloor} \frac{(-1)^{n+k}2^{n-2k}}{n-k}
%\begin{pmatrix}
%  n-k \\
%  k
%\end{pmatrix} 
%\frac{\bar{\frak B}^{n}}{\left({\frak D}^{\mp}_{\bar{\frak A},\mp p}\right)^{n} \left( {\frak D}^{\mp}_{\bar{\frak A}^2,\pm p}\right)^{n-k}} \nl
&=& -\sum_{n=1}^{\infty} \frac{(- e^{\pi i p})^n}{n} \left( 2\bar{\frak A}^{+n} + \bar{\frak A}^{-2n} \right) \nl
&& - 2\sum_{n=1}^\infty  \frac{(-1)^{n}}{n}
\sum_{\ell_1,\ell_2 = 0}^\infty \bar{\frak B}^{n} (-e^{\pm \pi i p})^{-\ell_1+\ell_2} \frac{(n)_{\ell_1} (n)_{2 \ell_2}}{\ell_1! (2\ell_2)!} \bar{\frak A}^{\mp (\ell_1+2 \ell_2)}, \label{eq:logD_TW}
\ee
where $(x)_n$ is the Pochhammer symbol and $\lfloor \bullet \rfloor$ is the floor function.
The first term in Eqs.(\ref{eq:logD_DW}) and (\ref{eq:logD_TW}) corresponds to the perturbative part, whereas the second term gives the non-perturbative part.
%In addition, $P_\ell$ in Eq.(\ref{eq:logD_DW}) is given by
%\be
%{P}_{\ell} := \sum_{m=0}^\ell e^{-\pi i p (\ell-2m)} =
%\begin{cases}
%  (-1)^{p \ell} (\ell+1)  & \mbox{if $p \ \in {\mathbb Z}$} \\
%  \frac{\sin[\pi p(\ell+1)]}{\sin(\pi p)} & \mbox{otherwise} 
%\end{cases}.
%\ee
It is notable that $\log {\frak D}^{\pm}$ corresponds to the Gutzwiller trace formula through $G^{\pm}(E)= -\pdv{E}\log {\frak D}^{\pm}(E)$.
%%%%%%%%%%%%%%%%%%%%%%%%%%%%%%%%%%%%%%%%%%%%%
%\subsection{Hidden topological angle and Maslov index}
%%%%%%%%%%%%%%%%%%%%%%%%%%%%%%%%%%%%%%%%%%%%%
$e^{\pm\pi ip}$ in Eqs.(\ref{eq:logD_DW}) and (\ref{eq:logD_TW}) is what is called the \textit{hidden topological angle} (HTA) \cite{Behtash:2015loa,Behtash:2015kna, Alireza2018}. 
This is the contribution of the fermion determinant $\pm \frac{\hbar p}{2} \frac{\pd W(x,\hbar)}{\pd x}$, but from the point of view of complex classical solutions of quantum deformed equations of motions, it can also be interpreted as an additional phase part of the action of the complex bion solution, which can be seen from\footnote{There are two complementary and equivalent perspective on HTA. One is to consider instanton-antiinstanton critical point at infinity. In that case, the quasi-moduli  steepest descent cycle (Lefschetz thimble) lives in the complex domain. And the phase arise from the integral over the quasi-zero mode direction. It is an invariant associated with the QZM  Lefschetz thimble integral \cite{Behtash:2018voa}. The second perspective is the one described above. Solving the second order equations of motions for the quantum deformed potential, one ends up with a complex solution.  The imaginary part of this bion action is given by hidden topological angle.}
\begin{align}
    \Im S_{\rm cb}&= \Im \int_0^T d\tau \qty(\frac{1}{2}\dot{x}_{\rm cb}(\tau)^2+V[\dot{x}_{\rm cb}(\tau)])\nl
    &=\frac{1}{2\hbar}\oint_C dx \sqrt{2E+W'^2+p\hbar W''}=ip \pi.
\end{align}
In terms of the Gutzwiller trace formula, it corresponds to a contribution that shifts the Maslov index associated with each periodic solution.

Let us derive a more familiar form for the non-perturbative part of the path integral\cite{Sato:2001ac}, which we denote by  $Z_{\rm np}$.
For the double-well case, from Eqs.(\ref{eq:Zpm}), (\ref{eq:logD_DW}) and (\ref{eq:Abar})-(\ref{eq:F_TDW}) one finds
\begin{align}
    Z_{\rm np}&=\frac{\beta}{2\pi i}\int_{\epsilon-i\infty}^{\epsilon+i\infty}\sum_{n=1}^\infty (-1)^n\frac{1}{n} \qty[\frac{\bar{\frak B}}{{\frak D}^{\mp}_{\bar{\frak A},-p} {\frak D}^{\mp}_{\bar{\frak A},+p}}]^n e^{-\beta E}dE\nl
    &=\frac{\beta}{2\pi i}\int_{\epsilon-i\infty}^{\epsilon+i\infty}\sum_{n=1}^\infty (-1)^n\frac{1}{n}\qty[e^{\mp 2\pi iE}\frac{e^{-S_{\rm B}/\hbar}}{2\pi}\Gamma\qty(\frac{1-p}{2}-E)\Gamma\qty(\frac{1+p}{2}-E)\qty(\frac{\hbar}{2})^{-2E}]^n e^{-\beta E}dE\nl
    &=e^{-\beta/2}\frac{\beta}{2\pi i}\int_{\epsilon-i\infty}^{\epsilon+i\infty}\sum_{n=1}^\infty \frac{1}{n}\qty[e^{\mp 2\pi is}\frac{e^{-S_{\rm B}/\hbar}}{2\pi}\Gamma\qty(-\frac{p}{2}-s)\Gamma\qty(\frac{p}{2}-s)\qty(\frac{\hbar}{2})^{-2s}\frac{2}{\hbar}]^n e^{-\beta s}ds. \label{eq:Znp_DW}
\end{align}
Here, $\epsilon$ should be chosen to make the ground state energy as the lowest one.\footnote{In principle, ${\frak D}^{\pm}$ includes physically irrelevant spectrum which is less than the ground state energy. $\epsilon$ should taken to pick up $E\ge(1-|p|)/2 + O(e^{-S_{\rm B}/\hbar})=E_0+O(e^{-S_{\rm B}/\hbar})$ for the double-well case.
For the triple-well case, $E \ge (1/2-p)/2 + O(e^{-S_{\rm B}/\hbar})$ if $p$ is positive, and $E \ge (1+p)/2 + O(e^{-S_{\rm B}/\hbar})$, otherwise.}
According to \cite{Sueishi:2019xcj}, \eqref{eq:Znp_DW} can be expressed as 
\begin{align}
  Z_{\rm np}&=e^{-\beta/2}\beta \sum_{n=1}^\infty\qty(\frac{e^{-S_{\rm B}/\hbar}}{2\pi}\frac{2}{\hbar})^n {\rm QMI}^{n}(p),
\end{align}
where the factor $2/\hbar$ comes from the fluctuation determinant around the bion solution. The form of quasi-moduli integral is
\begin{align}
   {\rm QMI}^n(p)&=\frac{1}{n}\qty(\prod_{j=1}^{2n}\int_0^\infty d\tau_j e^{-\mathcal{V}_j(\tau_j)})\delta\qty(\sum_{k=1}^{2n} \tau_k-\beta), \nl
  \mathcal{V}_j(\tau)&=\begin{cases}
    -\frac{2}{\hbar}e^{-\tau}+\frac{p}{2}\tau & \quad \mbox{for odd $j$}\\
    -\frac{2}{\hbar}e^{-\tau}-\frac{p}{2}\tau & \quad \mbox{for even $j$}
  \end{cases},
\end{align}
where the $O(\hbar^{-1})$ term is the classical interaction between the instanton and anti-instantons, and the $O(\hbar^{0})$ term is there due to quantum deformation of the potential.
It can be evaluated as
\begin{align}
   {\rm QMI}^n(p)=\frac{1}{2\pi i n}\int_{-i\infty}^{i\infty}ds\qty[\qty(e^{\pm i\pi (\frac{p}{2}-s)}\qty(\frac{\hbar}{2})^{\frac{p}{2}-s}\Gamma\qty(\frac{p}{2}-s))\qty(e^{\pm i\pi (-\frac{p}{2}-s)}\qty(\frac{\hbar}{2})^{-\frac{p}{2}-s}\Gamma\qty(-\frac{p}{2}-s))]^n e^{-s\beta},
\end{align}
which is nice consistent with Eq.(\ref{eq:Znp_DW}).
It is clear from the present derivation process that these two perspectives, coming from exact quantization condition and path integral carefully incorporating quasi-moduli integration,  are actually equivalent. In other words, the former perspective is that each cycle acquires a phase separately from the Maslov index due to the contribution of fermion, while the latter perspective is that the cycle is transformed into a gamma function and rewritten as the contribution of QMI.
This equivalence can only be obtained by writing down the Fredholm determinant and partition function using cycles in this way.

\begin{figure}
    \centering
    \includegraphics[width=11cm]{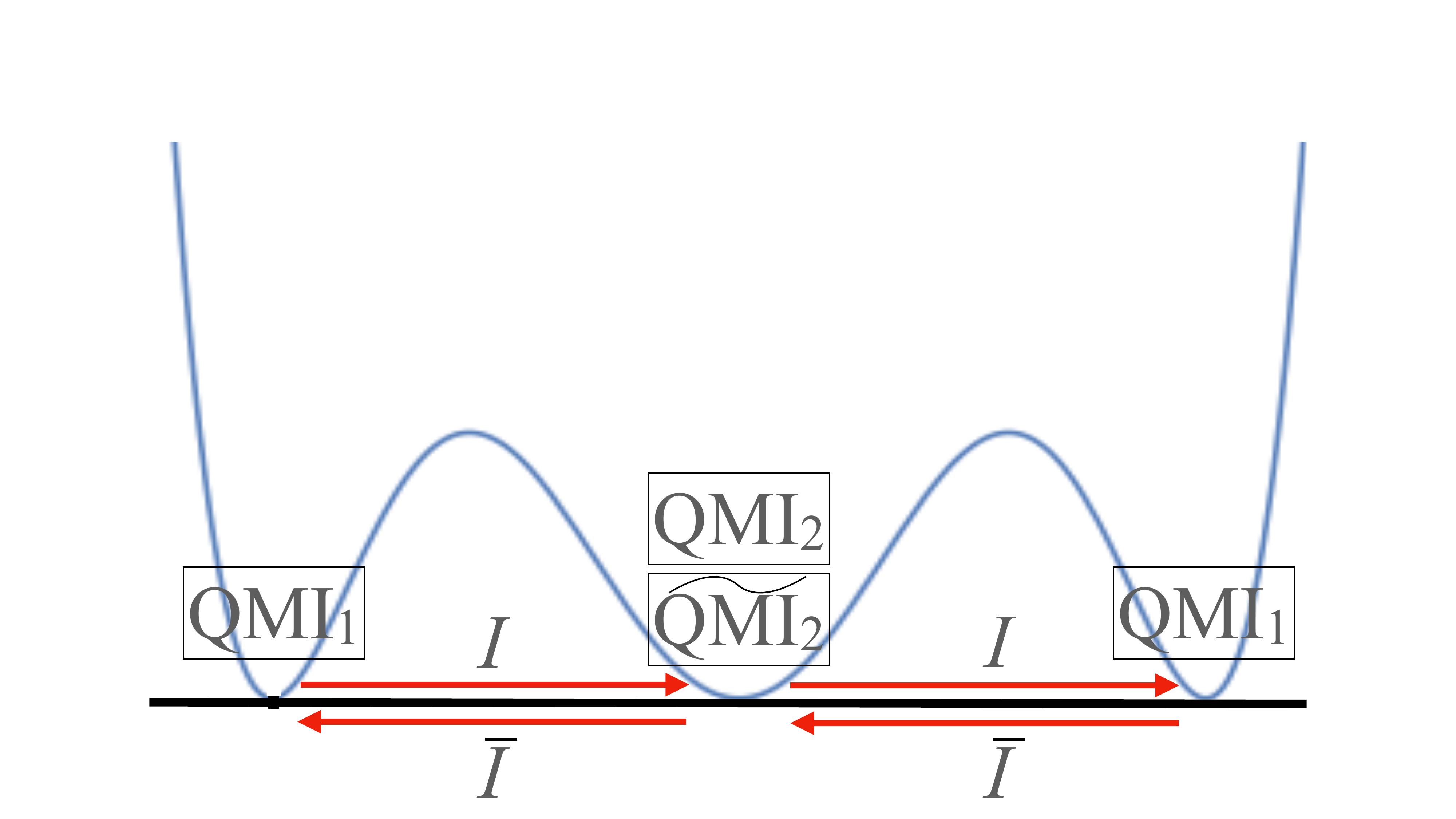}
    \caption{The each QMI exists between $I$ and $\bar{I}$ or $I(\bar{I})$ and $I(\bar{I})$. The effect of the deformed parameter $p$ is included in each QMI. }
    \label{fig:TripleWell_QMI}
\end{figure}

Now, let us describe  the triple-well case.  From Eqs.(\ref{eq:logD_TW}) and (\ref{eq:Abar})-(\ref{eq:F_TDW}), one finds
\be
    Z_{\rm np}&=&\frac{\beta}{2\pi i}\int_{\epsilon-i\infty}^{\epsilon+i\infty}\sum_{n=1}^\infty \frac{(-1)^n}{n} \left[\frac{2\bar{\frak B}}{{\frak D}^{\mp}_{\bar{\frak A},\mp p} {\frak D}^{\mp}_{\bar{\frak A}^2,\pm p}}  +  \frac{\bar{\frak B}^2}{\left( {\frak D}^{\mp}_{\bar{\frak A},\mp p} \right)^2 {\frak D}^{\mp}_{\bar{\frak A}^2,\pm p}}\right]^n e^{-\beta E}dE\nl
    &=&\frac{\beta}{2\pi i}\int_{\epsilon-i\infty}^{\epsilon+i\infty}\sum_{n=1}^\infty \frac{(-1)^n}{n} \left[2 e^{\mp 3\pi i E}\frac{e^{-S_{\rm B}/\hbar}}{2\pi} \Gamma\qty(\frac{1+p}{2}-E)\Gamma\qty(\frac{1-p}{2}-2E) \frac{\hbar^{-3 E}}{2^{E-p/2}} \right. \nl
    && \left. +e^{\mp6\pi iE}\qty(1+e^{\pm2\pi i(2E+p/2)})\frac{e^{-2S_{\rm B}/\hbar}}{(2\pi)^2}\Gamma\qty(\frac{1+p}{2}-E)^2\Gamma\qty(\frac{1-p}{2}-2E)^2 \frac{\hbar^{-6E}}{2^{2E-p}} \right]^n e^{-\beta E}dE\nl
    &=&\frac{\beta}{2\pi i}\int_{\epsilon-i\infty}^{\epsilon+i\infty}\sum_{n=1}^\infty \frac{1}{n}\left( 2[I\bar{I}]-[I\bar{I}]^2+[II][\bar{I}\bar{I}]\right)^n e^{-\beta E}dE \nl
    &=& -\frac{\beta}{2\pi i}\int_{\epsilon-i\infty}^{\epsilon+i\infty}\sum_{\substack{n,m=0\\n+m \ge 1}}^\infty \frac{\, _3F_2\left(1,1,1-n-m;\frac{3}{2}-\frac{n}{2}-m,2-\frac{n}{2}-m;1\right)}{2^{-2+n+2 m} \Gamma (1+m) \Gamma (3-n-2 m) \Gamma (n+m)} \nl 
    && \cdot (-1)^{n+m}[I\bar{I}]^n ([II] [\bar{I}\bar{I}])^m e^{-\beta E}dE, 
\ee
where $\,_pF_q(\{a_p\};\{b_q\};z)$ is the hypergeometric function.
Moreover, $[I\bar{I}]=\qty(\frac{e^{-S_{\rm B}/\hbar}}{2\pi\hbar}\frac{1}{\sqrt{2}})\mathrm{QMI}_1\mathrm{QMI}_2$, $[II] [\bar{I}\bar{I}]=\qty(\frac{e^{-S_{\rm B}/\hbar}}{2\pi\hbar}\frac{1}{\sqrt{2}})^2(\mathrm{QMI}_1)^2(\widetilde{\mathrm{QMI}_2})^2$, where  $\frac{e^{-S_{\rm B}/\hbar}}{2\pi\hbar}\frac{1}{\sqrt{2}}$ is one bion contribution including  fluctuation determinant. We also note the different signs in front of $[I\bar{I}], [I\bar{I}]^2, [II][\bar{I}\bar{I}]$ come from the contribution of the Maslov index. $[I\bar{I}]$ and $[II][\bar{I}\bar{I}]$ are counted as 1-cycle, which obtains $-1$ but $[I\bar{I}]^2$ is two-cycle, which obtains $(-1)^2$. The difference between $\mathrm{QMI}_2$ and $\widetilde{\mathrm{QMI}_2}$ comes from the fact that the force acting between instanton($I$) and instanton is repulsive, while the force acting between instanton and anti-instanton($\bar{I}$) is attractive (Fig.\ref{fig:TripleWell_QMI}). It makes the opposite signs in front of 
the classical interaction term $ \pm \frac{1}{\hbar} e^{-\tau}$ term  in the  QMI integrals for the triple-well system:
\begin{align}
    \mathrm{QMI}_1&=\int_0^\infty d\tau_1 e^{\qty(E-\frac{1+p}{2})\tau_1+\frac{1}{2\hbar}e^{-\tau_1}}=\Gamma\qty(\frac{1+p}{2}-E)\qty(\frac{2}{\hbar})^{E-\frac{1+p}{2}}e^{\pm\pi i\qty(\frac{1+p}{2}-E)}\nl
    \mathrm{QMI}_2&=\int_0^\infty d\tau_2e^{\qty(2E-\frac{1-p}{2})\tau_2+\frac{1}{\hbar}e^{-\tau_2}}=\Gamma\qty(\frac{1-p}{2}-2E)\qty(\frac{1}{2\hbar})^{2E-\frac{1-p}{2}}e^{\pm \pi i(\frac{1-p}{2}-2E)} \\
    \widetilde{\mathrm{QMI}_2}&=\int_0^\infty d\tau_2e^{\qty(2E-\frac{1-p}{2})\tau_2-\frac{1}{\hbar}e^{-\tau_2}}=\Gamma\qty(\frac{1-p}{2}-2E)\qty(\frac{1}{2\hbar})^{2E-\frac{1-p}{2}} \nn
\end{align}
Again, the partition function obtained by manipulating   exact quantization condition and 
path integral carefully incorporating quasi-moduli integration  are  equivalent. 

\begin{comment}
For NB
\small
\begin{align}
    &\mathrm{QMI}_1\mathrm{QMI}_2=\nl
    &\qty(\int_0^\infty d\tau_1 e^{\qty(E-\frac{1+p}{2})\tau_1+\frac{1}{2\hbar}e^{-\tau_1}})\qty(\int_0^\infty d\tau_2e^{\qty(2E-\frac{1-p}{2})\tau_2+\frac{1}{\hbar}e^{-\tau_2}}) = \Gamma\qty(\frac{1+p}{2}-E)\Gamma\qty(\frac{1-p}{2}-2E)\frac{\hbar^{-3E+1}}{2^{E-p/2}}\sqrt{2}e^{\mp\pi i(3E-1)} 
\end{align}
\normalsize
For $TB^2$
\small
\begin{align}
    &(\mathrm{QMI}_1)^2(\widetilde{\mathrm{QMI}_2})^2=\nl
    &\qty(\int_0^\infty d\tau_1 e^{\qty(E-\frac{1+p}{2})\tau_1+\frac{1}{2\hbar}e^{-\tau_1}})^2\qty(\int_0^\infty d\tau_2e^{\qty(2E-\frac{1-p}{2})\tau_2-\frac{1}{\hbar}e^{-\tau_2}})^2 = \Gamma\qty(\frac{1+p}{2}-E)^2\Gamma\qty(\frac{1-p}{2}-2E)^2\frac{\hbar^{-6E+2}}{2^{2E-p}}2e^{\mp\pi i(2E-p-1)} 
\end{align}
\normalsize
\end{comment}

\begin{comment}
Memo:... Changing $\hbar \rightarrow \hbar e^{\mp \pi i}$ and $2 \rightarrow 2 e^{\pm \pi i}$ coupling with $e^{-\tau}$ gives
\begin{align}
    \qty(\int_0^\infty d\tau_1  e^{\qty(s-\frac{1+p}{2})\tau_1+e^{\pm 2\pi i} \frac{2}{\hbar}e^{-\tau_1}})\qty(\int_0^\infty d\tau_2e^{\qty(2s-\frac{1-p}{2})\tau_2+\frac{1}{2\hbar}e^{-\tau_2}}) = \Gamma\qty(\frac{1+p}{2}-s)\Gamma\qty(\frac{1-p}{2}-2s)\frac{\hbar^{-3s}}{2^{s-p/2}}e^{\mp i\pi(s-p/2+1)} \nl
\end{align}
\end{comment}

%%%%%%%%%%%%%%%%%%%%%%%%%%%%%%%%%%%%%%%%%%%%%
\section{Median resummation of exact  quantization condition}
%%%%%%%%%%%%%%%%%%%%%%%%%%%%%%%%%%%%%%%%%%%%%
Let us reconsider the exact form of the quantization conditions.
Although the exact form is given by Eqs.  \eqref{eq:DpmDW2},   \eqref{eq:QcondTW2} and 
is invariant under the DDP formula
(\ref{eq:bar_DDP}), i.e, ${\cal S}_+[{\frak D}^+] = {\cal S}_-[{\frak D}^-]$,
it is not very  helpful expression because it is a function of ${\cal S}_{\pm} [{\frak A}_{\ell}]$, which has a non-perturbative contribution in it, and the solutions 
are not automatically in the Borel-Ecalle resummed form which is free of ambiguities. 
Now, let us express the exact form as a function of $({\frak A}_{\ell},{\frak B}_{\ell})$.

In order to do so, we consider the median resummation using the Stokes automorphism ${\frak S}$ which is defined as \cite{Ec1,2014arXiv1405.0356S,Dorigoni2019,Aniceto:2013fka}
\be
{\cal S}_{+} = {\cal S}_- \circ {\frak S}.
\ee
Stokles automorphism can also be expressed through the Alien derivative ${\Delta}_{w}$ as 
\be
&& {\frak S}^{\nu} = \exp \left[ \nu \sum_{w \in \Gamma} \bul{\Delta}_w \right] %= \exp \left[ \sum_{w \in \Gamma} e^{-w/\hbar} \Delta_{w}  \right]
= 1 + \sum_{k=1}^{\infty} \sum_{\{n_1,\cdots,n_k \ge 1\}} \frac{\nu^k}{k!} \prod_{s=1}^k \bul{\Delta}_{w_{n_s}} %\bul{\Delta}_{w_{n_1}}  \cdots \bul{\Delta}_{w_{n_k}}
, \\
&& \bul{\Delta}_w := e^{-w/\hbar} \Delta_w,
\ee
where $\Gamma$ is the set of singular points on the positive real axis in the Borel plane, $w_n \in \Gamma$ is the $n$-th singular point given by $w_n= n S_{\rm B}$ with $S_{\rm B}$ denoting  
the bion action  of $B$-cycles, and ${\frak S}=:{\frak S}^{+1}$. 
The alien derivative is a useful tool to probe the singularities in the Borel plane as described below.  
By defining the median resummation as
\be
{\cal S}_{\rm med}:= {\cal S}_{+} \circ {\frak S}^{-1/2} = {\cal S}_{-} \circ {\frak S}^{+1/2},
\ee
the quantization condition can be expressed by ${\frak D}_{\rm ex}$ as
\be
{\frak D}^{\pm} = {\frak S}^{\mp 1/2}[{\frak D}_{\rm ex}] \quad \Rightarrow \quad {\frak D}_{\rm ex} = {\frak S}^{\pm 1/2}[{\frak D}^{\pm}]. \label{eq:Dex}
\ee
Notice that ${\frak D}_{\rm ex}$ in Eq.(\ref{eq:Dex}) is still an asymptotic expansion, which is a function of $({\frak A}_{\ell},{\frak B}_{\ell})$.
But by replacing them with the Borel resummed form defined as
\be
 {\frak A}_{\ell} &\rightarrow& \widehat{\frak A}_{\ell}:= {\cal S}_{+} \circ {\frak S}^{-1/2}[{\frak A}_\ell] = {\cal S}_{-} \circ {\frak S}^{+1/2}[{\frak A}_\ell], \\
 {\frak B}_{\ell} &\rightarrow& \widehat{\frak B}_{\ell}:= {\cal S}_{+} \circ {\frak S}^{-1/2}[{\frak B}_\ell] = {\cal S}_{-} \circ {\frak S}^{+1/2}[{\frak B}_\ell],
\ee
one can obtain the exact form of the quantization condition.
It is notable that $\widehat{\frak A}_\ell$ does not have the imaginary ambiguity, and $\widehat{\frak A}_\ell(\hbar) \xrightarrow{\hbar \rightarrow 0_+} {\frak A}_\ell(\hbar)$.

The Alien derivative acting on cycles and their Stokes automorphism are given by
\be
\mbox{Double-well} &:& \quad \bul{\Delta}_{w_n}  {\frak A}_\ell =  \frac{(-1)^n}{n} {\frak A}_{\ell} {\frak B}^n, \qquad \bul{\Delta}_{w_n} {\frak B} = 0, \\
&& \quad {\frak S}^{\nu}[{\frak A}_\ell]= {\frak A}_{\ell}(1+{\frak B})^{-\nu}, \qquad {\frak S}^{\nu}[{\frak B}]={\frak B},\\ \nl
\mbox{Triple-well} &:& \quad \bul{\Delta}_{w_n}  {\frak A}_{2\ell-1} =  \frac{(-1)^n}{n} {\frak A}_{2\ell-1} {\frak B}^n_{\ell}, \qquad \bul{\Delta}_{w_n}  {\frak A}_{2} =  \frac{(-1)^n}{n} {\frak A}_{2} ({\frak B}^n_1 + {\frak B}^n_2), \nl
&& \quad \bul{\Delta}_{w_n} {\frak B}_{\ell} = 0, \\ 
&& \quad {\frak S}^{\nu}[{\frak A}_{2\ell-1}]= {\frak A}_{2\ell-1}(1+{\frak B}_{\ell})^{-\nu}, \qquad {\frak S}^{\nu}[{\frak A}_{2}]= {\frak A}_{2} \prod_{\ell=1}^2(1+{\frak B}_{\ell})^{-\nu}, \nl
&& \quad {\frak S}^{\nu}[{\frak B}_\ell]={\frak B}_\ell,
\ee
where $\ell \in \{1,2 \}$. These relations are  the hall-mark of resurgence, the fact that alien derivatives and Stokes automorphisms  of perturbative series in a given saddle  produce the series
around the other saddles and nothing else.    Also note that  
$\bul{\Delta}_{w_n} {\frak B} = 0 $  implies  the perturbative expansion of $B$ cycle  does not have singularities on positive real axis. Due to P-NP relation, we know that the $B$-cycle must also represent a divergent asymptotic expansion. The combination of these two facts implies that 
perturbation theory around the bion saddle must be Borel summable along positive real axis on Borel plane. On the other hand, it is expected to have singularities on the negative real axis.  In comparison,  perturbation theory around perturbative vacuum has infinitely many singularities on the positive real axis associated with bion configurations. 
As a result of these, the median resummation of quantization conditions
${\frak D}_{\rm ex}$  can be expressed as:
\be
\mbox{Double-well} &:& {\frak D}_{\rm ex} = 1 + \frac{{\frak A}_1}{{\frak A}_2} + \left( {\frak A}_1 + {\frak A}^{-1}_2 \right) \left( 1 + {\frak B} \right)^{1/2} \nl
&& \qquad = 1 + e^{-2 i \pi p} + \left( e^{- \pi i p} \bar{\frak A} + e^{- \pi i p} \bar{\frak A}^{-1}\right) \left( 1 + \bar{\frak B}\right)^{1/2}, 
\label{MR_DW}
\\
\mbox{Triple-well} &:& {\frak D}_{\rm ex} = {\frak A}_2^{-1} \prod_{\ell=1}^2 \left[ {\frak A}_{2 \ell -1} + \left( 1 + {\frak B}_{\ell} \right)^{1/2} \right] + \prod_{\ell=1}^2\left[ 1 + {\frak A}_{2 \ell -1}\left( 1+ {\frak B}_{\ell} \right)^{1/2}\right] \nl
&& \qquad = {\frak A}_2^{-1} \left[ {\frak A}_{1} + \left( 1 + {\frak B}_{1} \right)^{1/2} \right]^2 + \left[ 1 + {\frak A}_{1}\left( 1+ {\frak B}_{1} \right)^{1/2}\right]^2 \quad ({\frak A}_1 = {\frak A}_3, {\frak B}_1 = {\frak B}_2) \nl
&& \qquad = \prod_{\varepsilon \in \{-1,+1\}} \left[ e^{3\pi i p/2} - i \varepsilon  +  \left( e^{\pi i p/2} \bar{\frak A}^{-1} - i \varepsilon  e^{\pi i p} \bar{\frak A}\right) \left( 1 + \bar{\frak B} \right)^{1/2} \right].
\label{MR_TW}
\ee

Now, we can  obtain the non-perturbative contribution $\delta(\hbar)$ to the energy from ${\frak D}_{\rm ex}$.
For the double-well, it can be obtained as
\be
&& \mbox{Double-well :} \nl \nl
&& \qquad \delta_{p \in {\mathbb Z}}(\hbar) = {\cal P} \sqrt{\frac{{\frak B}_0}{\pi \hbar \Gamma (1+k) \Gamma (1+k-p)}} \left( \frac{\hbar}{2} \right)^{-k+\frac{p}{2}} \nl
&& \qquad \qquad \qquad \ - \frac{{\frak B}_0}{\pi \hbar \Gamma (1+k) \Gamma (1+k-p)} \left( \frac{\hbar}{2} \right)^{-2k+p} \Phi_0(k,p) + O({\frak B}_0^{3/2}), \label{eq:deltaDW_pint_ex}  \\ \nl
&& \qquad \delta_{p \notin {\mathbb Z}}(\hbar) = (-1)^{1+k} \frac{{\frak B}_0\Gamma(-k+p)}{\pi \hbar \Gamma(1+k)}  \left( \frac{\hbar}{2} \right)^{-2k+p}  \cos (\pi p) \nl
&& \qquad \qquad \qquad \ - \frac{2{\frak B}_0^2\Gamma(-k+p)^2}{\pi^2 \hbar^2 \Gamma(1+k)^2} \left( \frac{\hbar}{2}\right)^{-4k+2p} \cos^2(\pi p)  \cdot \left[ \Phi_0(k,p) - \frac{3\pi}{2} \tan (\pi p) \right]+ O({\frak B}_0^3),  \label{eq:deltaDW_pnonint_ex} 
\ee
where ${\cal P} \in \{-1,+1\}$ is parity, and $\Phi_n(k,p)$ is given by Eq.(\ref{eq:Phin}).

For the triple-well, we set $p=(2q+1)/3$.
The non-perturbative contribution to the energies is given by
\be
&& \mbox{Triple-well (inner-vacuum) :} \nl \nl
&& \qquad  \delta^{\varepsilon=(-1)^q}_{k+q \in 2 {\mathbb Z}+1}(\hbar) = 0, \label{eq:TW_delta1_in_Aborel}  \\ \nl
&& \qquad \delta^{\varepsilon=(-1)^{q+1}}_{k+q \in 2 {\mathbb Z}+1}(\hbar) = {\cal P} \sqrt{\frac{{\frak B}_0}{\pi \Gamma(1+k) \Gamma(\frac{1}{2}+\frac{k}{2}-\frac{q}{2})}}2^{-\frac{1}{4}(2+k-q)} \hbar^{-\frac{1}{4} \left(1+3k-q \right)} \label{eq:TW_delta2_in_Aborel} \\ \nl
&& \qquad \qquad \qquad \qquad  \quad - \frac{{\frak B}_0 2^{-\frac{1}{2}(2+k-q)} \hbar^{-\frac{1}{2} \left(1+3k-q \right)}}{2\pi \Gamma(1+k) \Gamma(\frac{1}{2}+\frac{k}{2}-\frac{q}{2})} \Psi^{(1)}_0 \left(k,\frac{2 q+1}{3}\right) + O({\frak B_0^{3/2}}),\cr \nl
&& \qquad \delta_{k+q \notin 2 {\mathbb Z}+1}(\hbar) =
-\frac{{\frak B}_0 2^{- \frac{1}{2}(2+k-q)}  \hbar ^{-\frac{1}{2} (1+3 k-q)} \Gamma(\frac{1}{2}-\frac{k}{2}+\frac{q}{2})}{\pi \Gamma(1+k)}  \sin \frac{\pi(k-q)}{2} \nl
&& \qquad \qquad \qquad \qquad  \quad - \frac{{\frak B}_0^2 2^{-(2+k-q)}  \hbar ^{-(1+3 k-q)} \Gamma (\frac{1}{2}-\frac{k}{2}+\frac{q}{2})^2}{\pi^2 \Gamma (1+k)^2} \sin^2 \frac{\pi(k-q)}{2} \nl
&& \qquad \qquad \qquad \qquad  \quad \cdot \left[ \Psi^{(1)}_0 \left(k, \frac{2q+1}{3}\right)- 3  \pi  \cot \frac{\pi(k-q)}{2} \right] + O({\frak B}_0^3), \label{eq:TW_delta3_in_Aborel}
\ee
\be
&& \mbox{Triple-well (outer-vacua) :} \nl \nl
&& \qquad \delta^{\varepsilon=(-1)^q}_{q \in {\mathbb Z}}(\hbar) = 0, \label{eq:TW_delta1_out_Aborel}  \\ \nl
&& \qquad \delta^{\varepsilon=(-1)^{q+1}}_{q \in {\mathbb Z}}(\hbar) = {\cal P} \sqrt{\frac{{\frak B}_0}{\pi \Gamma(1+k) \Gamma(2+2k+q)}}2^{-\frac{1}{4}(3+2k)} \hbar^{-\frac{1}{2} \left(2+3k+q \right)} \nl
&& \qquad \qquad \quad \quad \quad \quad \ \, \, - \frac{{\frak B}_0 2^{-\frac{1}{2}(3+2k)} \hbar^{- \left(2+3k+q\right)}}{2\pi \Gamma(1+k) \Gamma(2+2k+q)} \Psi^{(2)}_0 \left(k,\frac{2 q+1}{3}\right) + O({\frak B_0^{3/2}}), \label{eq:TW_delta2_out_Aborel} \\ \nl
&& \qquad \delta_{q \notin {\mathbb Z}}(\hbar) = -\varepsilon \frac{{\frak B}_0 2^{-\frac{1}{2}(3+2k)} \hbar^{-(2+3 k+q)}\Gamma (-1-2k-q)}{ \pi \Gamma (1+k)} \left( \tan \frac{\pi q}{2}\right)^\varepsilon \sin(\pi q) \nl
&& \qquad \qquad \qquad \ \, - \frac{{\frak B}_0^2 2^{-(3+2 k)} \hbar ^{-2 (3 k+q+2)} \Gamma (-1-2k-q)}{\pi^2 \Gamma (1+k)^2}  \left( \tan \frac{\pi q}{2}\right)^{2\varepsilon} \sin^2(\pi q) \nl
&& \qquad \qquad \qquad \ \, \cdot \left[ \Psi^{(2)}_{0}  \left(k,\frac{2q+1}{3}\right) - 3 \pi \varepsilon  \left( \cot \frac{\pi q}{2} \right)^{\varepsilon} \right] + O({\frak B}_0^3), \label{eq:TW_delta3_out_Aborel}
\ee
where $\varepsilon \in \{-1,+1\}$, and $\Psi^{(1,2)}_n(k,p)$ are given by Eqs.(\ref{eq:Psin1}) and (\ref{eq:Psin2}).

%\textcolor{red}{
Although these equations are very complicated,we can extract lots of nontrivial physical facts from them. From now on, we will show such facts one by one:
%}
Firstly, up to our knowledge, 
Eqs.(\ref{eq:TW_delta1_in_Aborel})-(\ref{eq:TW_delta3_out_Aborel}) are the first results of non-perturbative contribution to energies with already built-in resurgent cancellation. Let us briefly explain this.  
In the standard discussions of resurgence in the QM context, one finds a non-perturbative  ambiguity in the Borel resummation of 
perturbation theory, and then, one finds that the bion amplitudes are also two-fold ambiguous. 
Then, one shows that these two types of ambiguities cancel each other out  leading to an ambiguity free result. 
Our formalism, the median resummation of the exact quantization condition already takes care of this  process.  All discontinuities, left/right resummation of perturbation theory, two-fold ambiguous bion amplitudes, or equivalently,  the content of the  DDP formula is taken  into account.  The outcome  of the solution of median resummed exact quantization is already in an   
Borel-Ecalle resummed form.

The physical implications of  \eqref{eq:deltaDW_pint_ex} and \eqref{eq:deltaDW_pnonint_ex} are succinctly explained in  Fig.~\ref{fig:DWPNP}, and  itemized  in \S.\ref{sec:nettes}
for the quantum deformed double-well potential.  Indeed, we see that exact quantization conditions captures all non-trivial features of the whole spectrum, in particular the phenomenon of  appearance (for $p \in \Z$)  and disappearance (for $p \notin \Z$) of instanton contributions to the spectrum is neatly captured by the solution.

There is a very broad range of implications of Eqs.(\ref{eq:TW_delta1_in_Aborel})-(\ref{eq:TW_delta3_out_Aborel}) for  the spectrum of the quantum deformed triple-well potentials, and the result captures both perturbative and non-perturbative properties of different $q \in \mathbb R$ theories at once. 
We now describe them in turn \cite{Dunne:2020gtk,Brezin:1977gk}. 
%The cases of the classical potential and $q \in {\mathbb Z}$ are also discussed in Refs. \cite{Dunne:2020gtk,Brezin:1977gk}.

\begin{figure}[t]
    \centering 
\hspace{1.0cm}
    \includegraphics[width=13cm]{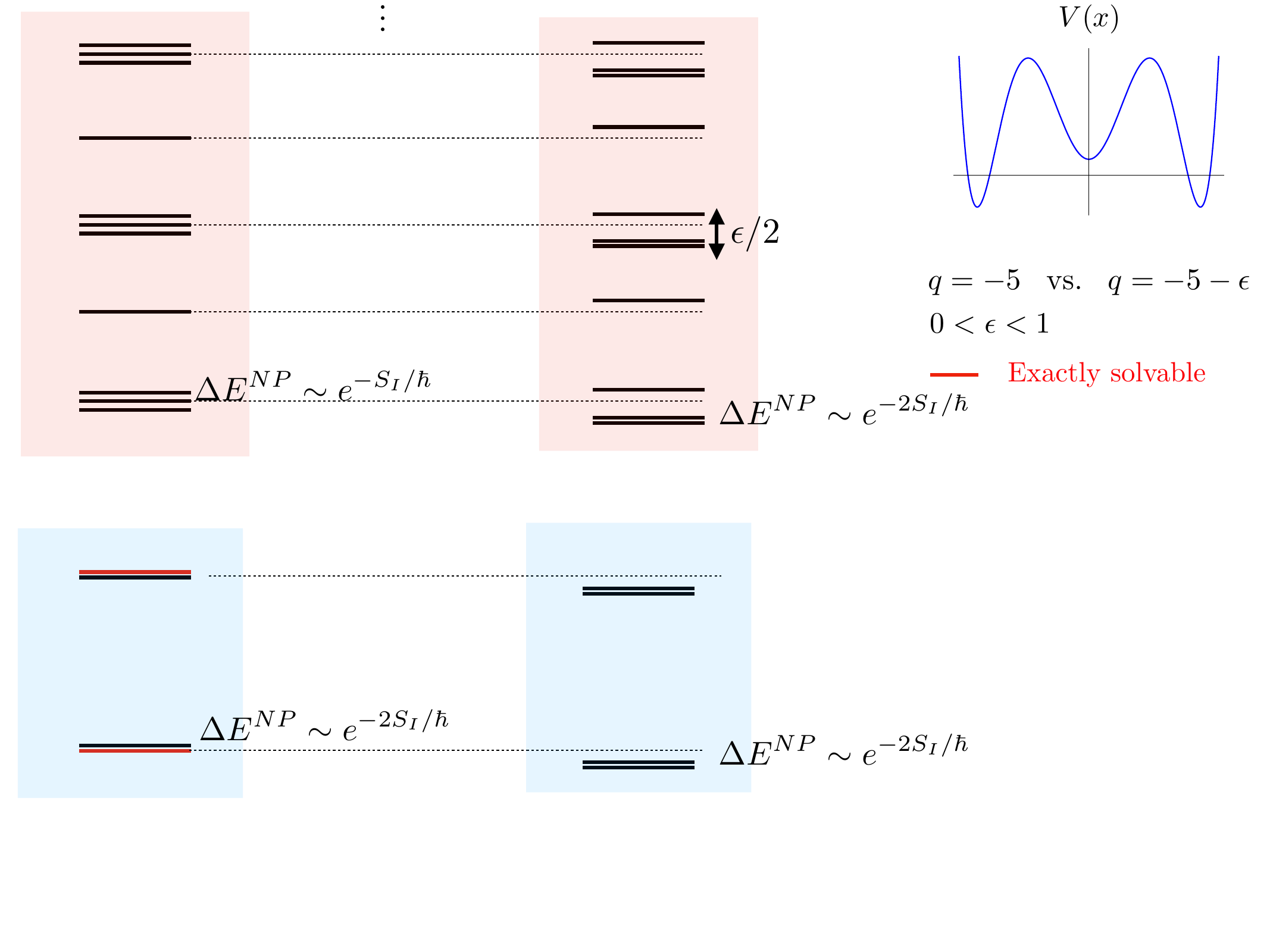}
      \vspace{-1.2cm}
    \caption{ Non-Perturbative spectrum of the triple-well potentials for $q=-5$ and $q=-5 - \epsilon$.  For $q=-5$, two states on  ${\cal H}_0$ (blue band) are exactly solvable, and they are separated from the perturbatively degenerate  state by a bion effect $e^{-2S_{\rm I}/\hbar}$. The triple degeneracy at the upper band is lifted by instanton effect. For $q=-5 - \epsilon$, where $\epsilon >   
    e^{-S_{\rm I}/\hbar}$, the perturbative  triplet  is split  to doublet (by $-\epsilon/3$)  and singlet 
    (by $\epsilon/6$). 
     All double degeneracies in the spectrum are lifted by bion effects  $e^{-2S_{\rm I}/\hbar}$. For $q \notin \mathbb Z$,  there are no single instanton effects in the spectrum.   For $\epsilon=1$, three states in the spectrum becomes exactly solvable, and this pattern continues.   
    }
     \label{fig:flow}
\end{figure}

We start with a reminder of a fact. The energy levels for inner and outer harmonic vacua  of quantum deformed triple-well systems is given in \eqref{harmonic}. Let us denote $p = \frac{2q+ 1}{3}$, and $q \in \mathbb R$.  
Perturbatively, there are two cases. 
\begin{itemize}
   \item For $q \in \mathbb Z$,  half of the  harmonic states in the central well are  aligned with the harmonic states in the outer wells,  as shown in Figs.\ref{fig:Pspec1} and \ref{fig:Pspec2}. This alignment holds to all orders in perturbation theory. We checked this  by obtaining perturbation theory from A-cycles and also, by studying perturbation theory by using Bender-Wu package \cite{Sulejmanpasic:2016fwr}. 
    \item For $q \notin \mathbb Z$,  the harmonic states in the central well are not aligned with the harmonic states in the outer wells. 
\end{itemize}
Therefore,  we expect  significant  differences in the spectral properties of the theories with $q \in \mathbb Z$ and $q \notin \mathbb Z$, and indeed, exact quantization condition reflects this.
Non-perturbatively, the outcome of the spectral properties of the  quantum deformed theory is depicted in Fig.\ref{fig:flow}.

\subsection{Unquantized  deformations or $q \notin \mathbb Z$ theories}
\label{sec:quantized}

 \noindent
 {\bf Classical potential:} First, note that the classical potential, corresponding to $p=0 \; (q=-1/2)$ is an example of  this class. In this case, the inner well energies  at leading harmonic order are 
    $E_{\rm in} = \frac{1}{2} \left( k_{\rm in} +  \frac{1}{2} \right) = \left \{ \frac{1}{4},\frac{3}{4}, \frac{5}{4}, \ldots  \right \} $ and outer well energies are $E_{\rm out} =  \left( k_{\rm out} +  \frac{1}{2} \right) = \left \{ \frac{1}{2},\frac{3}{2}, \frac{5}{2}, \ldots  \right \} $. The inner and outer levels are never aligned. Only the last one of Eqs. Eqs.(\ref{eq:TW_delta1_in_Aborel})-(\ref{eq:TW_delta3_out_Aborel}) apply. They imply that for the inner well, the leading non-perturbative contribution is bion, of order ${\frak B}_0$, and leads to mere shift of the energy. For the outer well, the leading NP contribution is again bion, extrapolating between the outer degenerate vacua, and  it leads to level splitting between parity even/odd states of order ${\frak B}_0$.

    The interesting point is that despite the fact that instantons are finite action saddles, there is no contribution to spectrum at the instanton level $\sqrt{{\frak B}_0}$. This comes about because the inner and outer wells are not  aligned. If one inspects the determinant of the fluctuation operator
    $F= -\frac{d^2}{d \tau^2} + V^{''}(x)|_{x_{\rm cl.} (\tau)}$ in an instanton background, it is infinite. As a result, the instanton amplitude is strictly zero.  On the other hand, the fluctuation operator in the background of bions is finite. 
    
     \noindent
 {\bf Other $q \notin \Z$ cases: } 
    Just like the classical case, for all $q \notin \Z$ cases, there are no    $\sqrt{{\frak B}_0} \sim e^{-S_{\rm I}/\hbar}$ contribution to energy spectrum.  Both energy shifts as well as level splittings (between degenerate states in the left/right outer vacua) are dictated by  
    $ {\frak B}_0 \sim e^{-2S_{\rm I}/\hbar}$.

     \subsection{Quantized  deformations or $q \in \mathbb Z$ theories}
    \label{sec:unquantized}
    \begin{figure}[t]
    \centering 
    \includegraphics[width=15cm]{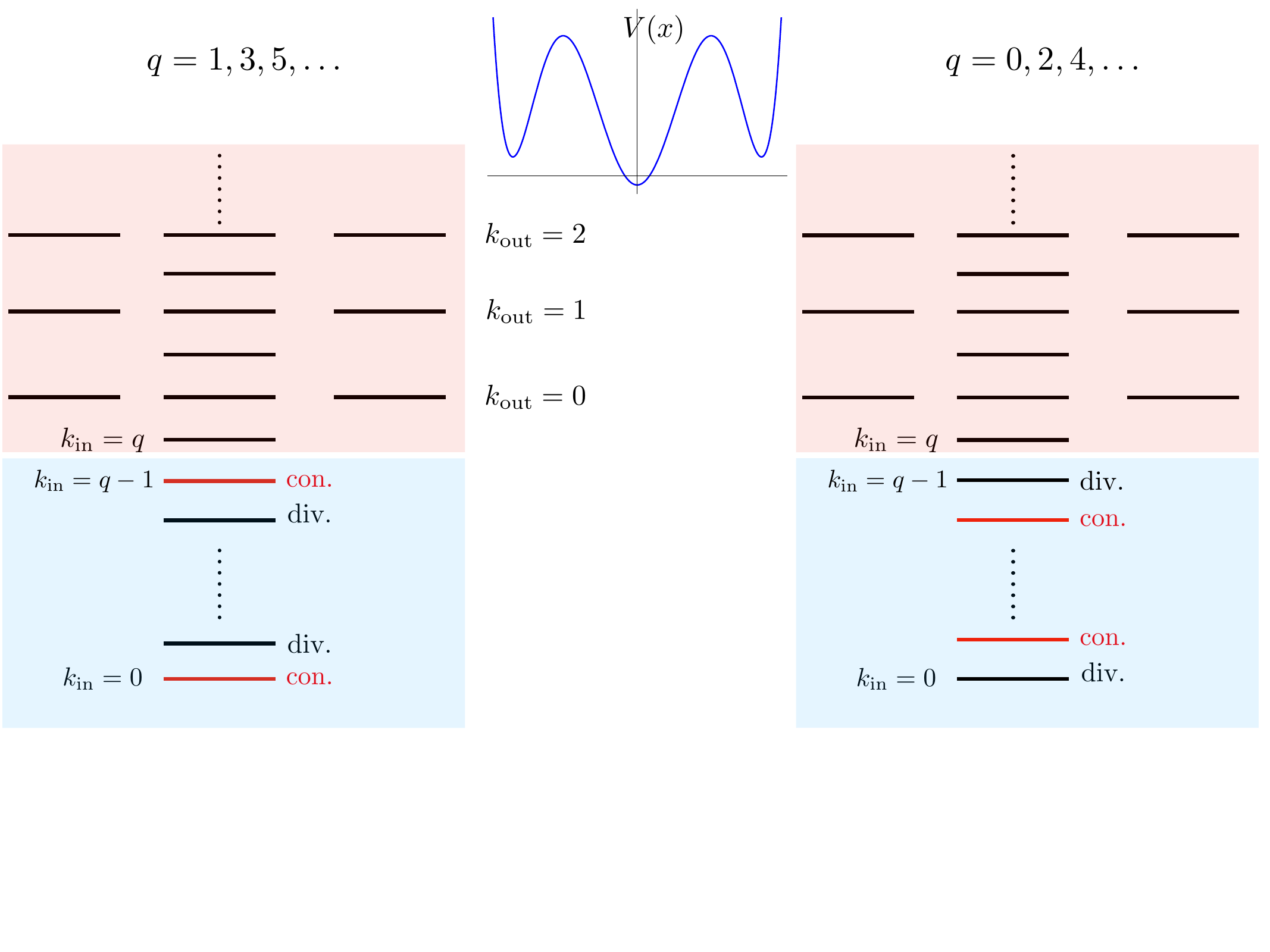}
      \vspace{-2cm}
    \caption{Perturbative spectrum of the triple-well potentials for non-negative integer values of $q$. The  three-fold degeneracies are exact to all orders in perturbation theory. 
    States in the blue shaded part have convergent/divergent pattern of perturbative expansion  and in the red  part, all perturbative expansion are asymptotic. Low levels only have bion non-perturbative contribution.  Higher singlet states only have bion contributions, and perturbatively treefold degenerate states have both instanton and bion contribution.  
    }
     \label{fig:Pspec1}
\end{figure}

\begin{figure}[t]
    \centering 
    \includegraphics[width=15cm]{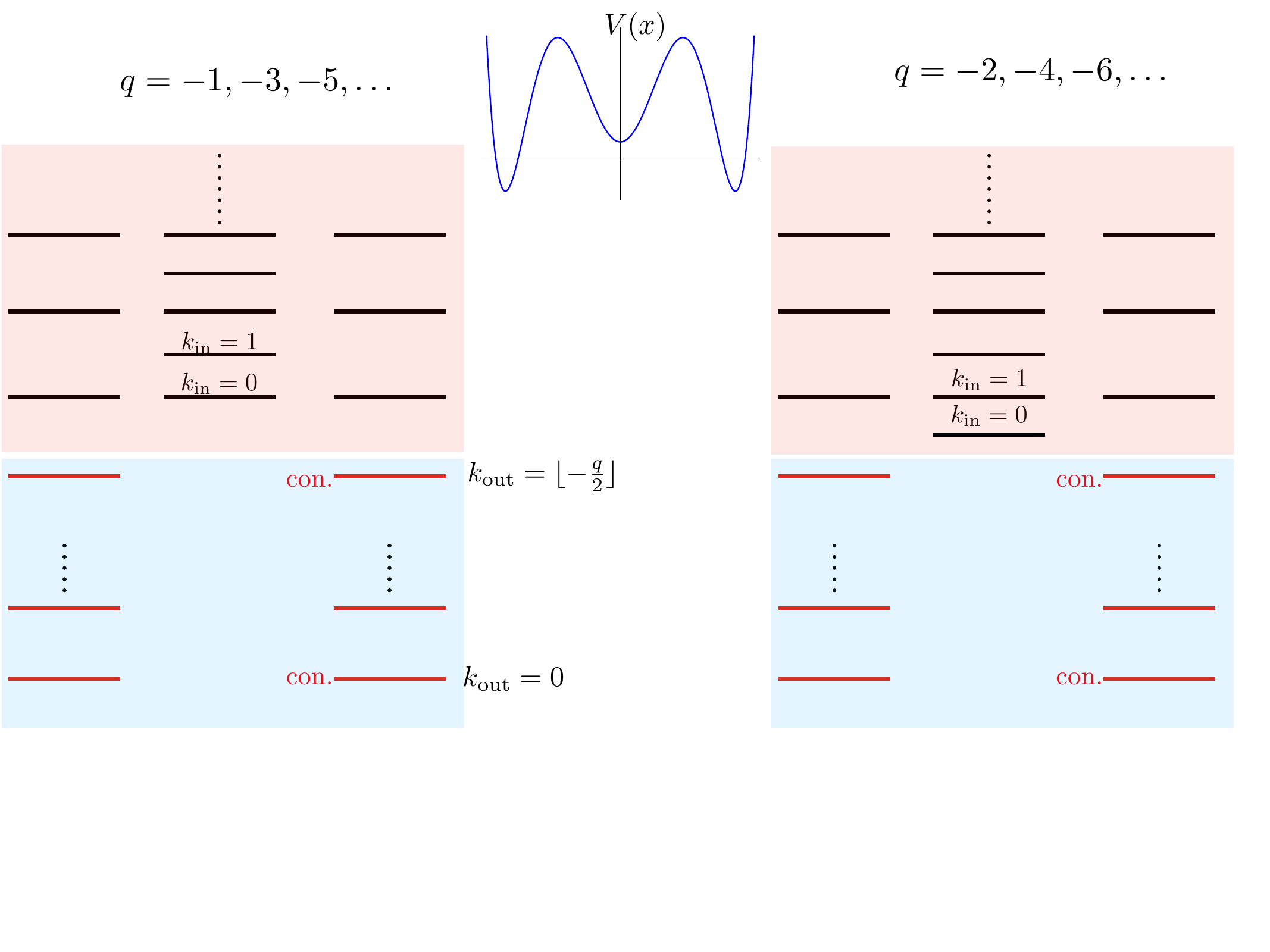}
      \vspace{-2cm}
    \caption{Perturbative spectrum of the triple-well potentials for negative integer values of $q$. The two-fold and three-fold degeneracies are exact to all orders in perturbation theory. 
    States in the blue shaded part have all convergent perturbative expansion  and in the red  part, all perturbative expansion are asymptotic. Low levels either have no NP contribution (and those states are exactly solvable QES states) or  bion NP contribution.  Higher singlet states  have bion contributions, and perturbatively treefold degenerate states have both instanton and bion contribution.  
    }
     \label{fig:Pspec2}
\end{figure}
  
     { \bf Special theories $ q \in \mathbb Z $:}  Triple well potentials for $q \in \Z$  are special. 
     For example, $q=1$ and $q=-2$ correspond to $p=\pm 1$ which are supersymmetric pairs. The 
     $q=2$ and $q=-3$ correspond to $p=\pm \frac{5}{3}$ which are QES systems. For integer $q$, 
     the states in the outer well become degenerate with every other state in the inner well, and this degeneracy is 
     an all orders statement in perturbation theory.  The perturbative spectrum for these special cases are shown in 
     Fig.~\ref{fig:Pspec1} and  Fig.~\ref{fig:Pspec2}.

     For $q=1,3, 5, \ldots$, calculating perturbation theory via either Bender-Wu package or using the perturbative $A$-cycle expansion
     yields an intriguing result. 
     Levels with  $k_{\rm in}=0, \ldots, q-1$  have a convergent/divergent 
     alternating pattern.  For   $k_{\rm in} \geq q$, states have divergent asymptotic perturbative expansion. 
     
   Inspecting   Eqs.(\ref{eq:TW_delta1_in_Aborel})-(\ref{eq:TW_delta3_in_Aborel}),  we indeed observe that due to the $\frac{1}{\Gamma(\frac{1}{2}+\frac{k_{\rm in}}{2}-\frac{q}{2})}$ factor, 
   instanton factor proportional to 
    $\sqrt{{\frak B}_0}$ disappears for  $k_{\rm in} \leq q-1$ and is present provided  $k_{\rm in} \geq q$ and $\varepsilon= (-1)^{q+1} $ states.  
    For $k_{\rm in}=q, q+2, \ldots$,  the single states in red shaded region, the instanton factor does not appear and complex bion contribution vanishes according to exact quantization condition.  Indeed, in the path integral formalism, we can actually show that the complex bion contribution appears as $ {\frak B}_0 e^{  \pm i \frac{\pi}{2}(q+ 1-k_{\rm in})}$ where the exponent is the hidden topological angle $\theta_{\rm HTA} $.  For $k_{\rm in } =q, q+2, \ldots $,  
    the real part of the complex bion contribution ${\rm Re}[{\frak B}_0 e^{  \pm i \frac{\pi}{2}(q+ 1-k_{\rm in})} ] =0$  is zero, there is a purely imaginary ambiguous  part ${\rm Im}[{\frak B}_0 e^{  \pm i \frac{\pi}{2}(q+ 1-k_{\rm in})} ] \neq 0$, and  
    that vanishes upon Borel-Ecalle resummation to cancel the ambiguity of perturbation theory.

    For $k_{\rm in}=q+1, q+3, \ldots$,  the   triple-degenerate state  states in red shaded region, the instanton factor  appears. Note that the degeneracy is valid to all orders in perturbation theory. Since $q$ is odd,  a harmonic central  state which is part of a triplet  is parity even. 
    We can discuss the linear combinations of the harmonic  states that appear as eigenstates of the Hamiltonian in simple terms.  At leading order in semi-classics,  
%The parity ${\cal P}$ in Eqs.(\ref{eq:TW_delta2_in_Bborel}) and (\ref{eq:TW_delta2_out_Bborel}) for the inner and outer energies can not be taken freely.
the transfer matrix defined by tunneling among each wavefunction for the basis $\psi:=(\psi_{\rm L},\psi_{\rm C},\psi_{\rm R})^{\top}$,
\be
\delta M: = 
\begin{pmatrix}
    0 &&  \delta_{\rm I} && 0 \\
    \delta_{\rm I}  && 0 &&  \delta_{\rm I} \\
    0 &&  \delta_{\rm I} && 0 
\end{pmatrix}, \qquad \delta_{\rm I}:=  e^{-S_{\rm I}/\hbar},
\ee
where $\psi_{\rm C}$ is the locally defined wavefunction around the central-vacuum, and $\psi_{\rm L/R}$ are the ones around the left/right outer-vacua.
The eigenvalues of $\delta M$ and their eigenvectors are given by
\be
&& {\rm Eigen}(\delta M)=\left\{   \pm \sqrt{2} e^{-S_{\rm I}/\hbar}, 0 \right\}, \\
&& 
v_{\pm}=(1,\pm\sqrt{2},1)^{\top}, \qquad v_{0}=(1,0,-1)^{\top}.
\ee
%\textcolor{red}{
%** added by SK **
The eigenfunction can be found by linear combination of $\psi_{\rm L,C,R}$ as $v^{\top} \psi$, and one can see the fact that the energy shift in (\ref{eq:TW_delta2_in_Aborel}) and (\ref{eq:TW_delta2_out_Aborel}) is caused by $\delta \psi_{\pm}:= v^{\top}_{\pm} \psi$.
Notice that $\psi_{\rm L}$ and $\psi_{\rm R}$ in $\delta \psi_{\pm}$ is symmetric, and thus the value of ${\cal P}$ is determined by the sign of $\pm \sqrt{2} \psi_{\rm C}$ in $\delta \psi_{\pm}$.
Unlike $\psi_{\rm L,C,R}$, $\delta \psi_{\pm}$ can not be defined around each local vacuum, so one can only say that ${\cal P}_{\rm in}=-{\cal P}_{\rm out}$ and that ${\cal P}=\pm 1$ arises by $\delta \psi_{\pm}$.
In the case of $\delta \psi_{0}:=v_{0}^{\top} \psi$, the $\psi_{\rm L}$ and $\psi_{\rm R}$ are antisymmetric each other and $\psi_{\rm C}$ does not enter into $\delta \psi_0$.
Therefore, only $\delta^{\varepsilon = (-1)^q}_{q \in {\mathbb Z}}$ in (\ref{eq:TW_delta1_out_Aborel}) has to be taken into account and $\delta^{\varepsilon = (-1)^q}_{k+q \in 2 {\mathbb Z}+1}$ in (\ref{eq:TW_delta1_in_Aborel}) is irrelevant to the zero-energy shift.
It is consistent with the fact that the triple-well has three-fold energy degeneracy.
In contrast to ${\cal P}$, $\varepsilon$ appears in the energy shift of the outer-vacua for any $q$ due to the parity symmetry of potential, i.e.  $V(x,\hbar)=V(-x,\hbar)$, and is also relevant to the energy splitting by bion contribution between the left and right vacua.
For $q \in {\mathbb Z}$, if $\varepsilon=(-1)^{q+1}$, then the wavefunctions of the left and right vacua reduce to the symmetric state giving the energy shift by the bion, otherwise the anti-symmetric state giving the zero-energy shift.
It is remarkable to state that all $\delta \psi_{\pm,0}$ can be unifiedly expressed through ${\cal P}$ and $\varepsilon$;
\be
&& \delta \psi_{\#} = \frac{1+ (-1)^{q+1} \varepsilon}{\sqrt{2}}{\cal P} \psi_{\rm C} + \psi_{\rm L}+ (-1)^{q+1} \varepsilon \psi_{\rm R}, \\
&& \# = 
\begin{cases}
+ & \mbox{if} \quad {\cal P}=+1 \quad \mbox{and} \quad \varepsilon=(-1)^{q+1} \\
- & \mbox{if} \quad {\cal P}=-1 \quad \mbox{and} \quad \varepsilon=(-1)^{q+1} \\
0 & \mbox{if} \quad \varepsilon=(-1)^{q} 
\end{cases}.
\ee
%}
As one can see from Eqs.(\ref{eq:TW_delta1_in_Aborel}) and (\ref{eq:TW_delta1_out_Aborel}), the non-perturbative contributions from the choice of $\varepsilon=(-1)^{q}$ with $q \in 2{\mathbb Z}+1$ and  $q \in {\mathbb Z}$ for the inner vacuum and outer-vacua in the triple-well, respectively, vanish by performing the Borel resummation.
These results are quite nontrivial in the sense that in general it is not quite enough to imagine the Borel resummed results only by taking look at the lower orders and their discontinuity (or imaginary ambiguity) in before-Borel resummed forms in Eqs.(\ref{eq:TW_delta1_in_Bborel}) and (\ref{eq:TW_delta1_out_Bborel}).
Eqs.(\ref{eq:TW_delta1_in_Bborel}) and (\ref{eq:TW_delta1_out_Bborel}) indeed have a part without discontinuities in $O({\frak B}^2_0)$, but it also vanishes by the Borel resummation, which means that a resurgence relation such as the DDP formula has to be needed to consider a next-leading term or higher in general.
In this sense, for a given function, the condition that it asymptotically gives a Borel non-summable perturbative power series  is in general  \textit{not} a sufficient condition 
%\textit{nor} a necessary condition 
for the existence of higher transmonomials such as an exponentially small factors in the expansion. 
\section{Summary and Prospects}
%%%%%%%%%%%%%%%%%%%%%%%%%%%%%%%%%%%%%%%%%%%%%
In this work, we generalized  the exact-WKB analysis  to quantum deformed potentials, $V(x,\hbar) =  V_0(x) + \hbar p  V_1(x)$, 
given in \eqref{cq}, where $V_0(x)$ is classical and  $\hbar   V_1(x)$ is quantum deformation. 
This class of systems are not rare. They are as typical as classical system, but received lesser attention. They arise naturally once the quantum mechanics of  a bosonic field $x(\tau)$ is coupled to $N_f$ Grassmann valued field $\psi_i(\tau)$ or to topological Wess-Zumino terms for internal spin-$S$.  In this  class of theories, so far only 
${\cal N}=1$ supersymmetric quantum mechanics received wide spread attention. However, all these theories have remarkable spectral properties, along with quite interesting non-perturbative aspects.  In this work, we examined quantum deformed  double- and triple-well  quantum mechanics.
Below, we summarize  physically the most important  outcome of our analysis  and state some open issues. 

\noindent
{\bf Median resummation of exact quantization condition:}  Exact quantization condition is 
already discussed in literature, and here, we provided a streamlined derivation for the generic polynomial potentials.  We argued that a more useful form of the exact quantization condition is its median resummed form in Eqs.(\ref{MR_DW})(\ref{MR_TW}). The idea of median resummation is used earlier  only in the important 
Pham et.al. \cite{DDP2} to show the reality of energy spectrum, but not used as a  computational tool.  In our work, solving the median resummed form, we obtained the non-perturbative correction to energy levels. The beauty of this formalism is that  the resurgent cancellations that are usually taken into account  only a posteriori and order by order (eg. between NP ambiguity of perturbation theory vs. two-fold ambiguity of bions, and then next level etc) is already built into the median resummed equation. The spectrum directly comes out to be as in Borel-Ecalle resummed form. Furthermore, the formalism also reveals some  all order cancellations that are not easy to see in the standard approach.

%Perhaps, physically the most interesting outcome of our analysis is following. 
\noindent
{\bf Fading vs. Robust contributions in semi-classics:}
Assume, as it is the case in our examples, there are  instanton 
solutions  to the classical system with potential $V_0(x)= \half W'(x)^2$.  These are also exact classical solutions in quantum deformed  
system, as the   instanton equations $\dot x(\tau) = \pm  W'(x)$ do not care about $\hbar$ deformation. Yet, we learn that the instanton contribution at leading order in semi-classics $e^{-S_{\rm I}/\hbar}$  only takes place in special cases, and generically it is not there.  
Its contribution requires an exact alignment,   that harmonic states in two consecutive wells must be exactly degenerate to {\it all-orders in perturbation theory}.  
For double-well, this happens only when $p \in \Z$ as shown in Eq.\eqref{eq:deltaDW_pint} and Eq.\eqref{eq:deltaDW_pnonint}, and for triple well, it happens only when  $q \in \Z$, 
where $ p= \frac{2q+1}{3}$ as shown in Eqs.(\ref{eq:TW_delta1_in_Bborel})-(\ref{eq:TW_delta1_out_Bborel}).    If the quantum deformation term $p  \in  \mathbb R \setminus \Z$ for double-well and 
$q  \in  \mathbb R \setminus \Z$ for triple-well, instantons do not contribute to the energy  spectrum at leading order $e^{-S_{\rm I}/\hbar}$.   This is a natural conclusion of the exact-WKB formalism applied to quantum-deformed potentials, and solutions of exact quantization conditions.    It would be nice to show this in full detail in path integral formulation. In particular, the quantization of the deformation parameter arises from the condition that harmonic levels in two consecutive wells must be the same.  
%is there to guarantee that that the harmonic levels coincides at leading order in harmonic analysis. 
In path integral formulation, what must happen is that the fluctuation determinant  in front of instanton amplitude  must blow up if the consecutive states are not aligned, hence leading to the vanishing of instanton amplitude.  
Indeed, in a special case, this is observed to be the case \cite{Dunne:2020gtk}.  On the other hand,  the energy spectrum  has contributions of order  
 $e^{-S_{\rm B}/\hbar} \sim  e^{-2 S_{\rm I}/\hbar}$ regardless of the value of the deformation parameter. In this sense, the contribution of instantons 
 is fragile or fading, depends on precise alignment of states, yet the contribution of bions is robust.

 \noindent
{\bf HTA vs. Maslov index:}  Our construction does not imply that bion contributions never disappear. In fact, for triple-well, 
in the ${\cal N}=1$ SUSY case, there exists a supersymmetric ground state, hence, $E_0=0$ non-perturbatively. In this case, there are
two types of bions, 
contributing to ground state and first excited state.  The amplitudes of these two types of bions are identical, except a relative phase in 
between. If the relative phase is odd/even multiple $\pi$, it leads to    destructive/constructive  interference between them. 
In fact, for QES systems corresponding to $q=-2, -4, \ldots$ in Fig.\ref{fig:Pspec2}, there are 
$2 \lfloor  - \frac{q}{2} \rfloor$ states in the low end of the spectrum (denoted as ${\cal H}_0$ in \eqref{split}) for which perturbation theory is convergent.  Half of these states are algebraically solvable with no non-perturbative contribution, and the other half receive NP bion contribution.  

%For these sub-set of states, bion contributions vanishes due to hidden topological angle. 

%{\color{red} continue Tomorrow}

 \noindent
{\bf Resurgent structure  for quantum deformed systems: }  
The exact resurgent structure, where all the imaginary ambiguities are canceled between the perturbative and non-perturbative sectors, 
was obtained for quantum deformed double- and triple-well systems.  For 
$q \in \Z$ (TW) and $p \in \Z$ (DW), 
states in ${\cal H}_0$ realizes Cheshire cat resurgence, namely, perturbation theory converges and ambiguities disappear, while states in 
${\cal H}_1$ realized standard resurgence where ambiguities are cancelled between the perturbative and non-perturbative sectors. 
If $q \notin \Z$ (TW) and $p \notin \Z$ (DW),  all states in ${\cal H}_0 \oplus {\cal H}_1$ realize standard resurgent cancellations.

 \noindent
{\bf P-NP  relation for bions:} We proved by explicit computations  that the perturbative fluctuations around the 
non-perturbative bion sectors is determined constructively by the perturbation theory around perturbative vacua for quantum deformed double- and 
triple-well systems. 
Since the symmetric triple well system can be practically reduced to a genus-1 system, and quantum deformation does not alter this property, 
the P-NP  relations for both inner- and outer-energy are available,  see  \eqref{DU2} and  \eqref{DU3}, even when there is 
no alignment of  the energy levels between the outer and inner well. 
One of the open interesting questions in this category is whether one can obtain a P-NP relation connecting different saddles constructively 
for higher genus potentials. 

 \noindent
\begin{acknowledgements}
    The authors especially thank O. Morikawa and students of E-lab for fruitful discussion on quantization conditions for $S^{1}$ quantum mechanics.
	T.\ M. is supported by the Japan Society for the Promotion of Science (JSPS) Grant-in-Aid for Scientific Research (KAKENHI) Grant Numbers 18H01217 and 19K03817.
	S.~K. is supported by the Polish National Science Centre grant 2018/29/B/ST2/02457. 
	The original questions related to the present work were posed in ``RIMS-iTHEMS International Workshop on Resurgence Theory" 
    at RIKEN, Kobe in 2017. The authors are grateful to the organizers and participants of the workshop.
	M.~U. acknowledges support from U.S. Department of Energy, Office of Science, Office of Nuclear Physics under Award Number DE-FG02-03ER41260.
	The authors thank Yukawa Institute for Theoretical Physics at Kyoto University. Discussions during the YITP-RIKEN iTHEMS workshop YITP-T-20-03 on "Potential Toolkit to Attack non-perturbative Aspects of QFT -Resurgence and related topics-" were useful to complete this work.
\end{acknowledgements}

%%%%%%%%%%%%%%%%%%%%%%%%%%%%%%%%%%%%%%%%%%%%%
\appendix
%%%%%%%%%%%%%%%%%%%%%%%%%%%%%%%%%%%%%%%%%%%%%
%%%%%%%%%%%%%%%%%%%%%%%%%%%%%%%%%%%%%%%%%%%%%
\section{Mellin transform} \label{sec:Mellin}
%%%%%%%%%%%%%%%%%%%%%%%%%%%%%%%%%%%%%%%%%%%%%

%%%%%%%%%%%%%%%%%%%%%%%%%%%%%%%%%%%%%%%%%%%%%
%\subsection{Basics}
%%%%%%%%%%%%%%%%%%%%%%%%%%%%%%%%%%%%%%%%%%%%%
Firstly, we briefly review the essence of the Mellin transform\cite{Zinn-Justin:2004qzw,DP1}.
The main point is that we transform a function $f(E)$ to a function of $s$ as $f(E) \rightarrow M(s)$ through an integration equipped with a measure $E^{-s-1}$.
$f(E)$ can be reproduced from $M(s)$ by an inverse transform:
\be
&&  M(s) = \int^{L}_0 E^{-s-1} f(E) dE, \qquad   f(E) = \frac{1}{2 \pi i} \int^{c+i \infty}_{c-i \infty} M(s) E^{s} ds.
\ee
For example, if $f(E)$ is a monomial of $E$ and $\log E$,\footnote{The value of $L$ depends on the functinal form of $f(E)$ and is taken such that the upper bound of integration does not contribute. For the analysis of $\log\bar{B}({\cal E},\hbar)$, we take $L=\infty$.}
\be
&& f(E)=E^n \quad (n>0): \nl
&& \qquad \qquad M(s) = \int^{L}_0 E^{n-s-1}  dE = \frac{L^{n-s}}{n-s} = \sum_{\ell=0}^{\infty} \frac{(\log L)^\ell}{\ell! (n-s)^{-\ell+1}} \nl
&& \qquad \qquad \qquad \ \, \rightarrow \quad \frac{1}{n-s} \quad \mbox{for \ $L=1$}, \\ \nl
%&&  \frac{1}{2 \pi i} \int^{c+i \infty}_{c-i \infty} M(s) E^{s} ds =  E^n.
%&& \quad \quad \ \ = \frac{1}{n-s} + \log L^{n-s} + O(n-s), \\ \nl
&& f(E)=E^n (\log E)^m \quad (n>0, m \in {\mathbb N}); \nl
&& \qquad \qquad M(s) = \int^{L}_0 E^{n-s-1} (\log E)^m  dE
%&& \qquad \ \, = \frac{L^{n-s}}{n-s} (\log L)^m  - \int^{L}_0 \frac{m}{n-s} E^{n-s-1} (\log E)^{m-1}  dE \nl
=   \sum_{\ell=0}^{\infty}  \sum_{k=0}^{m} \frac{m!}{k!} \frac{(-1)^{k+m}(\log L)^{\ell}}{(\ell-k)! (n-s)^{-\ell+m+1}} \nl
&& \qquad \qquad \qquad \ \, \rightarrow \quad \frac{(-1)^{m} m!}{(n-s)^{m+1}} \quad \mbox{for \ $L=1$}.
\ee
By looking at the location of poles and its order in $M(s)$, one can easily reproduce $f(E)$.

%%%%%%%%%%%%%%%%%%%%%%%%%%%%%%%%%%%%%%%%%%%%%
\subsection{Solutions of $M_n[\log \bar{B}](s)$}
%%%%%%%%%%%%%%%%%%%%%%%%%%%%%%%%%%%%%%%%%%%%%
In order to consider the reduction of a $B$-cycle from the Airy-type to the DW-type, one needs to consider the Mellin transform defined as
\be
M [\log \bar{B}](s,\hbar):= \oint_{\gamma_{\frak B}} dx  \, M[S^{\rm Airy}_{\rm odd}](x,s,\hbar) = \sum_{n=-1}^{\infty} \hbar^{n} M_n[\log \bar{B}](s), \label{eq:B_M}
\ee
where ${\bar B}({\cal E},\hbar)$ is a $B$-cycle given by the Airy-type as
\be
\log \bar{B}({\cal E},\hbar) := \oint_{\gamma_{\frak B}} dx \, S^{\rm Airy}_{\rm odd}(x,{\cal E},\hbar),
\ee
and ${\cal E}$ is the energy.
In addition, the integration contour of the $B$-cycle, $\gamma_{\frak B}$, is defined to be the same one as the case of the DW-type  by taking ${\cal E} \rightarrow 0_+$. 

We would write the solution of $M[\log \bar{B}](s,\hbar)$ in Eq.(\ref{eq:B_M}) below.

%%%%%%%%%%%%%%%%%%%%%%%%%%%%%%%%%%%%%%%%%%%%%
\subsubsection{Tilted double-well}
%%%%%%%%%%%%%%%%%%%%%%%%%%%%%%%%%%%%%%%%%%%%%
\begin{small}
\be
&& M_{-1}[\log \bar{B}](s) =
-\frac{2^{3 s-2}\sqrt{\pi} \Gamma (1-s) \Gamma (-s)}{\Gamma \left(-2s+\frac{5}{2}\right)}, \\
&& M_{0}[\log \bar{B}](s) = 0, \\
&& M_1[\log \bar{B}](s) = \frac{8^s \sqrt{\pi} \Gamma (-s-1) \Gamma (-s)}{ \Gamma \left(-2 s-\frac{1}{2}\right)} \left(- p^2+ \frac{2 s}{3} +1 \right), \\
&& M_{2}[\log \bar{B}](s) = 0, \\
&& M_3[\log \bar{B}](s) = -\frac{2^{3 s+1}\sqrt{\pi} \Gamma (-s-3) \Gamma (-s)}{\Gamma \left(-2 s-\frac{7}{2}\right)}\left( p^4 -  \frac{4 p^2 s}{3} + 6 p^2 + \frac{28s^2}{45} + \frac{152s}{45} +5\right), \\
&& M_{4}[\log \bar{B}](s) = 0, \\
&& M_5[\log \bar{B}](s) = \frac{8^{s+1}\sqrt{\pi}  \Gamma (-s-5) \Gamma (-s)^s}{\Gamma \left(-2 s-\frac{13}{2}\right)}\nl
&& \qquad \qquad \qquad \quad \cdot \left( -\frac{p^6}{3}+\frac{2 p^4 s}{3}+5 p^4-\frac{28 p^2 s^2}{45}-\frac{64 p^2 s}{9}-\frac{67 p^2}{3}+\frac{248 s^3}{945}+\frac{988 s^2}{315}+\frac{11866 s}{945}+\frac{53}{3} \right). \nl
\ee
\end{small}
%%%%%%%%%%%%%%%%%%%%%%%%%%%%%%%%%%%%%%%%%%%%%
\subsubsection{Tilted triple-well}
%%%%%%%%%%%%%%%%%%%%%%%%%%%%%%%%%%%%%%%%%%%%%
\begin{small}
\be
&& M_{-1}[\log \bar{B}](s) = -\frac{3\ 2^{s-1} \Gamma (1-s) \Gamma (2-s) \Gamma (-s)}{\Gamma (4-3 s)}, \\
&& M_{0}[\log \bar{B}](s) = 0, \\
&& M_1[\log \bar{B}](s) = \frac{2^{s-3} \Gamma (-s-1) \Gamma (-s)^2}{\Gamma (-3 s-1)} \left(-3 p^2+2 s+3\right), \\
&& M_{2}[\log \bar{B}](s) = -\frac{2^{s-3} \Gamma (-s-2) \Gamma (-s-1) \Gamma (-s)}{\Gamma (-3 (s+1))}p \left(p^2-1\right), \\
&& M_3[\log \bar{B}](s) = \frac{2^{s-6} (s+3) \Gamma (-s-3)^2 \Gamma (-s)}{15 \Gamma (-3 (s+2))} \nl
&& \qquad \qquad \qquad \quad \cdot \left( 45 p^4-60 p^2 s-270 p^2+28 s^2+152 s+225 \right), \\
&& M_{4}[\log \bar{B}](s) = \frac{2^{s-5} \Gamma (-s-4)^2 \Gamma (-s)}{3 (3 s+8) \Gamma (-3 (s+3))} p \left(p^2-1\right) (s+4) \left(-3 p^2+2 s+15\right), \\
&& M_5[\log \bar{B}](s) = \frac{2^{s-8} \Gamma (-s-5)^2 \Gamma (-s)}{(3 s+10) (3 s+11) (3 s+13) (3 s+14) \Gamma (-3 (s+5))} \nl
&& \qquad \qquad \qquad \quad \cdot \left( p^6 s+\frac{23 p^6}{9}-2 p^4 s^2-21 p^4 s-\frac{385 p^4}{9}+\frac{28 p^2 s^3}{15}+\frac{404 p^2 s^2}{15} \right. \nl
&& \left. \qquad \qquad \qquad \qquad +131 p^2 s+\frac{1781 p^2}{9}-\frac{248 s^4}{315}-\frac{4772 s^3}{405}-\frac{62338 s^2}{945}-\frac{470413 s}{2835}-\frac{473}{3}
\right). 
\ee
\end{small}

%%%%%%%%%%%%%%%%%%%%%%%%%%%%%%%%%%%%%%%%%%%%%
\subsection{Derivation of $\bar{G}(E,\hbar)$}
%%%%%%%%%%%%%%%%%%%%%%%%%%%%%%%%%%%%%%%%%%%%%

%%%%%%%%%%%%%%%%%%%%%%%%%%%%%%%%%%%%%%%%%%%%%
\subsubsection{Tilted double-well} \label{sec:derivation_G_DW}
%%%%%%%%%%%%%%%%%%%%%%%%%%%%%%%%%%%%%%%%%%%%%

We consider how to calculate $\bar{G}(E,\hbar)$ in $\bar{\frak B}(E,\hbar)$ below.
We begin with the Airy-type Schr\"{o}dinger equation with the energy ${\cal E}$.
The $B$-cycle can be calculated as through the Mellin transform and is given by 
\be
\log \bar{B}({\cal E},\hbar) &:=&\oint_{\gamma_{\frak B}} dx \, S^{\rm Airy}_{\rm odd}(x,{\cal E},\hbar) \nl
&=& -\frac{S_{\rm B}}{\hbar} + \sum_{n=0}^\infty \sum_{\ell=-n+1}^{\infty} {\cal E}^{\ell} \hbar^{n-1}  \left( \alpha_{n,\ell} + \beta_{n,\ell}   \log {\cal E} \right).
\ee
The first few non-zero coefficients are computed as
\be
&& \alpha_{0,1} = 2 + 2\log 2, \quad \alpha_{0,2} = -17+6 \log 2 , \quad \alpha_{0,3} = - 236 + 70 \log 2 , \nl
&& \beta_{0,1} = -2, \quad \beta_{0,2} = -6, \quad \beta_{0,3} = -70, \nl
&& \alpha_{2,-1} = \frac{1}{12}-\frac{p^2}{4}, \quad \alpha_{2,0} = \frac{3 p^2}{2}-\frac{p^2}{2} \log 2 -\frac{11}{6}+\frac{1}{2} \log 2, \nl
&& \alpha_{2,1} =  \frac{109 p^2}{4}-\frac{15 p^2}{2} \log 2 -\frac{605}{12}+\frac{25}{2} \log 2, \nl
&& \beta_{2,-1} = 0, \quad \beta_{2,0} = \frac{p^2}{2}-\frac{1}{2}, \quad \beta_{2,1} = \frac{15 p^2}{2}-\frac{25}{2}, \nn
\ee
and one finds $S_{\rm B}=\frac{1}{3}$.
Replacing the energy with ${\cal E}=E\hbar$ gives
\be
\log \bar{B}(E \hbar,\hbar) &=& - \frac{S_{\rm B}}{\hbar} + \sum_{n=0}^\infty \sum_{\ell = -n+1}^\infty E^{\ell} \hbar^{n+\ell-1}  \left[ \alpha_{n,\ell}  + \beta_{n,\ell} \log (E\hbar) \right] \nl
&=& - \frac{S_{\rm B}}{\hbar} + \sum_{n=0}^\infty \sum_{\ell = 0}^\infty E ^{n-\ell+1} \hbar^{n}  \left[ \alpha_{\ell,n-\ell+1} + \beta_{\ell,n-\ell+1}   \log (E\hbar) \right]. \label{eq:B_Ai}
\ee
In order to obtain $\bar{G}$ from Eq.(\ref{eq:B_Ai}), we rewrite Eq.(\ref{eq:B_DW}) as
\be
\log \bar{\frak B}(E,\hbar) &=& -\bar{G}(E,\hbar)  - \sum_{s \in \{-1,+1\}} \log \Gamma \left(\frac{1+sp}{2}-\bar{F}(E,\hbar) \right) \nl
&& + \log (2 \pi) + 2 \bar{F}(E,\hbar) \log  \left( \frac{\hbar}{2} \right), \label{eq:B_DW2}
\ee
and take the asymptotic expansion firstly for $E \rightarrow +\infty$, and then  for $\hbar \rightarrow 0_+$.
Since the asymptotic expansion of the $\Gamma$ function gives
\be
&& \log  \Gamma(z + p) = \left( z + p - \frac{1}{2}\right) \log z - z + \frac{1}{2} \log (2 \pi) 
 + \sum_{k=1}^{\infty} \frac{(-1)^{k+1} B_{k+1}(p)}{k(k+1)} z^{-k}, \\
&& B_n(p) = \sum_{k=0}^{n}
\begin{pmatrix}
  n \\
  k
\end{pmatrix} B_k p^{n-k},
\ee
where $B_n(x)$ is the Bernoulli polynomial with the Bernoulli number $B_n=B_n(0)$,
it can be expressed by
\be
\log \Gamma\left( \frac{1 \pm p}{2} - \bar{F}(E,\hbar) \right)
&=& \log \Gamma\left( p_\pm - \bar{F}(E,\hbar) \right) \qquad \left (p_\pm :=\frac{1 \pm p}{2} \right) \nl
&=& \left( -\bar{F}(E,\hbar)  \pm \frac{p}{2} \right) \log (-\bar{F}(E,\hbar)) + \bar{F}(E,\hbar) \nl
&& +\frac{1}{2} \log (2 \pi) + \sum_{k=1}^{\infty} \frac{(-1)^{k+1} B_{k+1}(p_\pm)}{k(k+1)} (-\bar{F}(E,\hbar))^{-k}.
\ee
Thus,
\be
&& \sum_{s \in \{-1,+1\}} \log \Gamma \left(\frac{1+sp}{2}-\bar{F}(E,\hbar) \right) \nl
&=&  -2\bar{F}(E,\hbar)   \log (-\bar{F}(E,\hbar)) + 2\bar{F}(E,\hbar) \nl
&& + \log (2 \pi) + \sum_{k=1}^{\infty} \frac{(-1)^{k+1} (B_{k+1}(p_+) + B_{k+1}(p_-))}{k(k+1)} (-\bar{F}(E,\hbar))^{-k}.
\ee
As a result, Eq.(\ref{eq:B_DW2}) can be written down as
\be
\log \bar{\frak B}(E,\hbar) &=& -\bar{G}(E,\hbar) + 2 \bar{F}(E,\hbar) \log  \left( \frac{\hbar}{2} \right) \nl
&& +  2\bar{F}(E,\hbar)   \log (-\bar{F}(E,\hbar)) - 2\bar{F}(E,\hbar) \nl
&& + \sum_{k=1}^{\infty} \frac{(-1)^{k} (B_{k+1}(p_+) + B_{k+1}(p_-))}{k(k+1)} (-\bar{F}(E,\hbar))^{-k}. \label{eq:B_DW3}
\ee

Finally, we connect Eqs.(\ref{eq:B_Ai}) and (\ref{eq:B_DW3}) to each other.
By considering the asymptotic expansion in terms of $E$ for $\bar{F}(E,\hbar)$ as
\be
\log \left(-\bar{F}(E,\hbar) \right) 
&=& \log E - \sum_{n=1}^\infty \frac{1}{n} \left( \frac{\bar{F}_{> 0}(E,\hbar)}{E} \right)^n, \nl
(-\bar{F}(E,\hbar))^{-k} 
&=& \frac{1}{E^{k}} \left[ \sum_{\ell=0}^{\infty} \left( \frac{\bar{F}_{> 0}(E,\hbar)}{E} \right)^\ell \right]^k,
\ee
where $\bar{F}_0(E)=-E, \bar{F}_{> 0}(E,\hbar):= \bar{F}(E,\hbar) - \bar{F}_0(E)$,
it can be written as
\be
&& - \frac{S_{\rm B}}{\hbar} + \sum_{n=0}^\infty \sum_{\ell = 0}^\infty E^{n-\ell+1} \hbar^{n}  \left[ \alpha_{\ell,n-\ell+1}   + \beta_{\ell,n-\ell+1}  \log (E\hbar) \right] \nl
&=&
 -\bar{G}(E,\hbar)  +  2\bar{F}(E,\hbar) \left[ \log \frac{E \hbar}{2} - \sum_{n=1}^\infty \frac{1}{n} \left( \frac{\bar{F}_{> 0}(E,\hbar)}{E} \right)^n -1 \right] \nl % - 2\bar{F}(E,\hbar) \nl
&& + \sum_{k=1}^{\infty} \frac{(-1)^{k} (B_{k+1}(p_+) + B_{k+1}(p_-))}{k(k+1)} \frac{1}{E^{k}} \left[ \sum_{\ell=0}^{\infty} \left( \frac{\bar{F}_{> 0}(E,\hbar)}{E} \right)^\ell \right]^k.
\ee
We put the ansatz for $\bar{G}(E,\hbar)$ as
\be
\label{eq:NPfunc2}
\bar{G}(E,\hbar) =  \sum_{n=-1}^{\infty} \bar{G}_n(E)\hbar^{n}.
\ee
By comparing terms coupling with $\log (E\hbar)$ in both hand sides, one finds that ${\beta}_{n,\ell}$ is essentially identical to $\bar{F}(E,\hbar)$,
\be
2 \bar{F}(E,\hbar) =  \sum_{n=0}^\infty \sum_{\ell = 0}^\infty  \beta_{\ell,n-\ell+1} E ^{n-\ell+1} \hbar^{n}.
\ee
Therefore,
\be
\label{eq:NPfunc3}
 \bar{G}(E,\hbar) &=& \frac{S_{\rm B}}{\hbar} - \sum_{n=0}^\infty \sum_{\ell = 0}^\infty \alpha_{\ell,n-\ell+1} E^{n-\ell+1} \hbar^{n} \nl
 && -  2\bar{F}(E,\hbar) \left[ 1 + \log 2 + \sum_{n=1}^\infty \frac{1}{n} \left( \frac{\bar{F}_{> 0}(E,\hbar)}{E} \right)^n  \right] \nl 
&& + \sum_{k=1}^{\infty} \frac{(-1)^{k} (B_{k+1}(p_+) + B_{k+1}(p_-))}{k(k+1)} \frac{1}{E^{k}} \left[ \sum_{\ell=0}^{\infty} \left( \frac{\bar{F}_{> 0}(E,\hbar)}{E} \right)^\ell \right]^k.
\ee

%%%%%%%%%%%%%%%%%%%%%%%%%%%%%%%%%%%%%%%%%%%%%
\subsubsection{Tilted triple-well} \label{sec:derivation_G_TW}
%%%%%%%%%%%%%%%%%%%%%%%%%%%%%%%%%%%%%%%%%%%%%

In the similar way to the double-well, $\bar{G}(E,\hbar)$ is available by using $B$-cycle given by the Airy-type and the Mellin transform.
It is given by
\be
\log \bar{B}(E \hbar,\hbar) &=& - \frac{S_{\rm B}}{\hbar} + \sum_{n=0}^\infty \sum_{\ell = -n+1}^\infty E^{\ell} \hbar^{n+\ell-1}  \left[ \alpha_{n,\ell}  + \beta_{n,\ell} \log (E\hbar) \right] \nl
%&=& - \frac{S_{\rm B}}{\hbar} + \sum_{n=0}^\infty \sum_{\ell = 0}^\infty \hbar^{\ell}  \left[ \alpha_{n,\ell-n+1} E^{\ell-n+1} + \beta_{n,\ell-n+1} E ^{\ell-n+1} \log (E\hbar) \right] \nl
&=& - \frac{S_{\rm B}}{\hbar} + \sum_{n=0}^\infty \sum_{\ell = 0}^\infty E ^{n-\ell+1} \hbar^{n}  \left[ \alpha_{\ell,n-\ell+1} + \beta_{\ell,n-\ell+1}   \log (E\hbar) \right]. \label{eq:B_Ai_T}
\ee
where $S_{\rm B} = \frac{1}{4}$.
We rewrite Eq.(\ref{eq:B_TDW}) as
\be
\log \bar{\frak B}(E,\hbar) &=& -\bar{G}(E,\hbar)  -  \log \Gamma \left(\frac{1-p}{2}-\bar{F}(E,\hbar) \right)  -  \log \Gamma \left(\frac{1+p}{2}-2\bar{F}(E,\hbar) \right) \nl
&& + \log (2 \pi) +  \left( \bar{F}(E,\hbar) + \frac{p}{2} \right) \log 2 + 3 \bar{F}(E,\hbar) \log \hbar. \label{eq:B_TDW2}
\ee
Since the gamma function can be written by
\be
&&  \log \Gamma \left(\frac{1-p}{2}-\bar{F}(E,\hbar) \right) + \log \Gamma \left(\frac{1+p}{2}-2\bar{F}(E,\hbar) \right) \nl
&=& -3\bar{F}(E,\hbar)  \log (-\bar{F}(E,\hbar)) + \left( -2\bar{F}(E,\hbar)  + \frac{p}{2} \right) \log 2 + 3\bar{F}(E,\hbar) \nl
&& + \log (2 \pi) + \sum_{k=1}^{\infty} \frac{(-1)^{k+1}}{k(k+1)} \left( B_{k+1}(p_-) (-\bar{F}(E,\hbar))^{-k} + B_{k+1}(p_+) (-2\bar{F}(E,\hbar))^{-k} \right), \nl
\ee
one finds that Eq.(\ref{eq:B_TDW2}) can be written down as
\be
\log \bar{\frak B}(E,\hbar) &=&
-\bar{G}(E,\hbar)   +  3\bar{F}(E,\hbar)  \left[  \log (-2\bar{F}(E,\hbar)\hbar)  - 1 \right] \nl
&& + \sum_{k=1}^{\infty} \frac{(-1)^{k}}{k(k+1)} \left( B_{k+1}(p_-) (-\bar{F}(E,\hbar))^{-k} + B_{k+1}(p_+) (-2\bar{F}(E,\hbar))^{-k} \right). \nl
\label{eq:B_TDW3}
\ee
By using the symbol as $\bar{F}_0(E)=-E, \bar{F}_{> 0}(E,\hbar):= \bar{F}(E,\hbar) - \bar{F}_0(E)$, it can be expressed as
\be
&& - \frac{S_{\rm B}}{\hbar} + \sum_{n=0}^\infty \sum_{\ell = 0}^\infty E^{n-\ell+1} \hbar^{n}  \left[ \alpha_{\ell,n-\ell+1}   + \beta_{\ell,n-\ell+1}  \log (E\hbar) \right] \nl
&=&
 -\bar{G}(E,\hbar)  +  3\bar{F}(E,\hbar) \left[ \log (2E \hbar) - \sum_{n=1}^\infty \frac{1}{n} \left( \frac{\bar{F}_{> 0}(E,\hbar)}{E} \right)^n -1 \right] \nl 
 && + \sum_{k=1}^{\infty} \frac{(-1)^{k}}{k(k+1)} \left( \frac{B_{k+1}(p_-)}{E^{k}} + \frac{B_{k+1}(p_+)}{(2E)^{k}} \right) \left[ \sum_{\ell=0}^{\infty} \left( \frac{\bar{F}_{> 0}(E,\hbar)}{E} \right)^\ell \right]^k.
\ee
We put the ansatz for $\bar{G}(E,\hbar)$ as
\be
\bar{G}(E,\hbar) =  \sum_{n=-1}^{\infty} \bar{G}_n(E)\hbar^{n}.
\ee
By comparing terms coupling with $\log (E\hbar)$ in both hand sides, one finds that ${\beta}_{n,\ell}$ is essentially identical to $\bar{F}(E,\hbar)$,
\be
3 \bar{F}(E,\hbar) =  \sum_{n=0}^\infty \sum_{\ell = 0}^\infty  \beta_{\ell,n-\ell+1} E ^{n-\ell+1} \hbar^{n}.
\ee
Therefore,
\be
\bar{G}(E,\hbar) &=& \frac{S_{\rm B}}{\hbar} - \sum_{n=0}^\infty \sum_{\ell = 0}^\infty \alpha_{\ell,n-\ell+1} E^{n-\ell+1} \hbar^{n}  \nl %+ \left( 2\bar{F}(E,\hbar)  - \frac{p}{2} \right) \log 2 \nl 
 && -  3\bar{F}(E,\hbar) \left[ 1 - \log 2 + \sum_{n=1}^\infty \frac{1}{n} \left( \frac{\bar{F}_{> 0}(E,\hbar)}{E} \right)^n  \right] \nl 
 && + \sum_{k=1}^{\infty} \frac{(-1)^{k}}{k(k+1)} \left( B_{k+1}(p_-) + \frac{B_{k+1}(p_+)}{2^{k}} \right) \frac{1}{E^k}\left[ \sum_{\ell=0}^{\infty} \left( \frac{\bar{F}_{> 0}(E,\hbar)}{E} \right)^\ell \right]^k.
\ee

\bibliographystyle{utphys}
\bibliography{EWKB.bib}

\end{document}